    \documentclass[fleqn,usenatbib,useAMS]{mnras}

    \usepackage{graphicx}    
    \usepackage{amsmath}    
    \usepackage{amssymb}    
		
    \usepackage{newtxtext,newtxmath}
		
		\usepackage{color}
    \newcommand{\Msun}{\mbox{$M_{\odot}$}}

		\title[A two-stage outflow in NGC\,1068]{A two-stage outflow in NGC\,1068}
    \author[D. May et al.]{
    D. May $^{1,2}$\thanks{E-mail: dmay@usp.br},
    J.E. Steiner$^{2}$
    \\
    $^{1}$Laborat\'orio Nacional de Astrof\'isica/MCTI, 37530-000, Itajub\'a, MG, Brazil\\
		$^{2}$Instituto de Astronomia, Geof\'isica e Ci\^encias Atmosf\'ericas, Universidade de S\~ao Paulo, 05508-090, S\~ao Paulo, SP, Brazil}
	
\begin{document}

    \date{Draft for internal use only}

    \pagerange{\pageref{firstpage}--\pageref{lastpage}} \pubyear{2017}

    \maketitle

    \label{firstpage}

    \begin{abstract}
    We present an analysis of the Seyfert\,2 galaxy NGC\,1068 of archive data from the SINFONI-VLT, in the $HK$ bands with pixel scales of 0.1 (data set 1 - DS1) and 0.025 (DS2) arcsec. The data are revisited with a sophisticated data treatment, such as the DAR correction, the application of a Butterworth filtering and deconvolution. The gain in the process is quantified by a significant improvement in the Strehl ratio and it shows that an unprecedented high spatial resolution is achieved.
		For DS1, a detailed study of the H$_{2}$, [Fe\,{\sc ii}] and [Si\,{\sc vi}] emission lines reveals a three-phase gas morphology: (1) the low-velocity [Fe\,{\sc ii}] emission representing the glowing wall of an hourglass structure, (2) the high-velocity compact blobs of low and high ionization emissions filling the hourglass volume, and (3) the distribution of H$_{2}$ molecular gas defines the thick and irregular walls of a bubble surrounding a cavity.
		Both the hourglass and the molecular emissions have an asymmetry caused by the fragmentation of the northeastern molecular wall, closest to the AGN, resulting in high-velocity compact blobs of ionized gas outside the bubble. The southwestern part of the bubble is excavated by the jet, where the blobs remain confined and are blown along the bubble's inner boundary. 	
		We propose that those blobs are driven by a hot ``secondary wind'' coming from the spot where the jet interacts and injects its energy in the molecular gas. The combination of a primary wind launched by the central source and the secondary wind is what we call a two-stage outflow.
		For DS2, we detected a [Si\,{\sc vi}] outflow nearly coplanar to the maser disc and orthogonal to the CO outflow found by a previous study. Such unexpected scenario is interpreted as the interaction between the central radiation field and a two-phase gas density torus.

    \end{abstract}


    \section{Introduction}
    \label{sec:intro}

    The Narrow Line Regions (NLRs) of Active Galactic Nuclei (AGNs), frequently associated with outflowing gas (\citealt{Veilleux91,Moore96,Fraquelli03,Muller11,Davies14792,Genzel14796}), are an indicator of a central ionizing source. Its overall conical shape is also the indirect evidence of a collimating structure surrounding the vicinity of a supermassive back hole (SMBH), a torus of gas and dust, from where the radiation/wind of the accretion disc escapes and ionizes/ejects the material into the interestellar medium (ISM). Depending on the orientation of our line of sight (LoS), the central source may be visible or obscured by a dusty torus and two main subclasses of AGNs arise: type 1 AGNs show broader permitted emission lines, vis-\`a-vis the forbidden lines; and type 2 present similar widths, with the broad components hidden by the torus. The uniqueness of the nature of both classes was decisively proven with the detection of broad lines through polarimetric observations of the Seyfert 2 galaxy NGC\,1068, by \citet{Antonucci85}, which proposed the Unified Model for AGN (see also \citealt{Antonucci93} and \citealt{Urry95}).    
    NGC\,1068 ((R)SA(rs)b) is the closest (14.4 Mpc, \citealt{Tully88}) and the brightest Seyfert 2 galaxy in the sky, providing the most detailed view for the study of the NLR dynamics and excitation. At this distance, 1 arcsec $\sim$70 pc (for z=0.003793 and $H_{o}$=75 km $s^{-1}$ Mpc$^{-1}$.

    NGC\,1068 has been a frequent target for a multi-wavelength approach; high resolution data (<0.1 arcsec) have been taken in the radio \citep{Wilson83,Gallimore96,Burillo16}, in optical spectroscopy and imaging \citep{Cecil90,Evans91} and in the infrared \citep{Raban09,Muller09,Gonzaga14}. These data were used to peer the influence of the jet on the ionized gas, and occasionally the role of the gas in bending the jet, as well as the nature of the torus and the extended emission of molecular gas and dust.

    \citet{Cecil90} reported intrinsic gas velocities deprojected to $\sim$1500 km $s^{-1}$ and discarded the origin of the clouds as being ejected from the broad line region (BLR). According to these authors, the clouds are likely to be accelerated by the AGN wind (0.1c), which interacts with the disc and is redirected above the galactic plane. The blown clouds are massive enough to remain stable along the NLR while they are lifted through this mechanism.
    Another issue concerns the highly extended clouds, which could reach velocities as the ones found by \citet{Cecil90}, and can be accelerated as far as $\sim$100 pc, when a turnover occurs and the clouds start to decelerate. Despite robust kinematic modeling (\citealt{Das06,Barbosa14}), such underlying acceleration mechanisms still remain an open question. A good reason for this are the poorly understood physical processes involved in the interaction of the NLR material with the intrinsic AGN outflow.

    The analysis of the jet-gas interaction in the NLR of Seyfert galaxies has long gathered a number of observational evidences indicating that it has a major role in producing emission lines up to hundreds of parsecs from the nucleus \citep{Capetti96a}.
    Shocks generated in the vicinity of the jet are known to expand and ionize the gas around the radio emission (\citealt{Capetti99,Axon98}), from where spatially distinct emissions, seen between the highly ionized lines and the jet, arise. However, this strong interaction accounts only for the kinematics near the jet. The combination of the energetics injected by a jet and the AGN wind is, in general, hard to unravel.
    \citet{Gallimore96} studied in detail the sub-arcsecond structure of each knot in the radio structure close to the nucleus, identifying a thermal emission attributed to the inner edge of a torus (knot \textbf{S1}), and the location where the jet is abruptly deflected by the interaction with the molecular gas (knot \textbf{C}), where a shock front is created. Based on the spectral indexes of VLBA (Very Long Baseline Array) and VLA/MERLIN, \citet{Gallimore04} suggested that the synchrotron particles undergo a reacceleration when they later interact with ISM material at knot \textbf{NE}.

    The location attributed to the nucleus, the \textbf{S1} component, coincides with the positionally resolved H$_{2}$O maser emission, which is not perpendicular to the jet before bending \citep{Gallimore96b,Gallimore96c,Gallimore01}, with a schematic diagram shown in \citet{Gallimore04}, in their Fig.8.
    The torus, imaged by ALMA (Atacama Large Millimiter/submillimiter Array) observations \citep{Burillo16}, is an extension of the detected maser emission and is also not perpendicular to the jet, or to the [O\,{\sc iii}] optical ionization cone seen with HST observations \citep{Evans91}. Mid-infrared observations performed by \citet{Bock00} show an emission correlated to the detected clumps in the optical emission and is coincident with the north-west wall of the ionization cone. In fact, the region mapped in the mid-infrared and optical depicts an unexpected place where highly ionized lines (as the coronal lines of Silicon \citealt{Mazzalay13a}), coexist with the molecular gas and dust \citep{Gonzaga14}.

    The relation between neutral and ionized gas is rather clarifying in the works of \citealt{Thaisa12,Mazzalay13a,Riffel14b,Barbosa14}, using NIFS (Near-Infrared Integral-Field Spectrograph) data. \citet{Mazzalay13a} focused on the coronal emission lines and their close relation to the ionization by the jet interaction, while the latter authors found a wider and a more extended ionization cone shown by the [Fe\,{\sc ii}] line, which in turn is better aligned with the maser emission. The partially ionized emitting regions of [Fe\,{\sc ii}] give, in fact, a more de-reddened view of the NLR structure when compared to the [O\,{\sc iii}] emission. The authors also favored the scenario where the molecular gas (associated with the youngest stars around the nucleus, of $\sim$30 Myr) is distributed in a ring-like structure in the galactic disc, with outflowing clouds near the nucleus.

    In light of this context, we re-analyzed data from SINFONI (Spectrograph for INtegral Field Observations in the Near Infrared) in the VLT (Very Large Telescope) in the $HK$-bands (partially analyzed by \citealt{Muller09}) and from the NIFS-Gemini North \citep{Thaisa12,Riffel14b,Barbosa14}, covering a total field of view (FoV) of $\sim$4.5 arcsec$^{2}$. After a meticulous procedure of image processing, we found more general and strict correlations between the emission lines, which leads us to propose a self-consistent scenario with an alternative accelerating mechanism for the NLR clouds.

    This paper is organized as follows. In Section~\ref{sec:2}, we describe the observations, data reductions and treatment, with attention to the deconvolution process; in Section~\ref{sec:results}, we present the results, starting with the [Fe\,{\sc ii}] line, continuing to other ionized emission lines and finishing with the molecular emission. The correlations and the origin of the NLR clouds are described in Section~\ref{sec:arc}, and we then present our discussion based on the new scenario for the NLR in Section~\ref{sec:discussion}. Finally, we draw the conclusions of this work in Section~\ref{sec:conclusions}.

    \section{Observations, reductions and data treatment}
    \label{sec:2}
    \subsection{Archival data from SINFONI-VLT}

    The data presented here were obtained with the adaptive-optics-assisted NIR (Near Infrared) integral field spectrograph SINFONI \citep{Bonnet04,Eisenhauer03}, on the Very Large Telescope (VLT) UT4. We used archive data with pixel scale of 0.05 arcsec~$\times$~0.1 arcsec (hereafter data set 1 - DS1), taken on the nights of 10-2005 with an individual exposure time of 50 sec, and with the pixel scale of 0.0125 arcsec~$\times$~0.025 arcsec (data set 2 - DS2), taken on 11-2006 with an individual exposure time of 200 sec. Both data sets were observed under programme 076.B-0098(A), with R.Davies as PI. The adaptive-optics system MACAO (Multi-Application Curvature Adaptive Optics) used as reference the galaxy nucleus in both observations. The resulting two-dimensional field has 64$\times$32 spatial pixels (spaxels), later arranged to 64$\times$64 spaxels, with FoV of 3.2 arcsec~$\times$~3.2 arcsec and 0.8 arcsec~$\times$~0.8 arcsec, respectively. Both data sets were taken in the $H+K$ bands, ranging from 1.45-2.45 $\mu$m, at a spectral resolution R$\sim$2400, corresponding to FWHM$\sim$125 km $s^{-1}$.

    The data were reduced using the EsoReflex software, which includes standard procedures such as bad pixel removal, correction for the linearity of the detector, flat-field correction, spatial rectification, wavelength calibration, sky subtraction and data cube reconstruction.
    For DS1, we performed the flux calibration using the G2V star HIP 5607, with magnitudes of 7.44 and 7.53 in the $K$ and $H$ bands, respectively; and for DS2 the G2V star HIP 17897, with magnitudes of 7.62 and 7.70. After the flux calibration, the data cubes were combined again in one single data cube comprising the $H$ and $K$ bands.

\begin{figure}
    \begin{minipage}{0.45\textwidth}
      \resizebox{\hsize}{!}{\includegraphics{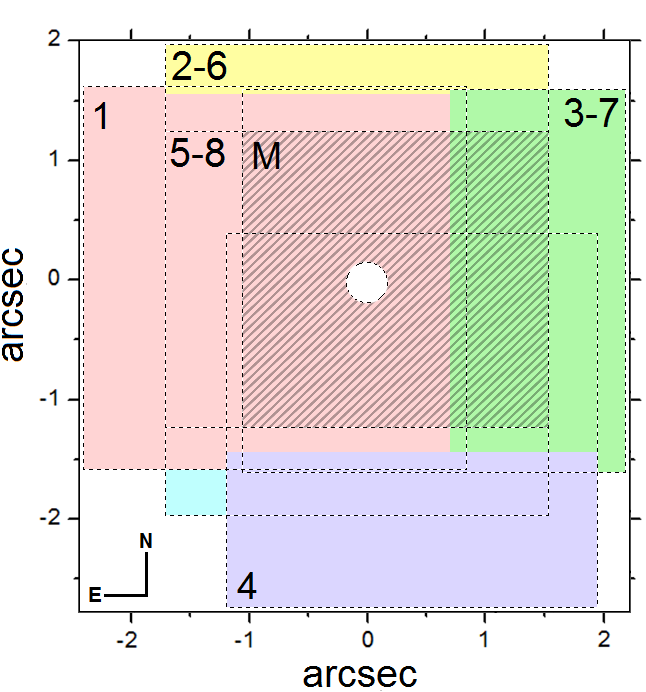}}
      \end{minipage}
     \begin{minipage}{0.45\textwidth}
      \resizebox{\hsize}{!}{\includegraphics{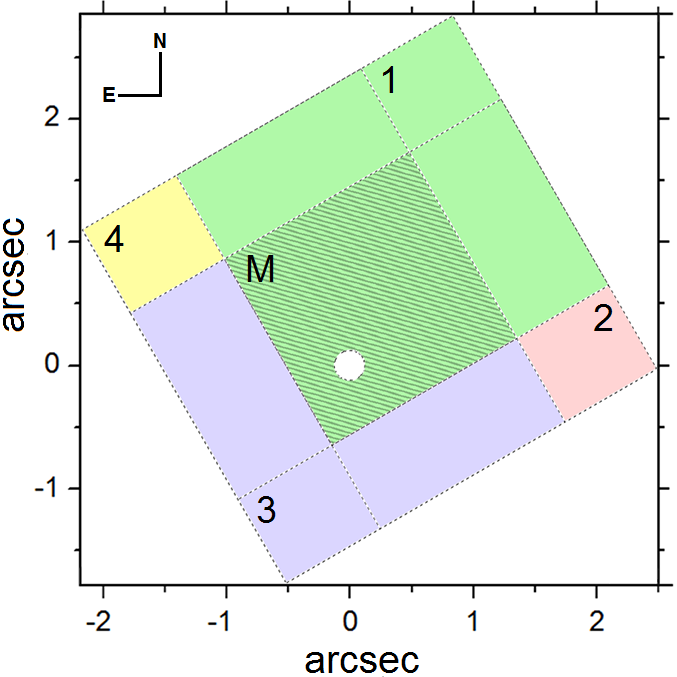}}
     \end{minipage}
     \caption{Top: spatial arrangement of the SINFONI data cubes from DS1, represented by dotted squares. The numbers 1 to 5 denote the ones used in the image mosaics, with the colours representing the area each one covers. The data cubes numbered 2, 3, 5, 6, 7 and 8 were combined in a single median data cube identified by the letter \textit{M} (dashed square). Bottom: the same for the NIFS data cubes, where all four exposures where used both for the image mosaics and to produce the median data cube. The open circles denote the masked region with radius of 0.18 arcsec, and the cross is centred on the respective nuclei, with north on top.}
    \label{fig:dataset}
    \end{figure}

    A total of eight observations from DS1 were used in this paper. Fig.~\ref{fig:dataset} (top panel) shows the arrangement of these data cubes. The first five individual exposures were combined in a mosaic, with each one corresponding to a specific area in the final image, as shown by the colours. Aiming to keep, at the same time, the largest FoV and high signal-to-noise (S/N), the data cubes numbered as 2, 3, 5, 6, 7 and 8 were combined, through a median, into a new data cube denoted by the letter \textit{M} (hereafter median data cube - MDC). The MDC, free from instrumental artifacts on the CCD and from cosmic rays (Sec.~\ref{sec:dt}), was used to apply PCA (Principal Component Analysis) tomography (see definition on \citealt{Steiner09} and applications of the method in \citealt{Menezes15}). Given the brightness of the nuclear region, the noise level severely reduce the contrast between the continuum and the emission lines in the centre, for this reason a central aperture of radius $\sim$0.18 arcsec was masked to highlight the structures of emission lines around the nucleus. The DS2 will complement the analysis of the central region with a smaller FoV.     

    For DS2, six data cubes were combined through a median, with a maximum dithering of $\sim$0.1 arcsec. Instead of single data cubes to produce the image mosaics, the resulting FoV is the one used to show all the emission line images, giving the small dithering.

    \subsection{Archival data from NIFS-Gemini North}

    With the aim of comparing with the DS1 SINFONI observations, specially after our image treatment procedure, we also reduced archive data from the NIFS-Gemini North Telescope \citep{McGregor03}. Both instruments operate with AO and have similar pixel sizes, but NIFS has twice the spectral resolution and, in this case, almost double the exposure time.

    NIFS operates with the adaptive optics module ALTAIR (ALtitude conjugate Adaptive optics for the InfraRed) and the data we used were obtained during the night of December 13, 2006, under program GN-2006B-C-9, with Storchi-Bergmann as PI. The pixel size of the instrument is 0.103 arcsec~$\times$~0.043 arcsec, between and along the slitlets, respectively, with a FoV of $\sim$3 arcsec~$\times$~3 arcsec. We used four observations, of 90 s each, taken with the $K_{long}$ setting of the $K$ grating ($2.11-2.52~\mu$m), with a spectral resolution of 5290 ($\approx 60$ km $s^{-1}$). 

    The data were reduced using tasks of the NIFS package in IRAF environment. The procedure included trimming the images, flat-fielding, sky subtraction, correcting for spatial distortions and wavelength calibration. We removed the telluric bands and calibrated the flux using the A0 star HIP 18863, with $K$-band magnitude of 6.86. 
    At the end of the data reduction process, the IFU (Integral Field Unit) data cubes were generated by the \texttt{nifcube} task, which re-sampled them to spaxels of $\sim$0.05 arcsec~$\times$~0.05 arcsec.

    \subsection{Data treatment}
\label{sec:dt}
    After the reduction, we performed a data treatment procedure described in more detail in \citet{Menezes14,Menezes15} and \citet{Dmay16}. The Differential Atmospheric Refraction (DAR), although known to have a small effect on the NIR, can displace the data cube centroids by up to 0.2 arcsec. Since this displacement is of the order of the spatial resolution obtained, such correction is relevant, for instance, when comparing different images of emission lines. Here, the DAR effect was empirically corrected in all data sets by fitting third degree polynomials through the spatial location of the centroids along the data cube, one for each spatial dimension, to maintain them at the same position in each wavelength. At the end of the correction, all the centroids, measured from the peak in the image of the stellar continuum, remained the same with a precision of 0.01 arcsec. This empirical approach is more precise to remove the DAR effect, since the theoretical curves (described in \citealt{Filippenko82}) do not reproduce the spatial displacements properly along the spectral axis. The exact cause of this difference is not known, but one hypothesis is that it could be caused by instrumental effects.

    The next step was the spatial re-sampling of the data, followed by a quadratic interpolation (Lsquadratic). This procedure, which preserves the surface flux of the images, aims to improve the visualization of the contours of the structures. But when followed by the deconvolution process, the re-sampling/interpolation leads to better resolution. The new sampling is 0.025 arcsec~$\times$~0.025 arcsec~($\sim$2 pc/pixel) for DS1 and 0.00625 arcsec~$\times$~0.00625 arcsec~($\sim$0.5 pc/pixel) for DS2. The later procedure introduces high spatial frequency components, which can be seen in the Fourier transform of the images. These components can be removed by the Butterworth spatial filtering in the frequency domain, without changing more than 2\% of the point spread function (PSF).

    \subsubsection{Deconvolution and emission line images}
    \label{sec:dec}

    Since the analyzed data are not seeing-limited, some description is needed of the impact of AO correction in both instruments.
    High resolution observations ($\sim$0.1 arcsec) with AO result in a complex PSF profile, not always described by a combination of distribution functions (i.e., Gaussian, Lorentzian, Moffat, Voigt and Airy profiles). In fact, if we take into account some of the PSF images, they also present asymmetric distributions, originated mainly by distinct pixel dimensions \citep{Dmay16} and/or instrumental features introduced by the AO correction. In this sense, a real PSF (extracted from the same observation) is highly desirable, although available only in a few cases.

\begin{figure*}
    \resizebox{0.80\hsize}{!}{\includegraphics{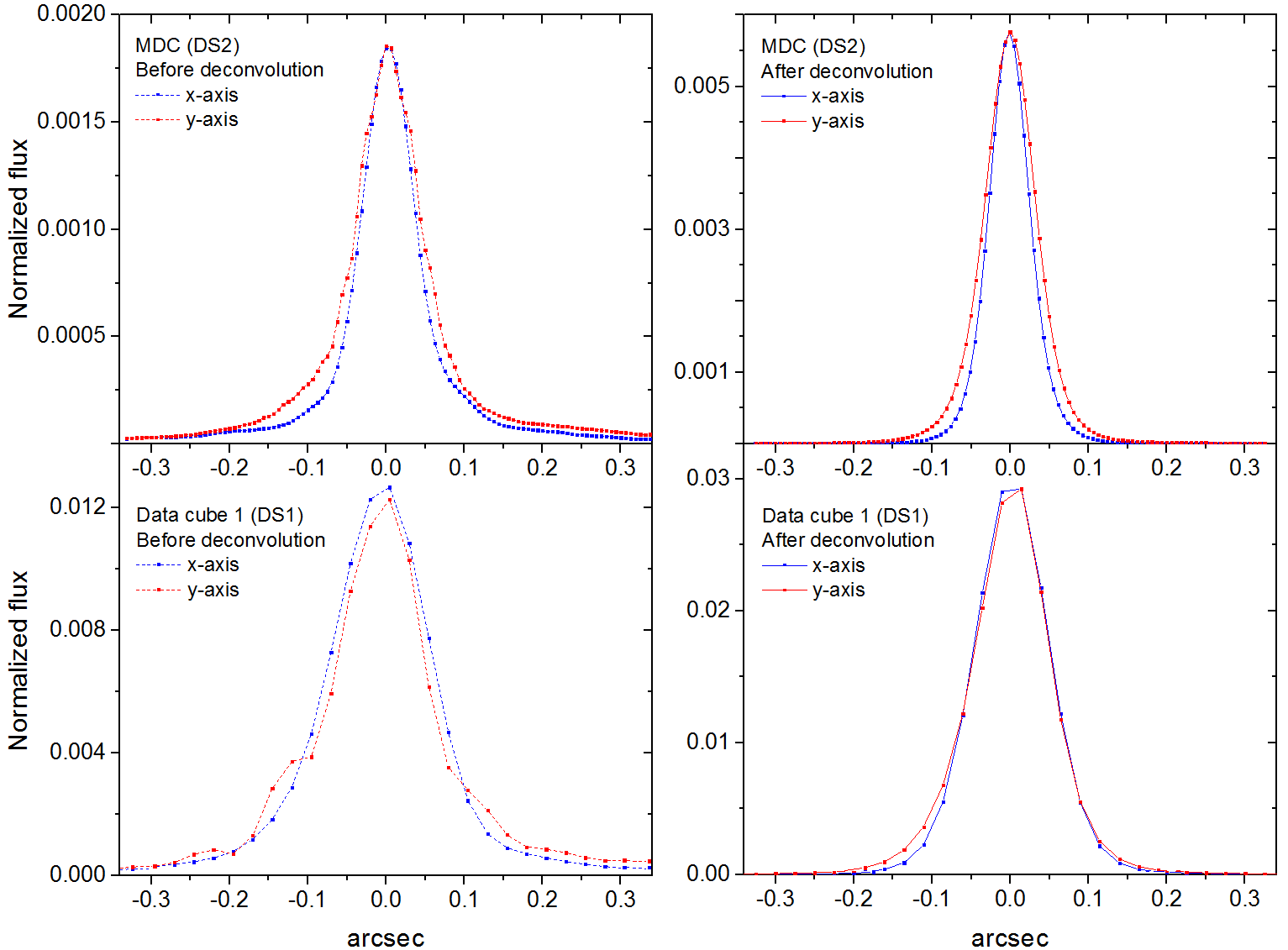}}
    \caption{Top: PSF profiles for the DS2 MDC along both dimensions, before (left) and after (right) deconvolution. Bottom: the same for the data cube 1 of DS1, the one which covers the larger area in the image mosaics (Fig.~\ref{fig:dataset}, top panel). These spatial profiles are representative of the others data cubes used here, and the FWHM of each one is shown in Table~\ref{table:psf}.}
     \label{fig:psf}
    \end{figure*}

    Luckily, NGC\,1068 shows a strong spectral signature of a nuclear and hot dust emission ($>$800 K) (unresolved in DS1, and at the resolution limit in DS2), which is seen in the $K$-band continuum, and thus may be tested as a PSF. The emission coming from this dusty structure is attributed to the inner part of a torus surrounding the AGN, and mainly emitted around the 2\,$\mu$m spectral range. In NGC\,1068, the size of the torus was measured as 1.35\,pc long \citep{Raban09}. This size corresponds to 0.02 arcsec, significantly less than the scale given by diffraction limit for the data we used, $\sim$0.06 arcsec, or three times lower. Thus, we can isolate the continuum image of the emission attributed to the inner part of the torus and use this as a PSF. The procedure to define the continuum was done fitting simultaneously the $H$ and $K$ bands with a cubic spline function. In Fig.~\ref{fig:psf} we plot the PSF profiles for the MDC of DS2 and the cube 1 of DS1, in both $x$ and $y$ directions, before and after deconvolution. The measured FWHM for all the data cubes used here is shown in Table~\ref{table:psf}.

    We draw attention to the small bump seen in the PSF profile along the $Y$ axis, for data cube 1, before applying the deconvolution (Fig.~\ref{fig:psf}, bottom left panel). This feature is not recurrent in all data cubes of DS1 and, more relevant, is not seen in the smaller pixel scale, of 25\,mas (DS2), a strong indication that it is not a real feature, but was probably introduced by the AO correction and/or the asymmetric pixel scale in this direction. As a matter of completeness, we performed an additional deconvolution with the corrected profile (without the bump), and the same result was obtained.

    We could not discard, however, a possible contribution of the featureless continuum to the PSF, or even an extra-nuclear contribution of thermal emission. To check if we are extracting, in fact, only the unresolved nuclear image of the continuum, we applied PCA tomography \citep{Steiner09} to the cube 1 of DS1 masking the emission and absorption lines. The first eigenvector obtained (99.7\% of variance) shows a very reddened continuum shape, corresponding to a tomogram with a point-like structure, with FWHM=0.16 arcsec.
    Thus, we deconvolved the MDC with the PSFs extracted both directly from the continuum and from tomogram 1, and obtained an image with the same FWHM=0.09 arcsec. Such result discard any relevant possibility of a contribution from the stellar emission to the PSF and confirms the validity of the use of the continuum image as a PSF.

    Since we had a reliable PSF, the next step was to decide which strategy to adopt to deconvolve the data cubes and have the best  choice to produce the emission line images. Since the PSF is wavelength dependent, the ideal is to obtain a PSF corresponding to each wavelength frame of the data. 
		
		Individual frames are not always free from cosmic rays and CCD defects, so the solution was to extract a median PSF from the continuum data cube taking into account only the spectral region of the emission lines of interest. Doing that, we obtained one PSF for a small wavelength interval, which is more representative than a PSF taken along the entire spectral dimension of the data cube. On the other hand, the data cube to be deconvolved by a PSF extracted in this way may also include cosmic rays, which had been removed by replacing the fluxes of the contaminated spectral pixels by the average value of the adjacent ones. This procedure was performed in all data cubes used in the image mosaics, in the spectral ranges of interest.         

    To better quantify the improvement obtained after deconvolution, we used the Strehl ratio, which measures the relative intensity peaks between the normalized PSF of the data and the diffraction-limited Airy profile applied to the configuration of the respective telescope. Thus, this ratio ranges from zero to one.

    \begin{table}
    \begin{center}
    \caption[strehl]
    {The FWHM of the PSFs extracted along the $x$ and $y$ axis, before (B) and after (A) deconvolution. The PSF profiles for the DS1 are shown in Fig.~\ref{fig:psf}.}
    \begin{tabular}{|c|c|c|c|c|}
    \hline \hline
    Data Cube & \multicolumn{4}{c}{PSF $K$-band FWHM (arcsec)} \\ \hline
    & \multicolumn{2}{c}{$x-axis$} & \multicolumn{2}{c}{$y-axis$} \\
    SINFONI & B & A & B & A \\ \hline
    1 & 0.13 & 0.097 & 0.14 & 0.093 \\
    2 & 0.13 & 0.098 & 0.12 & 0.092 \\
    3 & 0.13 & 0.093 & 0.12 & 0.091 \\
    4 & 0.14 & 0.11 & 0.12 & 0.107 \\
    MDC (DS1) & 0.16 & 0.11 & 0.16 & 0.13 \\
    MDC (DS2) & 0.08 & 0.05 & 0.10 & 0.07 \\ \hline
    \end{tabular}
    \label{table:psf}
    \end{center}
    \end{table}

    Table~\ref{table:strehl}~shows the results for the data cubes used in the image mosaics throughout this paper and is organized with the measurements made before and after deconvolution, in the $K$ and $H$ bands, both for the NIFS and SINFONI instruments. The last column presents the measured values for the seeing at the time of each observation and, despite the small sample presented here, there is no indication of the expected correlation between the seeing and the initial Strehl ratio (May D. \& Steiner J. E., in preparation).
    On average, the improvement in resolution for the $K$-band is 116\% for the SINFONI data cubes and 72\% for the NIFS, although their average seeing is almost the same and their final Strehl ratio for the $K$-band is only slightly different. For the SINFONI $H$-band, we still had a significant improvement of 65\% on average, but not in the case of NIFS data, which was only 8\%. In principle, the AO correction worked better for the VLT configuration, in this case. At this point, we justify the choice for using SINFONI data cubes, which reach higher spatial resolution. Basically, for our analysis, it is more important to resolve spatially the velocity components than to resolve velocity channels without knowing well their spatial distribution. Along this work we compare the mosaics for the coronal and molecular emissions, respectively, in both instruments to directly address the differences in spatial resolution and to check the consistency between the data after using the same methodology for treating data cubes.
    To check the validity of the deconvolution method in DS2, in Sec.~\ref{sec:ds2} we compare the smaller pixel scale with ALMA observations of the torus structure.

    \begin{table*}
    \begin{center}
    \caption[strehl]
    {The Strehl ratio for the SINFONI and NIFS data cubes used in the image mosaics shown and numbered in Fig.~\ref{fig:dataset}. The ratio is measured before (B) and after (A) deconvolution. When not shown, the errors worth $\pm$0.002, or less.}
    \begin{tabular}{|c|c|c|c|c|c|}
    \hline \hline
    Data Cube & \multicolumn{4}{c}{Strehl} & Seeing \\ \hline
    & \multicolumn{2}{l}{$K$-band $\lambda$21218 \AA} & \multicolumn{2}{c}{$H$-band $\lambda$16440 \AA} & \\
    SINFONI & B & A & B & A &\\ \hline
    1 & 0.107 & 0.241 & 0.043 & 0.082 & 0.63 \\
    2 & 0.084$^{1}$ & 0.183$^{1}$ & 0.037 & 0.061 & 0.63 \\
    3 & 0.098 & 0.227 & 0.044 & 0.079 & -- \\
    4 & 0.079 & 0.153 & 0.032 & 0.036 & 0.91 \\
    MDC (DS1) & 0.10$\pm$0.02 & 0.22$\pm$0.01 & 0.06$\pm$0.01 & 0.08$\pm$0.01 & -- \\
    MDC (DS2) & 0.34$\pm$0.7 & 0.54$\pm$0.05 & 0.15$\pm$0.05 & 0.21$\pm$0.05 & 0.53 \\ \hline
    NIFS & B & A & B & A &\\ \hline
    1 & 0.123 & 0.203 & 0.018 & 0.020 & 0.90 \\
    2 & 0.101 & 0.166 & 0.009 & 0.009 & 0.56 \\
    3 & 0.123 & 0.208 & 0.015 & 0.017 & 1.06 \\
    4 & 0.100 & 0.191 & 0.011 & 0.012 & 0.47 \\
    \hline
    \end{tabular}
    \begin{minipage}{9cm}
      Notes:
      1: Strehl measured for the [Si\,{\sc vi}] $\lambda$19634 \AA~line, since this data cube is not used in the H$_{2}$ mosaic.
    \end{minipage}
    \label{table:strehl}
    \end{center}
    \end{table*}

    \section{Results}
    \label{sec:results}

    The following sections investigate the individual properties of the emission lines.  

    \subsection{The [Fe\,{\sc ii}] emission}
    \label{sec:FeII}
    \subsubsection{The hourglass: the low ionization NLR}

\begin{figure*}
    \resizebox{0.85\hsize}{!}{\includegraphics{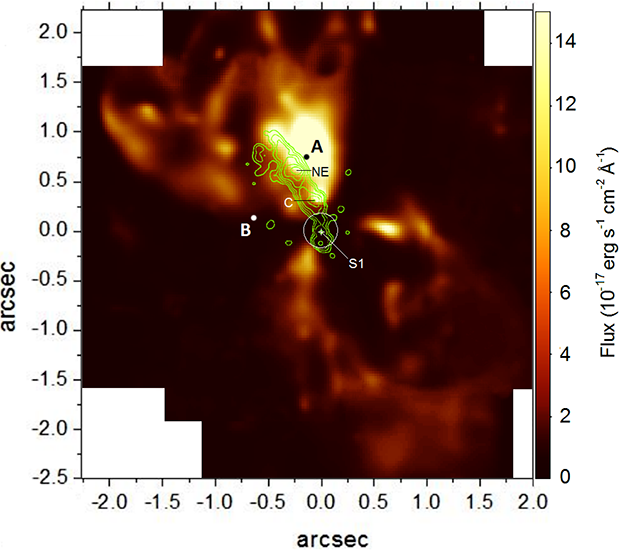}}
    \caption{Mosaic of the [Fe\,{\sc ii}] $\lambda$16440 \AA~emission line overlaid with the MERLIN 5 GHz radio emission (green contours). The \textbf{NE} and \textbf{C} knots are named according to \citet{Gallimore96}.}
     \label{fig:FeII}
    \end{figure*}

    Based on the mosaic of Fig.~\ref{fig:dataset} (upper panel), Fig.~\ref{fig:FeII} shows the image for the [Fe\,{\sc ii}] $\lambda$16440 \AA~emission line extracted from the continuum-subtracted SINFONI data cubes (as explained in Sec.~\ref{sec:dec}). The contrast was adjusted to show the faintest emission and, as the nucleus is extremely bright in the NIR, the central part will be explored in Sect~\ref{sec:ds2}. At first sight, some remarkable features can be identified, such as the shape of the emission boundaries of the two ionization cones, resembling an ``hourglass'', as already noted by \citet{Riffel14b} and \citet{Barbosa14} (their Figs.4 and 2, respectively). The position angle of the northeast cone is slightly different than the one in the southwest cone (PA=30\textdegree$\pm$2\textdegree~and PA=39\textdegree$\pm$2\textdegree, respectively), with an average of PA=34\textdegree$\pm4$\textdegree. In addition, the apex of each cone does not coincide, neither which each other nor with the nucleus. The northeast cone is in the near side of the galaxy and presents an asymmetric emission with respect to the major axis of the cones, with the emission peak at point \textbf{A}. The southwest part seems to form a closed oval-shaped ring/bubble structure, delimiting a cavity, while the northeast cone displays an open cone-shaped geometry. Despite the previous identification of the overall [Fe\,{\sc ii}] structure, it is the first time that such asymmetries are clearly detected.  
		
		To overlap the MERLIN 5\,GHz radio emission on the [Fe\,{\sc ii}] mosaic, an accurate position for the AGN centre has to be defined. Later in Sec.~\ref{sec:ds2}, with a smaller pixel scale, we show that it is possible to identify the kinematic centre for the central [Si\,{\sc vi}] emission, adopted as the AGN centre. The displacement between the [Si\,{\sc vi}] kinematic centre and the dust continuum emission peak is less than the spatial resolution of DS1. Therefore, we adopted the dust emission peak as the AGN centre for DS1 and as the reference to overlap the radio knot \textbf{S1} (already identified as the AGN position by \citealt{Gallimore96}).
		 
    Similarly to what is seen in the [O\,{\sc iii}] $\lambda$5007 \AA~line \citep{Gallimore96}, the [Fe\,{\sc ii}] emission is more intense at the edges of the radio emission, near the knots \textbf{NE} and \textbf{C}. The position angle for the [Fe\,{\sc ii}] is equivalent to that of the jet after the bending from PA$\sim$11\textdegree~to PA$\sim$33\textdegree, with the radio emission following the major axis of the cones.

    The shape of the [Fe\,{\sc ii}] emission associated with the hourglass is a strong suggestion of a collimating structure, admittedly defined by the axis and the open angle of the torus. In fact, the hourglass and bubble shape geometry have been identified in numerous systems, of a wide range of sizes, like the ``Teacup'' galaxy J1430+1339 ($\sim$10 kpc), the nebulae S\,106 ($\sim$500 pc) and NGC\,6302 ($<$1 pc), suggesting a common structure of collimation.
    In Sec.~\ref{sec:torus} we describe the orientation of the torus in light of our observations and the literature.

\begin{figure*}
    \resizebox{\hsize}{!}{\includegraphics{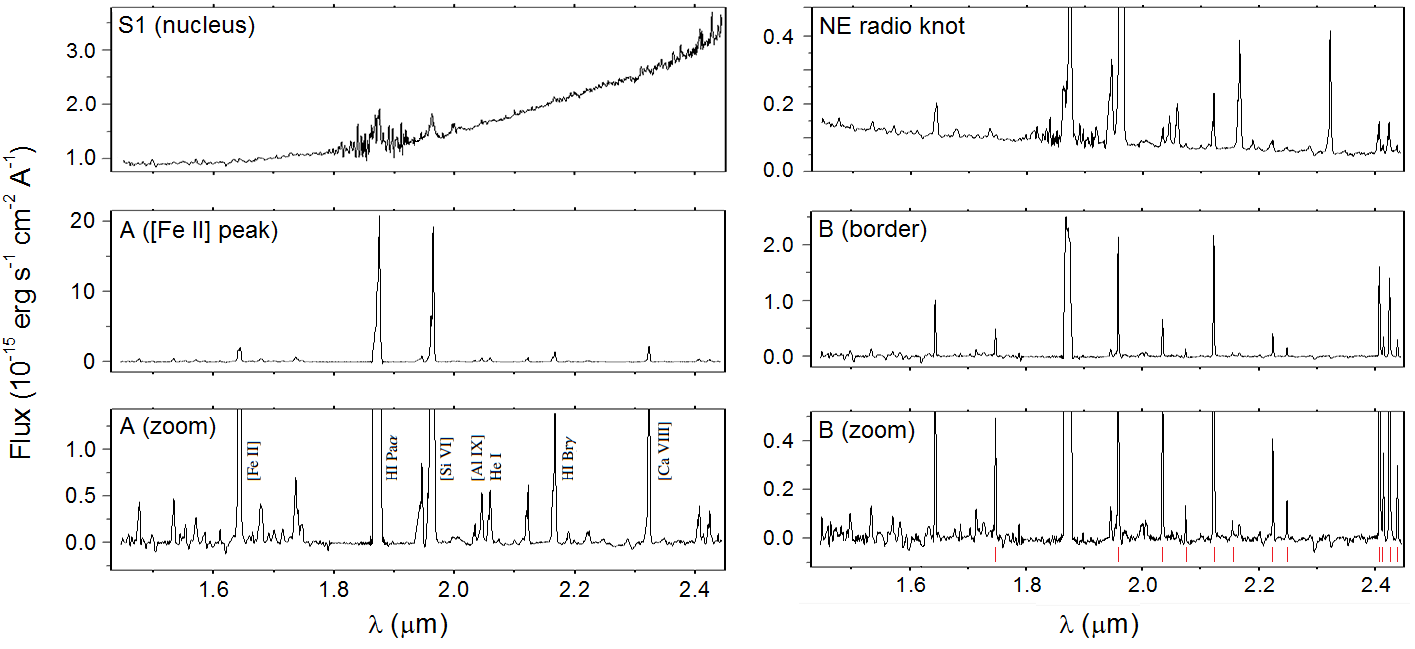}}
    \caption{Spectra taken from four regions marked in Fig.~\ref{fig:FeII}. Region A represents the [Fe\,{\sc ii}] emission peak, B the ionization cone border (close to the molecular emission peak) and \textbf{NE} the farthest radio knot, with a circular aperture radius of 0.1 arcsec. For the nucleus (centred in the S1 knot) we used a larger aperture radius of 0.25 arcsec. The bottom panels show the zoomed spectrum for regions A and B, with the fitted ionized lines identified in the left panel and the H$_{2}$ lines marked in red in the right panel.}
     \label{fig:spectra}
    \end{figure*}

				We extracted the spectra from circular regions, with a radius of 0.1 arcsec, denoted by \textbf{A}, \textbf{B} and \textbf{NE}. The spectrum of the nucleus, denoted by \textbf{S1}, was also extracted, but from a circular region with a radius of 0.25 arcsec. The extraction regions and the extracted spectra are shown in Figs.~\ref{fig:FeII} and \ref{fig:spectra}, respectively. The spectrum from the central region was extracted before applying the mask and the continuum subtraction, showing a red continuum, which indicates a dominant emission from the hot dust \citep{Gonzaga14}. Region \textbf{A} is located at the [Fe\,{\sc ii}] line emission peak, near the border of the detected radio emission associated with the jet. Region \textbf{B} is at the border of a [Fe\,{\sc ii}] emission wall, showing a narrower [Fe\,{\sc ii}] profile, close to the molecular gas emission peak (see Sect~\ref{sec:h2}).

    Table~\ref{table:flux}~displays the measured fluxes for the spectra shown in Fig.~\ref{fig:spectra}.

    \begin{table*}
    \begin{center}
    \caption[linefluxes]
    {Measured emission line fluxes for a circular aperture radius of 0.1 arcsec in regions \textbf{A}, \textbf{B} and \textbf{NE} and 0.25 arcsec at the \textbf{S1} knot, marked in Fig.~\ref{fig:FeII} and with the spectra shown in Fig.~\ref{fig:spectra}. All values are in units of $10^{-15}$~erg s$^{-1}$~cm$^{-2}$. Blended lines were fitted by multiple Gaussian components.}
    \begin{tabular}{ccccccc}
    \hline \hline
    $\lambda_{vac}$~(\AA) & Ion & Line ID ($J_{i} - J_{k}$) & \textbf{A} ([Fe\,{\sc ii}] peak) (n)$^{1}$ & \textbf{B} (border) (n)$^{1}$ & \textbf{NE} knot (n)$^{1}$ & \textbf{S1} (nucleus) \\ \hline
    16 440 & [Fe\,{\sc ii}] & $a^{4}F_{9/2} - a^{4}F_{7/2}$ & 12.83$\pm$0.04 (2) & 0.25$\pm$0.02 (1) & 0.65$\pm$0.01 (3) & - \\
    17 480 & H$_{2}$ & $1 - 0 S(7)$ & 0.55$\pm$0.04 (2) & 0.52$\pm$0.03 (1) & 0.50$\pm$0.05 (2) & -- \\
    18 751 & H\,{\sc i}\,Pa$\alpha$ & $3 - 4$ & 30.75$\pm$0.03 (3) & 0.92$\pm$0.08 (1) & 6.06$\pm$0.04 (3) & 70$\pm$30 \\
    19 576 & H$_{2}$ & $1 - 0 S(3)$ & -- & 2.13$\pm$0.01 (1) & 1.08$\pm$0.02 (1) & -- \\
    19 641 & [Si\,{\sc vi}] & $^{2}Po_{3/2} - ^{2}Po_{1/2}$ & 42.0$\pm$2 (2) & -- & 11.46$\pm$0.08 (2) & 106$\pm$5 \\
    20 338 & H$_{2}$ & $1 - 0 S(2)$ & 0.37$\pm$0.02 (2) & 0.69$\pm$0.01 (1) & 0.27$\pm$0.03 (2) & -- \\
    20 450 & [Al\,{\sc ix}] & $^{2}Po_{3/2} - ^{2}Po_{1/2}$ & 0.7$\pm$0.1 (2) & -- & 0.30$\pm$0.05 (2) & 14$\pm$1 \\
    20 587 & He I & $^{1}So - ^{1}Po_{1}$ & 1.1$\pm$0.1 (2) & 0.05$\pm$0.0 (2) & 0.66$\pm$0.04 (2) & -- \\
    20 735 & H$_{2}$ & $2 - 1 S(3)$ & 0.09$\pm$0.01 (2) & 0.14$\pm$0.2 (2) & 0.05$\pm$0.01 (2) & -- \\
    21 218 & H$_{2}$ & $1 - 0 S(1)$ & 0.87$\pm$0.03 (2) & 2.26$\pm$0.02 (2) & 0.67$\pm$0.02 (2) & -- \\
    21 542 & H$_{2}$ & $2 - 1 S(2)$ & 0.012$\pm$0.004 (2) & 0.05$\pm$0.01 (2) & 0.021$\pm$0.007 (2) & -- \\
    21 661 & H\,{\sc i}\,Br$\gamma$ & $4 - 7$ & 2.8$\pm$0.3 (2) & 0.05$\pm$0.02 (2) & 1.30$\pm$0.03 (3) & 48$\pm$1 \\
    22 233 & H$_{2}$ & $1 - 0 S(0)$ & 0.4$\pm$0.1 (3) & 0.38$\pm$0.02 (2) & 0.10$\pm$0.02 (2) & -- \\
    22 477 & H$_{2}$ & $2 - 1 S(1)$ & 0.12$\pm$0.03 (2) & 0.13$\pm$0.02 (2) & 0.05$\pm$0.02 (2) & -- \\
    23 211 & [Ca\,{\sc viii}] & $^{2}Po_{1/2} - ^{2}Po_{3/2}$ & 3.0$\pm$0.1 (2) & -- & 0.67$\pm$0.03 (2) & -- \\
    24 066 & H$_{2}$ & $1 - 0 Q(1)$ & 0.70$\pm$0.02 (2) & 1.83$\pm$0.02 (2) & 0.35$\pm$0.01 (3) & -- \\
    24 131 & H$_{2}$ & $1 - 0 Q(2)$ & 0.17$\pm$0.01 (2) & 0.36$\pm$0.01 (2) & 0.10$\pm$0.01 (2) & -- \\
    24 237 & H$_{2}$ & $1 - 0 Q(3)$ & 0.68$\pm$0.01 (2) & 1.75$\pm$0.01 (2) & 0.38$\pm$0.01 (3) & -- \\
    24 375 & H$_{2}$ & $1 - 0 Q(4)$ & 0.12$\pm$0.01 (2) & 0.32$\pm$0.03 (1) & 0.06$\pm$0.01 (1) & -- \\
    \hline
    \end{tabular}
    \begin{minipage}{15cm}
      Notes:
      1: Number of detected Gaussian components for each emission line.
    \end{minipage}
    \label{table:flux}
    \end{center}
    \end{table*}

    \subsubsection{The low- and high-velocity NLR}
    \label{sec:felh}

    The nuclear gas kinematics of NGC\,1068 is known to be very complex, as it presents redshifted and blueshifted velocities in both sides of the cones. This is the result of a huge amount of gas reservoir in the NLR, which is, in principle, efficiently transported by the AGN wind and perturbed by the presence of a jet (\citealt{Cecil90, Gallimore96, Crenshaw00, Das06}).

\begin{figure*}
    \resizebox{\hsize}{!}{\includegraphics{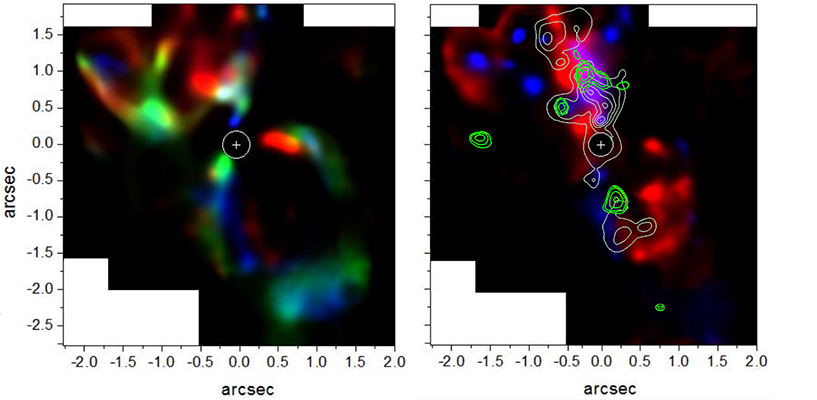}}
    \caption{Mosaic of the [Fe\,{\sc ii}] $\lambda$16440 \AA~emission line for the low-velocity range between -310 km s$^{-1}<v<237$~km s$^{-1}$ (left panel) and the high-velocity regime for -1951 km s$^{-1}<v<-310$~km s$^{-1}$~and 237 km s$^{-1}<v<1514$~km s$^{-1}$ (right panel), which basically fills the hourglass volume. Overlaid with the high-velocity regime are the contours of the coronal line emission [Si\,{\sc vi}] (white) and the high-velocity molecular gas (green). The circle denotes the masked region and the cross indicates the nucleus.}
     \label{fig:Felh}
    \end{figure*}

    If the gas structures are resolved, we can start to disentangle the multi-components of the emission line profiles as spatially distinct clouds and filaments, allowing a better understanding of the innermost region. By analyzing the velocity frames of the [Fe\,{\sc ii}] line, it is possible to identify a two-fold behaviour for the NLR (Fig.~\ref{fig:Felh}): the low-velocity (-310 km s$^{-1}<v<237$~km s$^{-1}$) [Fe\,{\sc ii}] emission shows the wall of a ``glowing hourglass'' structure, which can be identified as the walls of the two ionization cones (phase-1), and the high-velocity (-1951 km s$^{-1}<v<-401$~km s$^{-1}$~and 328 km s$^{-1}<v<1514$~km s$^{-1}$) emission, which corresponds only to 9\% of the total [Fe\,{\sc ii}] brightness (not accounting the masked region). The high-velocity gas distribution displays, mostly, compact blobs in blueshift surrounded by cloud's filaments in redshift in the northeast cone and mainly a concentration of clouds in redshift in the southwest part (part of phase-2), filling the hourglass volume. 	
			
		In Fig.~\ref{fig:Felh} (right panel) we also show the contours of the [Si\,{\sc vi}] emission line (white) and of the high-velocity molecular gas (green), latter discussed in Sect.~\ref{sec:clb} and \ref{sec:h2}.

\begin{figure*}
    \resizebox{\hsize}{!}{\includegraphics{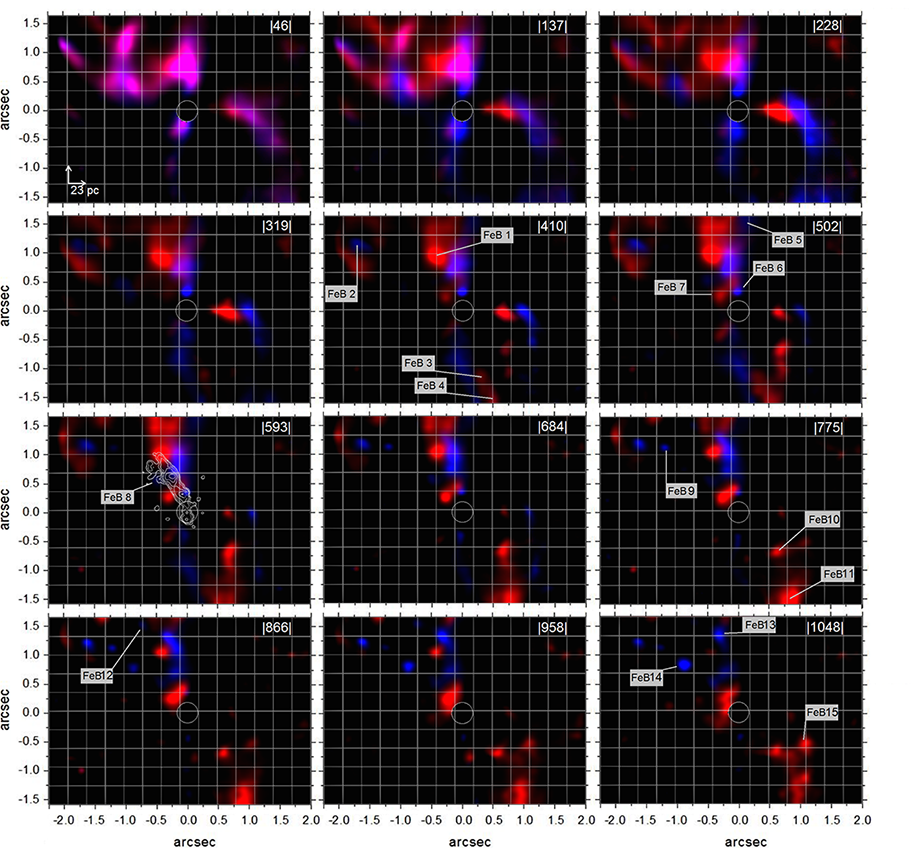}}
    \caption{BRV maps (blueshift and redshift velocity maps) for the [Fe\,{\sc ii}] $\lambda$16440 \AA~emission line. The velocity ranges from $|v|\lesssim$2000 km s$^{-1}$, with steps of $\sim$90 km s$^{-1}$. The grid has a small square of 23 pc$\times$23 pc and the the corresponding absolute velocity of each map is show at the top. The sixteen selected blobs are indicated as ``Fe Blob(number)'', with the measured peak velocities, velocity dispersions, distances from the nucleus and position angles shown in Table~\ref{table:Feb}. The circle denotes the masked region centred on the nucleus.}
     \label{fig:Ferbvmaps1}
    \end{figure*}

    The asymmetry of the hourglass is also seen in the spatial distribution of the kinematics of the NLR, where the high-velocity gas is confined to the southwest walls of the cone (except for the extended southwest emission), but not to the northeast part; instead, it seems to be located between internal filaments of low-velocity gas. In Fig.~\ref{fig:Ferbvmaps1} we show channel maps with frames of same absolute velocity (hereafter referred to blueshift and redshift velocity maps - BRV maps), where we identified the blobs with a near-discrete morphology to extract their peak and dispersion velocities, distance and position angles, shown in Table~\ref{table:Feb}. Along this work, the gas kinematics will be shown only in the form of BRV maps and RGB compositions. Because of the complexity of the emission line profiles, fitting Gaussian functions do not lead to any insightful kinematic map. In addition, BRV maps reduce by half the number of velocity panels when compared to traditional channel maps and help to visualize possible symmetries between redshifted and blueshifted blobs.
		
    \begin{figure*}
    \resizebox{\hsize}{!}{\includegraphics{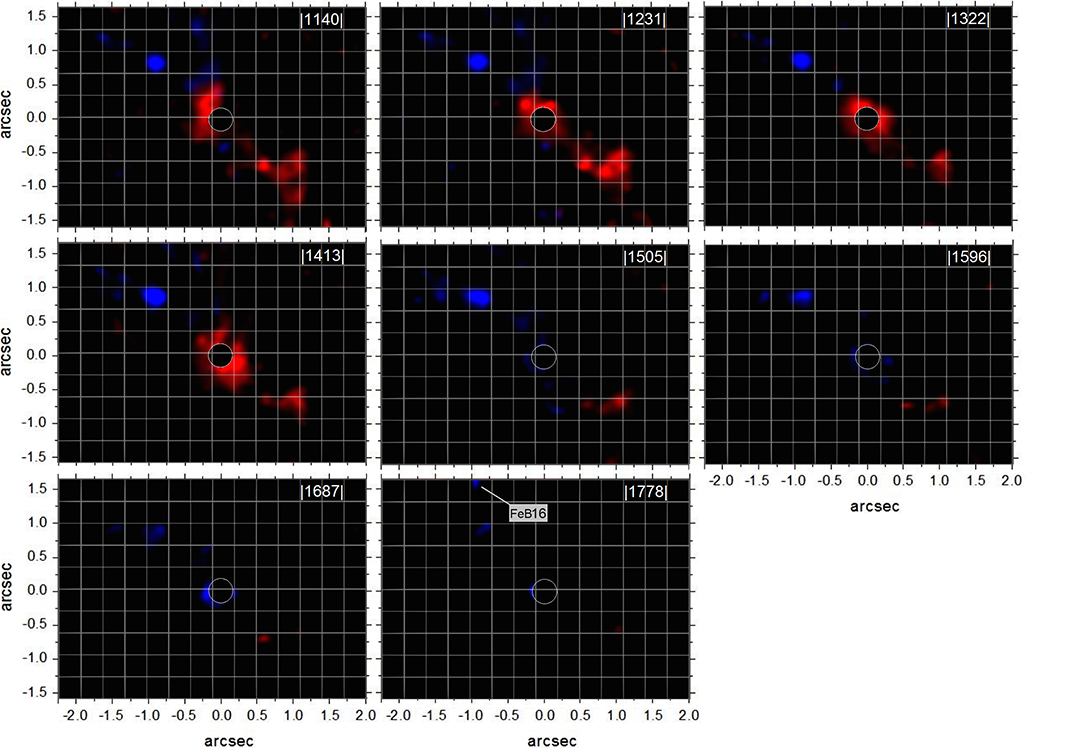}}
    \caption{Extension of Fig.~\ref{fig:Ferbvmaps1}. See caption in the previous figure.}
     \label{fig:Ferbvmaps2}
    \end{figure*}

    \begin{table}
    \begin{center}
    \caption[Feb]
    {Peak velocity, velocity dispersions, distance from the nucleus and position angle measured for the [Fe\,{\sc ii}] blobs shown in Figs.~\ref{fig:Ferbvmaps1} and \ref{fig:Ferbvmaps2}.}
    \begin{tabular}{cccccc}
    \hline \hline
    Blob ID & v$^{1}$ (km s$^{-1}$) & $\sigma^{2}$ (km s$^{-1}$) & Distance$^{3}$ (pc) & PA$^{4}$ & PA$^{5}$ \\ \hline
    FeB1 & 427 & 208 & -75 & 26 & -11 \\
    FeB2 & -826 & 247 & -154 & 57 & 20 \\
    FeB3 & 376 & 70 & 84 & -164 & -- \\
    FeB4 & 507 & 130 & 116 & -162 & -- \\
    FeB5 & -503 & 160 & -114 & -6 & -43 \\
    FeB6 & -350 & 200 & -23 & 0 & -37 \\
    FeB7 & 636 & 250 & -29 & 49 & 12 \\
    FeB8 & -596 & 102 & -57 & 44 & 7 \\
    FeB9 & -742 & 100 & -120 & 49 & 12 \\
    FeB10 & 560 & 105 & 69 & -135 & -- \\
    FeB11 & 773 & 107 & 126 & -149 & -- \\
    FeB12 & -863 & 138 & -122 & 27 & -10 \\
    FeB13 & -923 & 163 & -101 & 15 & -22 \\
    FeB14 & -1187 & 198 & -89 & 49 & 12 \\
    FeB15 & 1355 & 117 & 90 & -115 & -- \\
    FeB16 & -1836 & 130 & -137 & 32 & -5 \\
    \hline
    \end{tabular}
    \begin{minipage}{8cm}
      Notes:
			(1) The uncertainty in velocity is $\sim$10 km s$^{-1}$.
      (2) Corrected for instrumental broadening.
			(3) Distance offset of each blob from the bulge centre; negative numbers mean a northeast offset with respect to a line orthogonal to the cone's major axis.
      (4) Blob's position angle relative to the northeast and (5) relative to the cone's major axis, of 34\textdegree.
    \end{minipage}
    \label{table:Feb}
    \end{center}
    \end{table}

    Similar kinematic panels can be found in \citet{Barbosa14}, with data from NIFS-Gemini North. The velocity components shown by these authors are between -723 km s$^{-1}<v<$ 600 km s$^{-1}$, in agreement with the same interval in our maps. They modeled the outflow with the [Fe\,{\sc ii}] line and concluded that the lemniscate (hourglass) shape does not improve significantly the conical model from \citet{Das06} for the velocity field.

\begin{figure}
    \resizebox{\hsize}{!}{\includegraphics{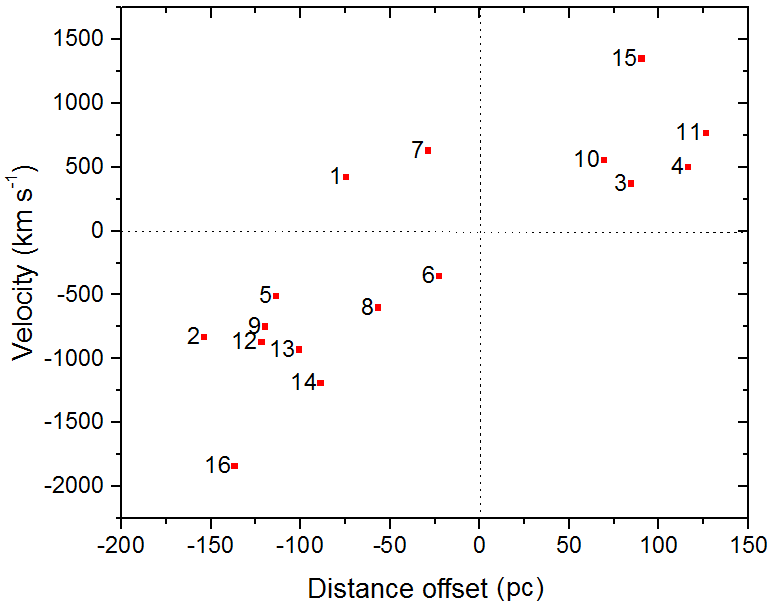}}
    \caption{Diagram for the position and velocity of the [Fe\,{\sc ii}] blobs identified in Fig.~\ref{fig:Ferbvmaps1}.}
     \label{fig:Feblobs}
    \end{figure}

    The BRV maps reveal that the blobs in blueshift and redshift reach velocities up to -1824 and 1368 km s$^{-1}$, respectively, with the highest values at more distant projected positions from the nucleus. In fact, there is evidence that [Fe\,{\sc ii}] blobs are being accelerated, in both directions of the cones, as can be seen in the graph of Fig.~\ref{fig:Feblobs}. Such behaviour has been known since the work of \citet{Cecil90} but is shown much clearer here.

    Looking to the southwest cone, the sequence of confined blobs 3-4-11-15 (see Fig.~\ref{fig:Ferbvmaps1} and the velocities in Table~\ref{table:Feb}) is increasing in velocity and follows the wall of the cone. Considering that the jet could be interacting with the southeast inner wall of the low-velocity hourglass, it seems natural to interpret that these blobs had the same origin and are accelerated along the same trajectory. The blobs also share similar velocity dispersions, of $107\pm5$ km s$^{-1}$.

    In the northeast cone, the blobs reach higher velocity dispersions, with a mean of $173\pm8$ km s$^{-1}$, tending toward higher values closer to the jet. The mean velocity dispersion for all the blobs is $152\pm55$ km s$^{-1}$. Such analysis will be dealt with more detail in Sect.~\ref{sec:arc}, after we present the results for the [Si\,{\sc vi}] and the molecular lines. 

    \subsection{The coronal line blobs}
    \label{sec:clb}

    Coronal lines in the NIR, with IP$>$100\,eV (reaching values of up to $\sim$300\,eV), are commonly broader than other forbidden transitions, and vanish due to collisional de-excitation for electron densities between 10$^{8}$~and 10$^{9}$~cm$^{-3}$. The coronal line (CL) with the highest critical density is found closest to the ionizing source and presents the broadest FWHM \citep{Ardila11}. There are two main competing mechanisms for their ionization: photo-ionization by the central source and by jet-induced shocks \citep{Dopita96,Bicknell98,Mazzalay13a}.

    In the later case, the term shock means that there is a significant production of X-ray photons associated with the free-free emission from the gas heated in the shock wave induced by the interaction with the radio jet. The forbidden transitions of highly ionized ions are then colisionally excited by free electrons. This mechanism is generally evoked when high ionization emission lines are found far from the nucleus and close to the radio emission. In some cases, the radial gradient of the ionization parameter is not consistent with photo-ionization by a central source, which may even increase for larger distances from the nucleus \citep{Capetti96a}. 

\begin{figure}
    \resizebox{\hsize}{!}{\includegraphics{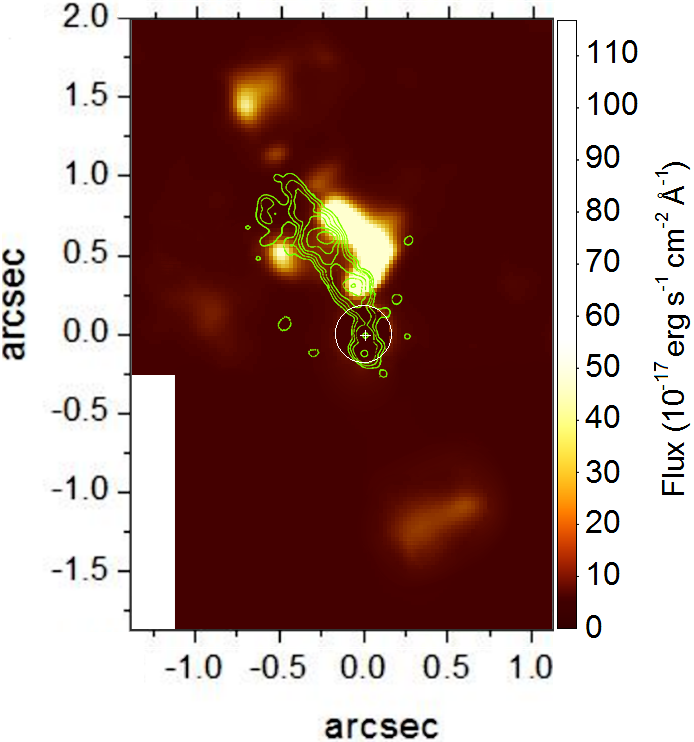}}
    \caption{Mosaic of the [Si\,{\sc vi}] $\lambda$19641 \AA~emission line overlaid with the MERLIN 5 GHz radio emission (green contours). The circle denotes the masked region and the cross indicates the nucleus.}
     \label{fig:SiVI}
    \end{figure}

    The most intense CL present in our spectra is the [Si\,{\sc vi}] line, which is described in more detail in this section. In Fig.~\ref{fig:SiVI}, the spatial distribution of the [Si\,{\sc vi}] $\lambda$19641 \AA~line (IP=168\,eV) shows a narrower morphology than that of the [Fe\,{\sc ii}] gas and is preferably oriented in the direction of the extended radio emission.
		
    An additional issue about this line is the presence of the blended line H$_{2}$ $\lambda$19576 \AA, preventing us, in principle, from presenting a ``pure'' intensity map of the [Si\,{\sc vi}] $\lambda$19641 \AA~line. Closer to the jet, for instance, we know that both emission lines are partially coincident (Fig.~\ref{fig:Felh}, right panel). To avoid this problem, we have to look for some region where they are not spatially coincident. Thus, the following procedure was performed: we verified for an indication of [Si\,{\sc vi}] emission around the molecular peak (region \textbf{B} of Fig.~\ref{fig:FeII}) by looking for other highly ionized emission lines (like [Al\,{\sc ix}], [Ca\,{\sc viii}] or even He\,{\sc i}) which surely are not blended to any other line. After verifying that these lines do not present any sign of emission in this region (some of them with lower IP), we assume that the [Si\,{\sc vi}] emission is also absent there. Then, based on the flux measured in a circular aperture radius of 0.25 arcsec, we scaled the H$_{2}$ $\lambda$21218 \AA~line, which is more intense, so it had the same flux of the H$_{2}$ line $\lambda$19576 \AA, not affected by the blending with [Si\,{\sc vi}] line at this specific region. The resulting image estimate the real H$_{2}$ $\lambda$19576 \AA~flux at the region where it is blended with the [Si\,{\sc vi}] line. This was done frame by frame for each velocity channel to cover the entire line profile. Finally, we subtracted the corresponding frames from those of the image taken with the blended profiles. The result is a [Si\,{\sc vi}] coronal line almost free from molecular emission.

    The [Si\,{\sc vi}] intensity map, after the subtraction of the molecular emission, is shown in Fig.~\ref{fig:SiVI}. The LUT was adjusted to show the faintest emissions at the border of the FoV, more than $100\times$ weaker than the peak emission near the centre. The [Si\,{\sc vi}] emission is mostly originated from discrete structures and its extended northeast-southwest emission presents a noticeable symmetry, with the gas extending roughly at the same distance and position angle. Considering that maybe not all radio emission is recovered in the MERLIN interferometric map, the only region which seems to be associated with the radio map is located right above the masked region, but in general, the coronal gas does not coincides with the regions of most intense radio emission. 
		
		\subsubsection{[Si\,{\sc vi}] kinematics}
    \label{sec:sik}
		
\begin{figure*}
    \resizebox{0.80\hsize}{!}{\includegraphics{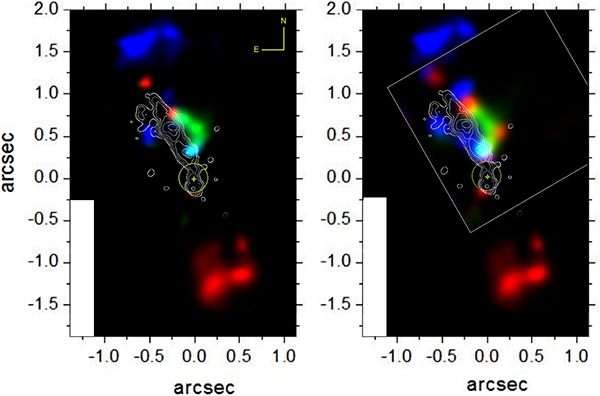}}
    \caption{Left panel: RGB composition of the [Si\,{\sc vi}] $\lambda$19641 \AA~emission line or the SINFONI DS1, with the central frame emission ranging between $\pm153$ km s$^{-1}$ (green). The MERLIN 5 GHz radio emission is shown in white contours. Right panel: comparison with one data cube from NIFS (gray square) for a similar velocity range (see Sec.~\ref{sec:comp}). The circle denotes the masked region and the cross indicates the nucleus.}
     \label{fig:SiVIk}
    \end{figure*}

		In the left panel of Fig.~\ref{fig:SiVIk} we show two central frames of the line profile, with the velocity ranges of $-154$ km s$^{-1}<v<154$~km s$^{-1}$, $-1588$ km s$^{-1}<v<-192$~km s$^{-1}$ and $192$ km s$^{-1}<v<954$~km s$^{-1}$ in green, blue and red, respectively. Like the [Fe\,{\sc ii}] distribution, we still find blobs both in redshift and blueshift in the northeast cone, although the blobs in redshift do not exceed 600 km s$^{-1}$. The right panel of Fig.~\ref{fig:SiVIk} will be discussed in Sec.~\ref{sec:comp}.

\begin{figure*}
    \resizebox{\hsize}{!}{\includegraphics{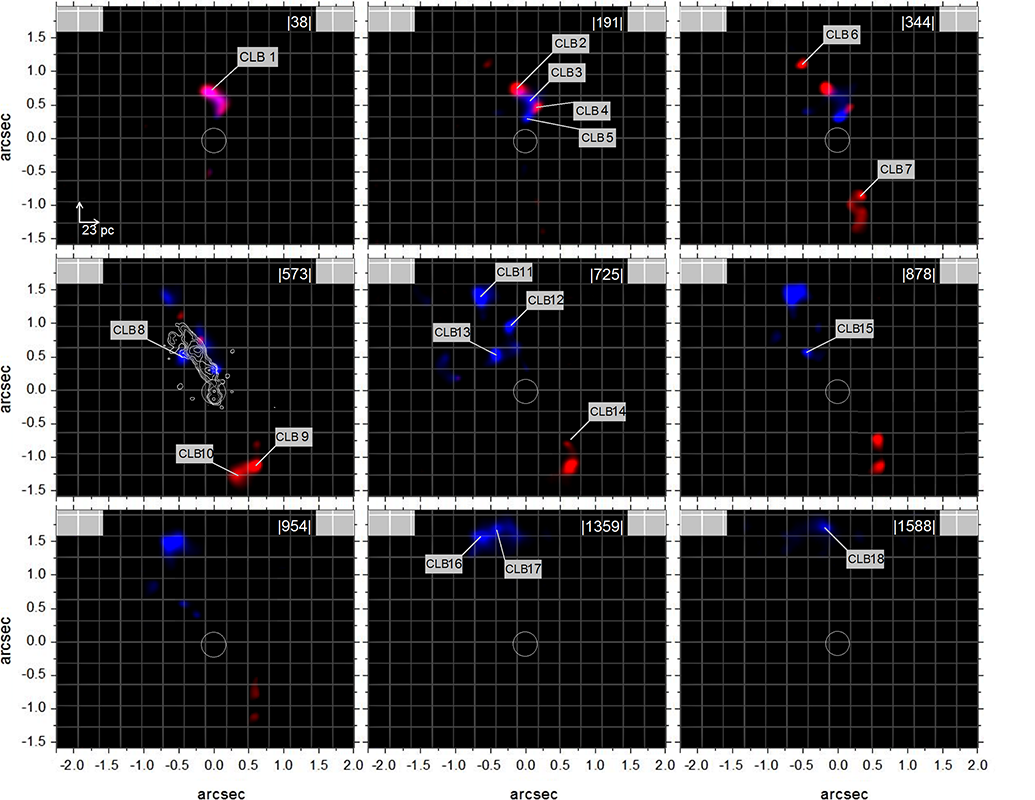}}
    \caption{BRV maps for the [Si\,{\sc vi}] $\lambda$19641 \AA~emission line. The velocity ranges from $|v|\lesssim$1600 km s$^{-1}$, with varying intervals, representative of significant changes in the gas distribution. The grid has a small square of 23 pc$\times$23 pc and the the corresponding absolute velocity of each map is shown on top. The eighteen selected coronal line blobs are indicated as ``CL Blob(number)'', with the measured peak velocity, velocity dispersions, distances from the nucleus and position angles shown in Table~\ref{table:Sib}. The circle denotes the masked region centred on the nucleus.}
     \label{fig:Sirbvmaps}
    \end{figure*}

    In Fig.~\ref{fig:Sirbvmaps} we show the BRV maps with the eighteen identified blobs. The quantities similar to those derived for the [Fe\,{\sc ii}] line are shown in Table~\ref{table:Sib}.

    \begin{table*}
    \begin{center}
    \caption[Sib]
    {Peak velocity, velocity dispersions, distance from the nucleus and position angle measured for the [Si\,{\sc vi}] blobs shown in Fig.~\ref{fig:Sirbvmaps}.}
    \begin{tabular}{ccccccc}
    \hline \hline
    Blob ID & v$^{1}$ (km s$^{-1}$) & $\sigma^{2}$ (km s$^{-1}$) & Distance$^{3}$ (pc) & PA$^{4}$ & PA$^{5}$ & [O\,{\sc iii}] cloud$^{6}$ \\ \hline
    CLB1 & -177 & 147 & -52 & 6 & -31 & -- \\
    CLB2 & 61 & 186 & -46 & 9 & -28 & D \\
    CLB3 & 270 & 122 & -44 & -8 & -45 & D \\
    CLB4 & 113 & 68 & -38 & -20 & -57 & E \\
    CLB5 & -421 & 140 & -24 & -6 & -43 & C \\
    CLB6 & -246 & 78 & -90 & 25 & -12 & -- \\
    CLB7 & 163 & 28 & 65 & -158 & -- & -- \\
    CLB8 & -543 & 141 & -52 & 43 & 6 & -- \\
    CLB9 & 520 & 78 & 93 & -151 & -- & -- \\
    CLB10 & 331 & 80 & 95 & -165 & -- & -- \\
    CLB11 & -870 & 178 & -115 & 25 & -12 & G \\
    CLB12 & -830 & 56 & -73 & 14 & -23 & -- \\
    CLB13 & -786 & 134 & -53 & 40 & 3 & F \\
    CLB14 & 565 & 111 & 95 & -149 & -- &  \\
    CLB15 & -985 & 70 & -55 & 38 & 1 & F \\
    CLB16 & -1401 & 127 & -125 & 23 & -14 & -- \\
    CLB17 & -1420 & 130 & -128 & 15 & -22 & -- \\
    CLB18 & -1592 & 133 & -126 & 6 & -31 & -- \\
    \hline
    \end{tabular}
    \begin{minipage}{15cm}
      Notes:
			(1) The uncertainty in velocity is $\sim$10 km s$^{-1}$.
      (2) Corrected for instrumental broadening.
			(3) Distance offset of each blob from the bulge centre; negative numbers mean a northeast offset with respect to a line orthogonal to the cone's major axis.
      (4) Blob's position angle relative to the northeast and (5) relative to the cone's major axis, of 34\textdegree.
      (6) [O\,{\sc iii}] clouds identified by \citet{Evans91} and shown in Fig.~\ref{fig:SiOIII}.
    \end{minipage}
    \label{table:Sib}
    \end{center}
    \end{table*}

    The low-velocity [Si\,{\sc vi}] emission differs from that of the [Fe\,{\sc ii}] gas in the sense that the coronal line morphology is distributed mainly in the form of discrete blobs, and the emission closer to the AGN seems to originate from a fragmented filament of gas perturbed by the jet, given the vicinity of the blobs. In fact, the dynamics of the CLR is likely to be closely related to the presence of the deflected jet, with the farthest blobs following its orientation, which is not the case for the [Fe\,{\sc ii}] blobs (see right panel of Fig.~\ref{fig:Felh}). The role of the jet as the excitation mechanism of the coronal lines was studied in detail in the work of \citet{Mazzalay13a}, which found a strong indication of photo-ionization produced by shocks between the gas and the jet, showing a higher [Si\,VII]/[Si\,{\sc vi}] line ratio where the jet interacts with the NLR gas. The mean velocity dispersion for all the [Si\,{\sc vi}] blobs is $111\pm43$ km s$^{-1}$.

\begin{figure}
    \resizebox{\hsize}{!}{\includegraphics{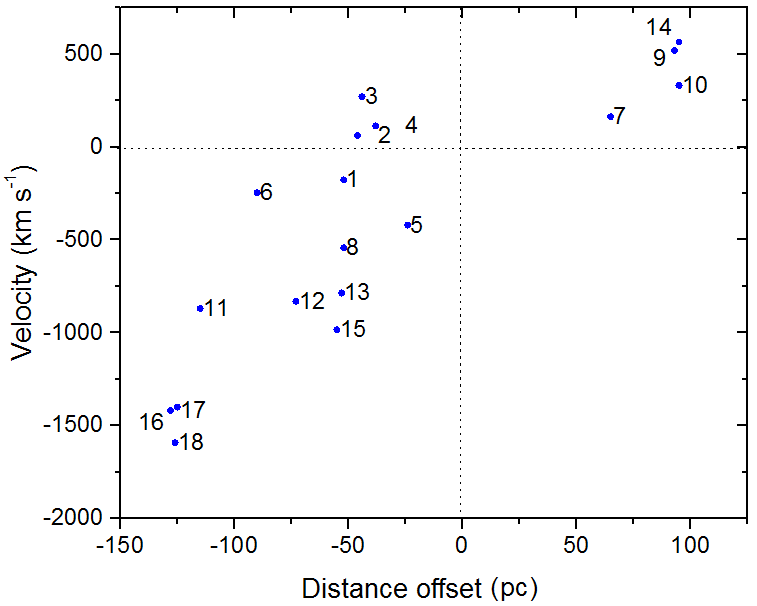}}
    \caption{Diagram for the position and velocity of the [Si\,{\sc vi}] blobs identified in Fig.~\ref{fig:Sirbvmaps}.}
     \label{fig:Siblobs}
    \end{figure}

    Although with a narrower opening angle, when compared to the [Fe\,{\sc ii}] morphology, the increasing radial velocity of the CLBs (Coronal Line Blobs) follows a similar behaviour as for the [Fe\,{\sc ii}] blobs. This is shown in Fig.~\ref{fig:Siblobs}. 
		
    \subsubsection{Comparison to the [O\,{\sc iii}] emission}
    \label{sec:OIII}      

\begin{figure}
    \resizebox{\hsize}{!}{\includegraphics{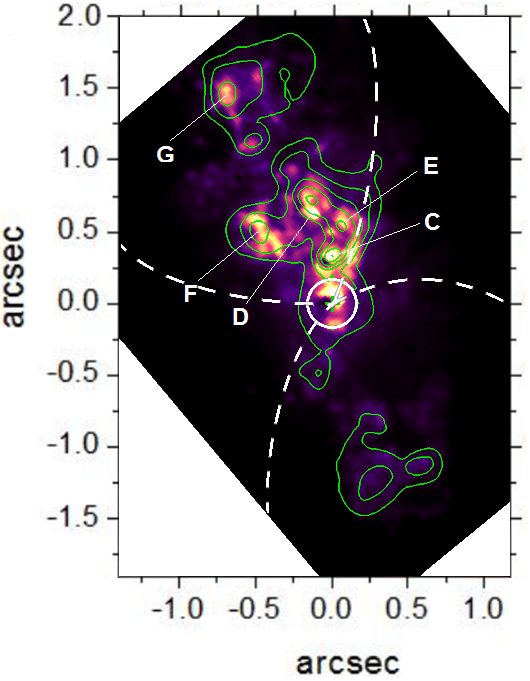}}
    \caption{Continuum-subtracted [O\,{\sc iii}] $\lambda$5007 \AA~image overlaid with the contours of the [Si\,{\sc vi}] emission. The white dashed line is the shape of the [Fe\,{\sc ii}] hourglass and the white circle denotes the masked region for the [Si\,{\sc vi}] line.}
     \label{fig:SiOIII}
    \end{figure}

    Although there is a correspondence with the [Fe\,{\sc ii}] emission along the jet orientation (Fig.~\ref{fig:Felh}), the CLBs have a stronger association with the narrow band (\textsc{F501N}) [O\,{\sc iii}] emission, as shown in Fig.~\ref{fig:SiOIII}.    

    The distribution of the [O\,{\sc iii}] emission is commonly mentioned as having a cone-shaped structure \citep{Evans91} (with PA$\sim$15\textdegree), because of its extended morphology, which fills the region between the clouds with higher luminosity. Looking at the radio emission, the location \textbf{S1} (Fig.~\ref{fig:FeII}) is the most probable position for the AGN (based on the analysis of the radio spectrum by \citealt{Gallimore96}), which is assumed to coincide with a faint cloud A (see Fig.2 of \citealt{Evans91}), considered as the apex of the cone. This reference was adopted to overlap the countours of the [Si\,{\sc vi}] and [O\,{\sc iii}] emissions in Fig.~\ref{fig:SiOIII}. The hourglass shape for the NLR is shown to compare with the [O\,{\sc iii}] cone morphology (Fig.~\ref{fig:SiOIII}); in addition, like the [Si\,{\sc vi}] emission, the northeast part of the cone in the optical is preferably located in the northwest side of the hourglass seen in [Fe\,{\sc ii}]. Both [O\,{\sc iii}] and [Si\,{\sc vi}] emissions, with the former presenting a more diffused distribution, present an remarkable correspondence between their clouds (as already noted by \citealt{Mazzalay13a}). Such agreement strongly discard the hypothesis of an asymmetric distribution for the optical emission caused by the presence of dust into the cone, or that the UV radiation may be partially absorbed by an excess of dust in the direction of the east side of the hourglass, near the nucleus.

    The asymmetry between the cones seen in the NIR and in the optical is most likely caused by a misalignment between the radiation collimated by the torus and by the plane of the accretion disc, respectively. In Sections.~\ref{sec:torus} and~\ref{sec:2phase} we look closer to the central structures and their orientations. 

    In the southwest cone, the [O\,{\sc iii}] emission is also in close agreement with the [Si\,{\sc vi}] emission and with the [Fe\,{\sc ii}] high-velocity blobs, filling the volume defined by the low-velocity [Fe\,{\sc ii}] walls (Fig.~\ref{fig:Felh}). The optical emission is fainter in the southwest cone, an effect that is naturally attributed to the presence of dust in the LoS, as we see this cone behind the galactic disc.
    The Chandra HRC (high resolution camera) image of the X-ray emission \citep{Wang12} also has a strong correspondence with the [O\,{\sc iii}] emission line, including the cone's open angle and the identified clouds.

    \subsection{Additional detected atomic emission lines}

\begin{figure}
    \resizebox{\hsize}{!}{\includegraphics{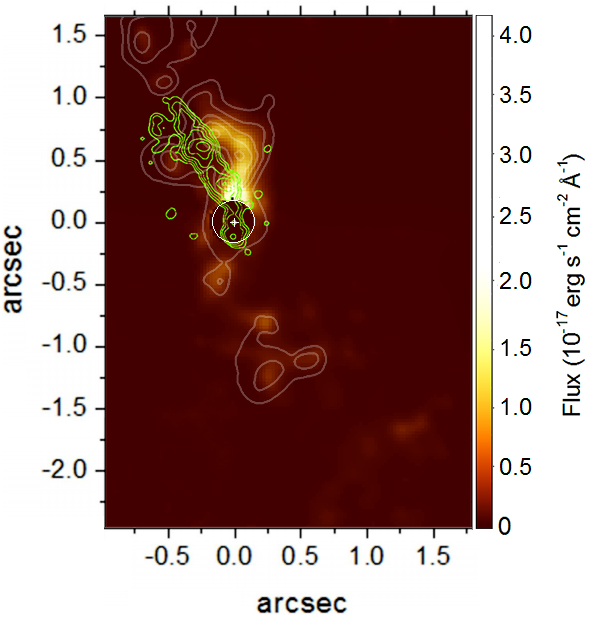}}
    \caption{Image of the Br$\gamma$ $\lambda$21661 \AA~emission line overlaid with the MERLIN 5 GHz radio emission (green contours) and the [Si\,{\sc vi}] emission (white contours). The circle denotes the masked region and the cross indicates the nucleus.}
     \label{fig:hbr}
    \end{figure}

\begin{figure*}
    \resizebox{0.80\hsize}{!}{\includegraphics{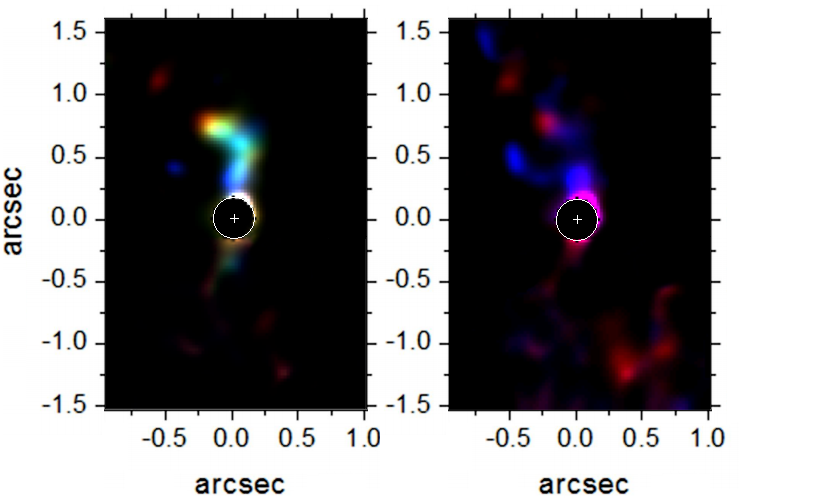}}
    \caption{RGB composition of the Br$\gamma$ $\lambda$21661 \AA~emission line for the low-velocity range between 235 km s$^{-1}>v>-318$~km s$^{-1}$ (left panel) and the high-velocity regime for 720 km s$^{-1}>v>235$~km s$^{-1}$ and -318 km s$^{-1}>v>-1200$~km s$^{-1}$ (right panel). The circle denotes the masked region and the cross indicates the nucleus.}
     \label{fig:hbrlh}
    \end{figure*}

\begin{figure*}
    \resizebox{0.80\hsize}{!}{\includegraphics{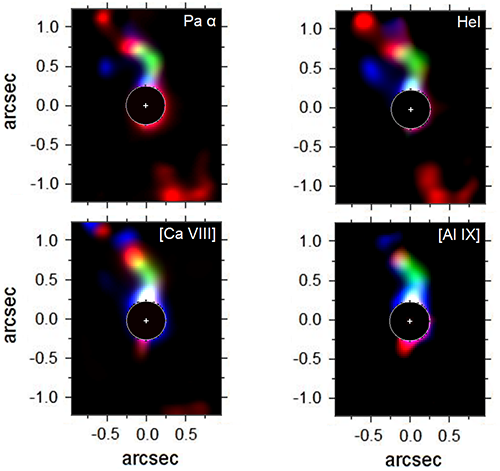}}
    \caption{Top: RGB compositions for the P$\alpha$ $\lambda$18571 \AA~(left) and He\,{\sc i} emission lines. Bottom: the same for the [Ca VII] $\lambda$23211 \AA~and [Al\,{\sc ix}] $\lambda$20450 \AA~emission lines. The velocity interval in green shows the three central frames, and the velocity range for the other colours can be found in Table~\ref{table:vel}. The circle denotes the masked region and the cross indicates the nucleus.}
     \label{fig:hion}
    \end{figure*}

    A variety of ionized elements are seen in the $HK$-bands of NGC\,1068. Besides the [Si\,{\sc vi}] emission line, a similar gas morphology and kinematics is found in five more emission lines, namely, Br$\gamma$ $\lambda$21661 \AA~(Fig.~\ref{fig:hbr} and Fig.~\ref{fig:hbrlh}), P$\alpha$ $\lambda$18751 \AA, He\,{\sc i} $\lambda$20587 \AA, [Ca\,{\sc viii}] $\lambda$23211 \AA~and [Al\,{\sc ix}] $\lambda$20450 \AA, the last two lines also being CL emissions. Their RGB kinematic compositions are shown in Fig.~\ref{fig:hion} and the velocity ranges for each image are listed in Table~\ref{table:vel}. Their intensity measurements were calculated for some regions and are shown in Table~\ref{table:flux}. 

    \begin{table*}
    \begin{center}
    \caption[vel]
    {Velocity range from the RGB composition of all emission lines presented in this paper, as shown in Figs.~\ref{fig:FeII}, \ref{fig:SiVIk}, \ref{fig:hbrlh}, \ref{fig:hion} and \ref{fig:h2s}.}
    \begin{tabular}{cccccccc}
    \hline \hline
    $\lambda_{vac}$~(\AA) & Line & IP$^{1}$ & Centre$^{2}$ & FWHM$^{2}$ & v$_{blue}$$^{3}$ (km s$^{-1}$) & v$_{green}$$^{3}$ (km s$^{-1}$) & v$_{red}$$^{3}$ (km s$^{-1}$) \\ \hline
    16 440 & [Fe\,{\sc ii}] & 8.9 & 0.07 & 0.30 & -1951$<v<$-401 & -310$<v<$237 & 328$<v<$1514 \\
    18 751 & H\,{\sc i}\,Pa$\alpha$ & 12.8 & 0.04 & 0.24 & -831$<v<$-192 & -112$<v<$128 & 208$<v<$1072 \\
    19 641 & [Si\,{\sc vi}] & 168 & 0.01 & 0.23 & -1588$<v<$-192 & -154$<v<$154 & 192$<v<$954 \\
    20 450 & [Al\,{\sc ix}] & 285 & 0.02 & 0.26 & -1056$<v<$-410 & -337$<v<$117 & 190$<v<$557 \\
    20 587 & He I & 21.2 & 0.03 & 0.27 & -1165$<v<$-219 & -146$<v<$87 & 160$<v<$874 \\
    21 218 & H$_{2}$ & 0.65 & 0.14 & 0.24 & -820$<v<$-254 & -183$<v<$170 & 241$<v<$1201 \\
    21 661 & H\,{\sc i}\,Br$\gamma$ & 13.3 & 0.03 & 0.26 & -1661$<v<$-540 & -471$<v<$401 & 470$<v<$1093 \\
    23 211 & [Ca\,{\sc viii}] & 128 & 0.0 & 0.24 & -1020$<v<$-323 & -258$<v<$142 & 207$<v<$1111 \\
    \hline
    \end{tabular}
    \begin{minipage}{15cm}
      Notes:
      (1) Ionization potential for the specific line transition.
      (2) Parameters for the Gaussian fits shown in Fig.~\ref{fig:grad}.
			(3) The uncertainty in velocity is $\sim$10 km s$^{-1}$.
    \end{minipage}
    \label{table:vel}
    \end{center}
    \end{table*}

    The Br$\gamma$~emission line (Fig.~\ref{fig:hbr}) presents a similar morphology traced by the [Si\,{\sc vi}] line, as evinced by the contours of the same line, but with a fraction of low intensity gas, with a more fragmented and diffuse distribution into the southwest cone, more associated with the high-velocity emission (Fig.~\ref{fig:hbrlh}, right panel), similar to the [Fe\,{\sc ii}] emission of the high-velocity gas (Fig.~\ref{fig:Felh}, right panel). Moreover, an extended structure seems to be correlated with the low-velocity [Fe\,{\sc ii}], in the southwest wall of the cone. Such extended emission becomes more evident when seen by separated velocity ranges, as shown in Fig.~\ref{fig:hbrlh} (right panel). As noticed in the [Si\,{\sc vi}] emission, the low-velocity gas is associated with the intensity peak of the Br$\gamma$ line, preferably shifted to negative velocities. 

\begin{figure}
    \resizebox{\hsize}{!}{\includegraphics{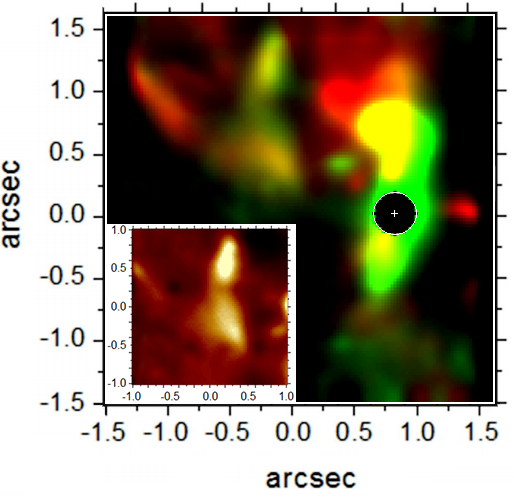}}
    \caption{Images of the low-velocity gas (Table~\ref{table:vel}) for the [Fe\,{\sc ii}] (red) and Pa$\alpha$ (green) lines. The small panel shows the upper-left zoomed region for the Pa$\alpha$ flux. The white circle denotes the masked region and the cross indicates the nucleus.}
     \label{fig:pafe}
    \end{figure}

    Despite the detection of an intense P$\alpha$~emission, strong signatures of atmospheric absorption bands in the wavelength range comprising the line detection make it difficult to analyze the emission after the telluric correction. We applied a Butterworth filter to the spectra shown in Fig.~\ref{fig:spectra}, with cut-off frequency \textsl{f}=0.45, to remove the high frequency noise and evince the P$\alpha$ emission at $\lambda$18571 \AA~(for a better explanation of the Butterworth filtering see \citealt{Menezes15,Menezes14}). However, we used the filtered spectrum to produce the images and measure the flux only for the P$\alpha$~line. As seen in Fig.~\ref{fig:hion} (top left panel), the flux map still shows a high-quality image and a strong similarity with others atomic emission lines, like the [Si\,{\sc vi}] emission (Fig.~\ref{fig:SiVI}), as much as the identified blobs and the velocities found for the presented FoV. We still note, however, a strong emission in region \textbf{B} (Fig.~\ref{fig:FeII}) with the respective spectrum shown in Fig.~\ref{fig:spectra}. Although it accounts for less than 1/10 of the flux peak, a narrow line profile with velocity close to zero is related to the same low-velocity emission found for the [Fe\,{\sc ii}] line (Fig.~\ref{fig:pafe}), which defines the filaments of the hourglass structure. Such equivalence is not seen for the Br$\gamma$ emission, maybe because it is too weak to be detected. Comparing to the [Fe\,{\sc ii}] velocity, the corresponding line profiles at the same locations are systematically shifted 5\,\AA~to the red. We checked for additional iron lines in the same spectral range and none was found, ensuring this emission comes from the Pa$\alpha$ line.

    The additional emission lines shown in Fig.~\ref{fig:hion} present little or no significant differences with respect to the [Si\,{\sc vi}] and Br$\gamma$~lines. The differences that we can easily identify are: (1) the He\,{\sc i} emission presents a smoother distribution between the blobs of gas, more similar to the Br$\gamma$ emission, which is to be expected given the low IP when compared to the other higher ionization lines; in fact, it will be shown in Sec.~\ref{sec:comp} that at least the [Si\,{\sc vi}] line has a fainter emission close to the jet when observed with a higher exposure time; (2) the southwest emission in redshift is weak in the [Ca\,{\sc viii}] line because the red wing of this line lies on a CO absorption band at $2.32~\mu$m and (3) the same emission in the southwest cone and the one relative to the blue blob located to the east of the \textbf{NE} point (Fig.~\ref{fig:FeII}) are intrinsically absent from the [Al\,{\sc ix}] line, which has the highest IP of 285\,eV. In the work of \citet{Izumi16}, a similar and more diffuse structure above the AGN is also seen in the continuum emission at v$_{rest}$=364\,GHz (their Fig.1), with spatial resolution $\sim3\times$ lower.

\begin{figure}
    \resizebox{\hsize}{!}{\includegraphics{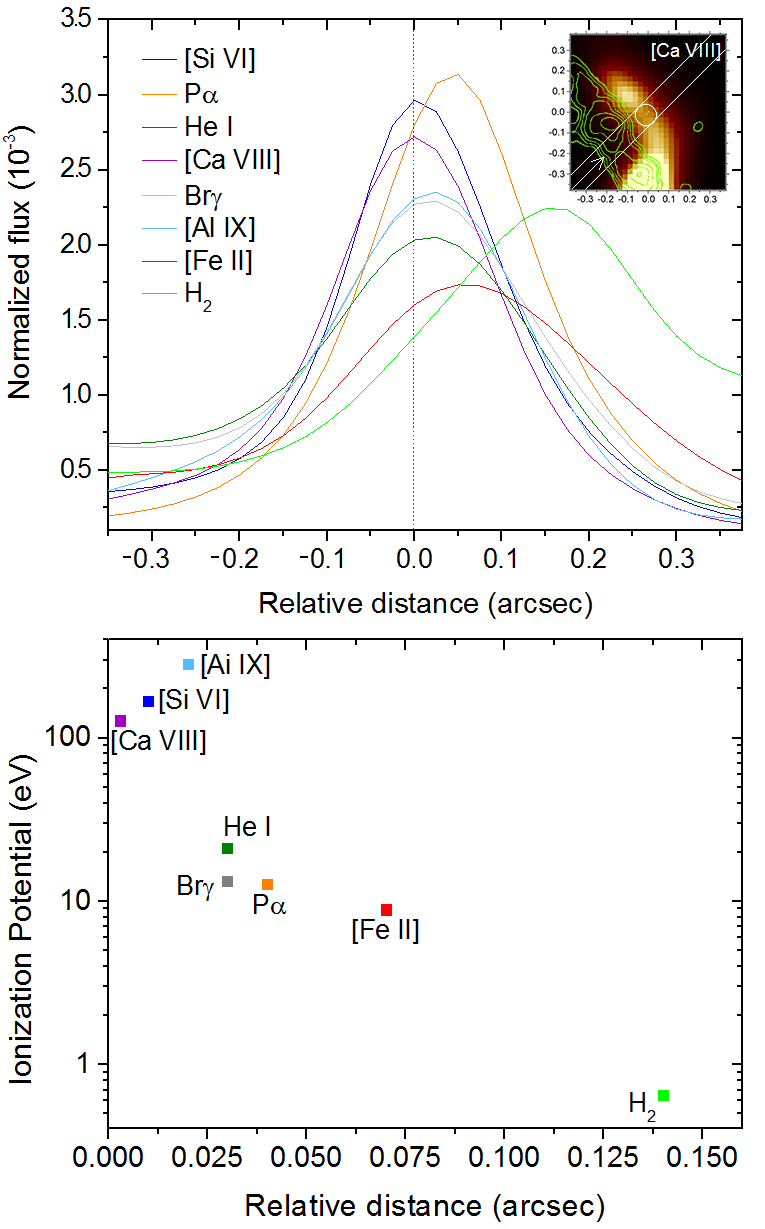}}
    \caption{Top panel: spatial profiles along the pseudo-slit shown at the top right of the same panel (with $\pm$0.1 arcsec of thickness) for the lines listed in Table~\ref{table:vel}. The reference centre is given by the [Ca\,{\sc viii}] line. Bottom: distances from the reference peak, derived by Gaussian fits, in function of the respective IPs of the lines.}
     \label{fig:grad}
    \end{figure}

    Considering the variety of IPs and the similarity in the morphology of some emission lines, we can ask if there is some ionization gradient along the peak emission close to the nucleus, in the vicinity of the jet. In order to check this assumption, in Fig.~\ref{fig:grad} (top panel) we plot the spatial profile for the eight analyzed emission lines, through the configuration shown in the zoom at the top-right of the panel, along the PA of 135\textdegree~and nearly perpendicular to the jet. All the profiles are well fitted by a Gaussian and have their parameters listed in Table~\ref{table:vel}, with the spatial reference given with respect to the Gaussian peak of the [Ca\,{\sc viii}] line (closest to the jet). Looking at the spatial profiles, the lines whose emission lies closer to the jet are those with high IPs. On the other hand, the molecular and the [Fe\,{\sc ii}] lines (with lower IP) lie more distant to the jet. The pixel scale for this fore-optics is 0.05 arcsec (with the spatial resolution improved by the deconvolution, as discussed in Sec.~\ref{sec:dec}), and the measured relative distances between these spatial profiles are as low as 0.01 arcsec, which is equivalent to a projected distance of $\sim$0.7\,pc. The precision of the fit is high enough to discriminate such small displacements, with a typical error of $\pm$0.002 arcsec for the Gaussian centroid.

    In the bottom panel of Fig.~\ref{fig:grad}, we plotted the Gaussian peak distances (relative to the [Ca\,{\sc viii}] emission) as a function of the IP. For the three lines with the highest IPs we found the smallest distances. We conclude that the emission lines with the highest IPs are located closer to the radio emission, but the cause of their ionization still cannot be discriminated between the jet and/or the higher exposure of the gas to the central ionizing source.     

    \subsection{The H$_{2}$~molecular emission}
    \label{sec:h2}

\begin{figure*}
    \resizebox{0.85\hsize}{!}{\includegraphics{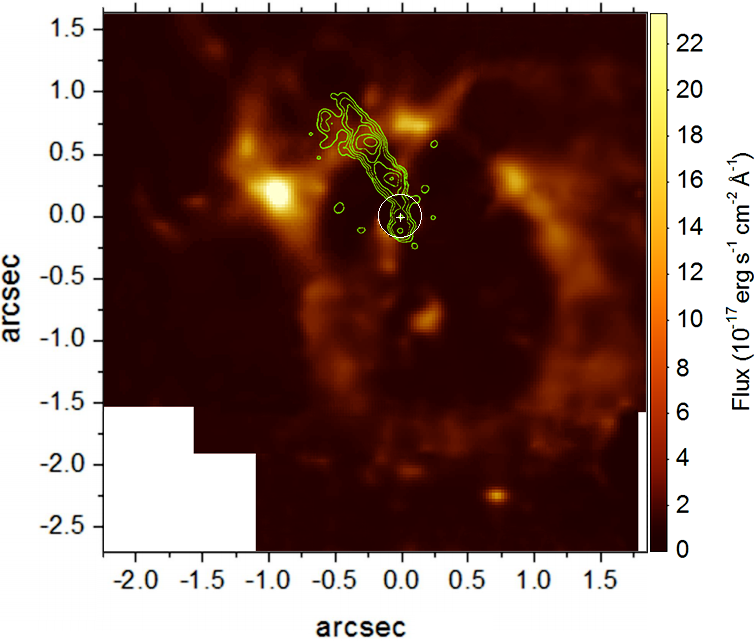}}
    \caption{Mosaic of the H$_{2}$ $\lambda$21218 \AA~line overlaid with the MERLIN 6 cm radio emission (green contours). The circle denotes the masked region and the cross indicates the nucleus.}
     \label{fig:h2}
    \end{figure*}

    As reported, molecular gas surrounding AGNs pervades a wide group of galaxies, irrespective of their morphological type or degree of activity. \citet{Ardila04,Ardila05} have shown that almost all active galaxies display H$_{2}$ emission in the inner 500\,pc. The molecular distribution in NGC\,1068 covers basically the same spatial extension shown by the other lines, but with a quite distinct morphology. The intensity map is shown in Fig.~\ref{fig:h2}, together with the contours of the radio emission.  
		
		The feature with the strongest intensity is located at $\sim$1 arcsec from the nucleus (close to point \textbf{B} of Fig.~\ref{fig:FeII}) with a PA of 81\textdegree, from where four unfolding H$_{2}$~arms seem to split apart from the central clump. Two of those arms, closest to the AGN, seem to form an asymmetric ring whose major axis has a PA of 30\textdegree$\pm$2\textdegree~(this PA is similar to that of the jet and of the low-velocity [Fe\,{\sc ii}] emission). These main features are the same found in \citet{Riffel14b} and \citet{Barbosa14} (both in their Figs.~4). Unlike the lines analyzed so far, the H$_{2}$~emission is not associated with the jet and its emission peak is located close to the inner border of the hourglass structure seen in [Fe\,{\sc ii}], i.e., still exposed to the radiation of the central source.		
		The $2 - 1 S(1)$/$1 - 0 S(1)$~vs. $1 - 0 S(2)$/$1 - 0 S(0)$~and $2 - 1 S(1)$/$1 - 0 S(1)$~vs. $1 - 0 S(3)$/$1 - 0 S(0)$~line ratios for region \textbf{B}, are (0.06, 1.8) and (0.06, 0.94) (Table~\ref{table:flux}), respectively, fully consistent with thermal X-ray excitation according to the model of \citet{Lepp83}. 
		In contrast, the same ratios calculated for regions \textbf{A} and \textbf{NE} are not well constrained by any single model, indicating contributions from both thermal and non-thermal excitation mechanisms.
		Possible origins for the H$_{2}$~clump will be discussed in Sec.~\ref{sec:arc}.  

    We draw attention to the small H$_{2}$ ring-like hole at the end of the radio emission with equivalent PA. This feature could suggest some interaction between the jet and the molecular arm in the region indicated as \textbf{NE} in Fig.~\ref{fig:FeII}, showing the spot where the jet hits and breaches the molecular arm. In fact, the identified knots \textbf{C} and \textbf{NE} coincides, in projection, with the molecular gas (Fig.~\ref{fig:h2}).

    The complex and multi-peaked line profiles in the NLR of this galaxy (as seen in Fig.~\ref{fig:spectra} - upper-right panel) are a longstanding and well-known subject of debate, as described by \citet{Cecil90} in the optical. In the NIR we noted that the line splitting starts where the jet intercepts the inner wall of the molecular arm (\textbf{NE} knot) and gradually breaks its profile in two, both increasing in absolute velocity, as far as it goes through the molecular arm. This arm eventually ends up in the H$_{2}$ hole, which is filled by a high ionized blob of [Si\,{\sc vi}] (Fig.~\ref{fig:h2s}), where the MERLIN 5\,GHz radio emission is no longer detected.

    \subsubsection{H$_{2}$ kinematics}
    \label{sec:h2kin}

\begin{figure*}
    \resizebox{\hsize}{!}{\includegraphics{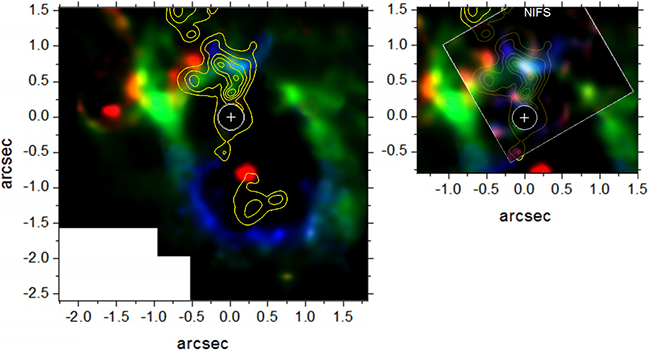}}
    \caption{RBG composition for the H$_{2}$~$\lambda$21218 \AA~emission line with the velocity ranges shown in Table~\ref{table:vel}, overlaid with the coronal line emission of [Si\,{\sc vi}] (yellow contours). Left panel: image for the SINFONI data. Right panel: comparison with one data cube from NIFS. The circle denotes the masked region and the cross indicates the nucleus.}
     \label{fig:h2s}
    \end{figure*}

    In the RGB composition of Fig.~\ref{fig:h2s} it is clear that the southwest molecular wall is dominated by velocities in blueshift and the northeast emission by lower velocities near the rest frame, with fragmented blobs showing the higher velocities. There is a region associated with the highly ionized emission of [Si\,{\sc vi}] and of the [O\,{\sc iii}] cone, preferably presenting blueshifted velocities, and a concentration of redshifted blobs distributed along the H$_{2}$ arm, closest to the molecular emission peak. The main picture of this kinematics agrees with the CO(6-5) velocity map shown by \citet{Burillo16} (their Fig.2b).   

\begin{figure*}
    \resizebox{\hsize}{!}{\includegraphics{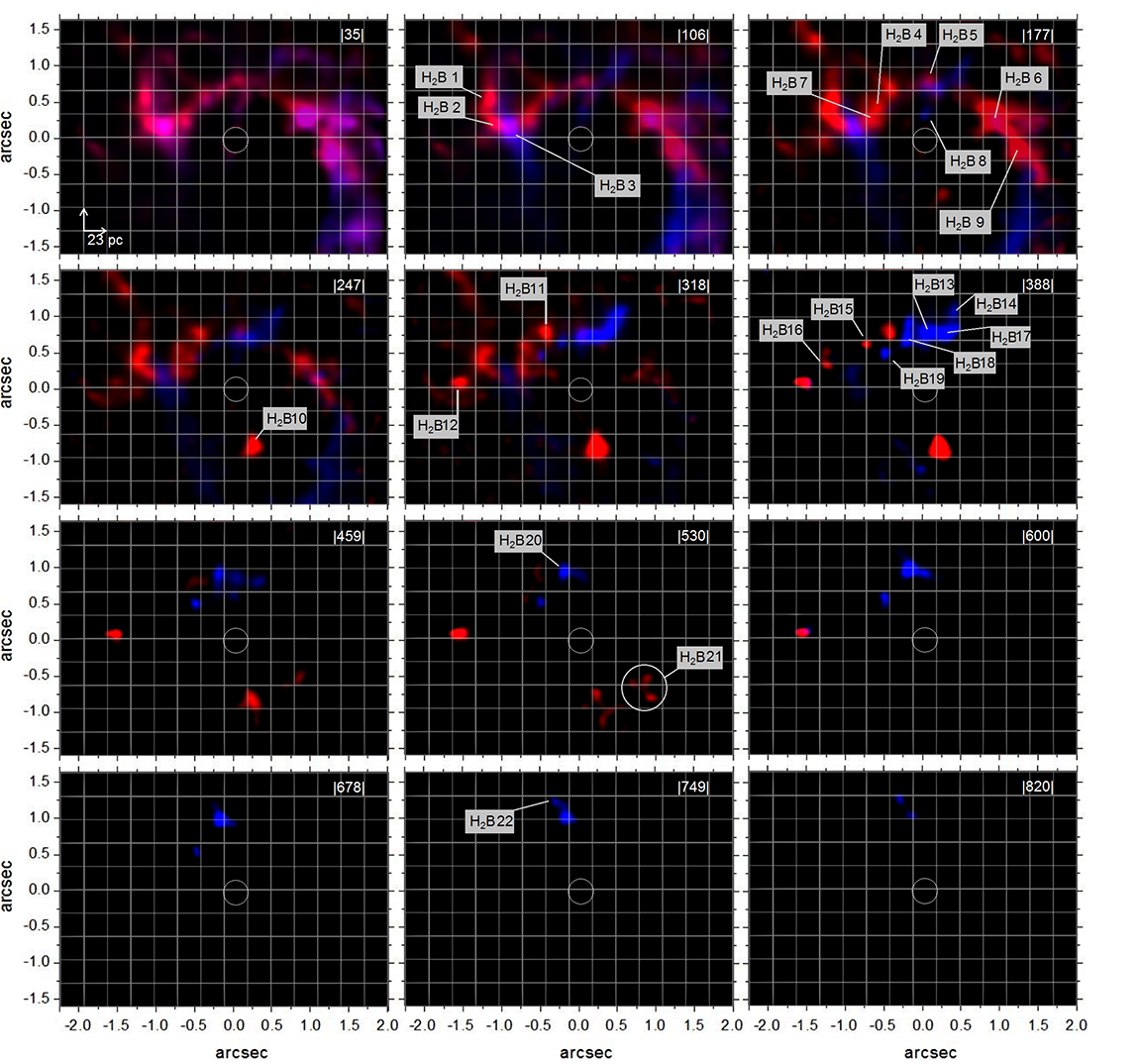}}
    \caption{BRV maps for the H$_{2}$~$\lambda$21218 \AA~emission line. The velocity ranges from -820 km s$^{-1}<v<593$~km s$^{-1}$, with varying intervals of velocities, representative of significant changes in the gas distribution. The southernmost blob, as discussed in this Section, is mentioned only in the text. The grid has a small square of 23 pc$\times$23 pc and the the corresponding absolute velocity of each map is show on top. The twenty two selected molecular blobs are indicated as ``H$_{2}$ Blob(number)'', with the measured peak velocity, velocity dispersions, distances from the nucleus and position angles shown in Table~\ref{table:H2b}. The circle denotes the masked region centred on the nucleus.}
     \label{fig:H2rbvmaps}
    \end{figure*}

    The large concentration of blobs in the northeast cone is seen as a fragmentation of two H$_{2}$~arms with those in redshift apparently uncorrelated with the radio emission. The high-velocity H$_{2}$~is shown in yellow contours in Fig.~\ref{fig:Felh} (right panel) and is mostly associated with the ionized gas, with the exception of the farthest eastern blob in the FoV (the H$_{2}$B12 blob in Fig.~\ref{fig:H2rbvmaps}) and the lowest clump (shown in Fig.~\ref{fig:Felh}, right panel).

    In the southwest part, there is an isolated redshifted molecular blob inside the cavity, apparently exposed to the AGN radiation and mostly preceding the highly ionized gas.
    This cavity is clearly surrounded by a moving gas, which is behind the galactic disc and mostly seen in blueshift, with velocities between -282 km s$^{-1}<v<-99$~km s$^{-1}$, which shows a decreasing blueshift for regions farther from the centre. This deceleration was also noted by \citet{Thaisa12}.

    The velocity of the molecular gas ranges from -820 km s$^{-1}<v<1201$~km s$^{-1}$. However, there is no emission between the interval of 593 km s$^{-1}<v<848$~km s$^{-1}$, and further there is a barely resolved single blob (FWHM=0.12 arcsec) with $v$=1017 km s$^{-1}$, $\sigma$=99$\pm$10 km s$^{-1}$ and distance of 170\,pc at a PA=-160\textdegree, located at the southern edge of the FoV. This emission is precisely co-spatial with another blob detected, with $v$=131 km s$^{-1}$~($\sigma$=68$\pm$46 km s$^{-1}$), and the line profile blended to the emission close to the rest frame. It is hard to suppose that an instrumental artifact would appear twice and at the same place along the spectra, even in a single data cube. If this feature is, in fact, real, little can be said about these co-spatial blobs. The large velocity gap between each of them, of 886 km s$^{-1}$, does not evoke any physical explanation considering that there is no ionized gas besides the molecular emission. If this detection represent a symmetrical kinematic phenomenon, it would imply an object with redshift of 574 km s$^{-1}$~and two spatially unresolved velocity components of $\mid v\mid\sim443$~km s$^{-1}$, suggesting an expanding structure with unknown morphology, resembling a supernova remnant.

    All the molecular blobs were identified in the BRV maps shown in Fig.~\ref{fig:H2rbvmaps}, with some properties listed in Table~\ref{table:H2b}. Looking at the velocity dispersion in the northeast cone, no distinction can be made from those close to the jet, with the mean value almost the same for the blobs in blueshift and in redshift (of 87 km s$^{-1}$~and 85 km s$^{-1}$, respectively). The mean velocity dispersion for all the molecular blobs is 95 km s$^{-1}$, lower than the previous calculations for the [Si\,{\sc vi}] (111 km s$^{-1}$) and [Fe\,{\sc ii}] blobs (152 km s$^{-1}$).

    As already noted by \citet{Barbosa14}, the H$_{2}$B10 and H$_{2}$B20 blobs seem to be symmetrically related, although each side of the cones presents noticeable asymmetry. This indicates a similar interaction and strength between the jet and the molecular gas on both sides of the cones, reinforcing the presence of the southwestern jet counterpart. These H$_{2}$~``bullets'' are at a projected distance of 70 and 62\,pc from the nucleus, and have velocities of -586$\pm$12 km s$^{-1}$~and 300$\pm$10 km s$^{-1}$ in the northeast and southwest cones, respectively.  

    \begin{table}
    \begin{center}
    \caption[H2b]
    {Peak velocity, velocity dispersions, distance from the nucleus and position angle measured for the molecular blobs shown in Fig.~\ref{fig:H2rbvmaps}.}
    \begin{tabular}{cccccc}
    \hline \hline
    Blob ID & v$^{1}$ (km s$^{-1}$) & $\sigma^{2}$ (km s$^{-1}$) & Distance$^{3}$ (pc) & PA$^{4}$ & PA$^{5}$ \\ \hline
    H$_{2}$B1 & 40 & 99 & -88 & 66 & 29 \\
    H$_{2}$B2 & 8 & 105 & -87 & 80 & 43 \\
    H$_{2}$B3 & -90 & 92 & -72 & 83 & 46 \\
    H$_{2}$B4 & 61 & 96 & -66 & 55 & 18 \\
    H$_{2}$B5 & 25 & 91 & -61 & -4 & -41 \\
    H$_{2}$B6 & 24 & 103 & -75 & -71 & -108 \\
    H$_{2}$B7 & 20 & 103 & -46 & 67 & 30 \\
    H$_{2}$B8 & -112 & 75 & -24 & -2 & -39 \\
    H$_{2}$B9 & 49 & 75 & 95 & -84 & -- \\
    H$_{2}$B10 & 300 & 47 & 59 & -162 & -- \\
    H$_{2}$B11 & 96 & 114 & -66 & 30 & -7 \\
    H$_{2}$B12 & 376 & 141 & -108 & 87 & 50 \\
    H$_{2}$B13 & 31 & 92 & -55 & -2 & -39 \\
    H$_{2}$B14 & -283 & 59 & -83 & -21 & -58 \\
    H$_{2}$B15 & 171 & 135 & -75 & 52 & 15 \\
    H$_{2}$B16 & 92 & 128 & -100 & 75 & 38 \\
    H$_{2}$B17 & -270 & 123 & -61 & -22 & -59 \\
    H$_{2}$B18 & -356 & 61 & -51 & 20 & -17 \\
    H$_{2}$B19 & -470 & 100 & -53 & 49 & 12 \\
    H$_{2}$B20 & -586 & 99 & -68 & 15 & -22 \\
    H$_{2}$B21 & 492 & 79 & 80 & -127 & -- \\
    H$_{2}$B22 & -757 & 79 & -92 & 16 & -21 \\
    \hline
    \end{tabular}
    \begin{minipage}{8cm}
      Notes:
			(1) The uncertainty in velocity is $\sim$10 km s$^{-1}$.
      (2) Corrected for instrumental broadening.
			(3) Distance offset of each blob from the bulge centre; negative numbers mean a northeast offset with respect to a line orthogonal to the cone's major axis.
      (4) Blob's position angle relative to the northeast and (5) relative to the cone's major axis, of 34\textdegree.
    \end{minipage}
    \label{table:H2b}
    \end{center}
    \end{table}

    A larger amount of molecular blobs is found when compared to the previous lines. We identified only those where a ubiquitous velocity could be extracted. This suggests a high degree of fragmentation of the molecular arms, with the blobs located mainly above them and not as much collimated as those ones identified in the ionized lines. 

\begin{figure}
    \resizebox{\hsize}{!}{\includegraphics{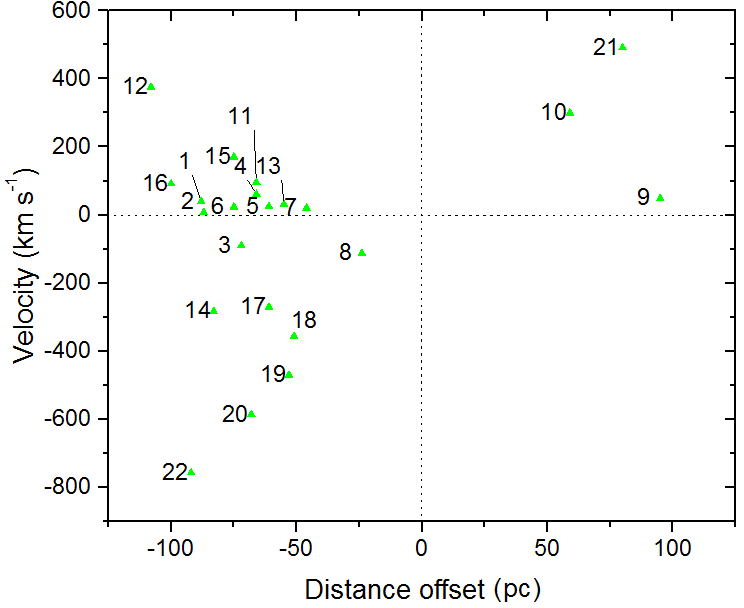}}
    \caption{Diagram for the position and velocity of the H$_{2}$~blobs identified in Fig.~\ref{fig:H2rbvmaps}.}
     \label{fig:H2blobs}
    \end{figure}

    In Fig.~\ref{fig:H2blobs} we plot the blobs velocities vs. its distance to the centre. The high number of blobs with low-velocity may be due to the fact that they are just at the start of the expanding process, along the molecular arms. Although with only 22 blobs, this plot resembles a conical diagram, typical of conical outflows \citep{Das06}, with the farthest blobs having the largest velocities.		

    We emphasize that it is the first time that the blobs in the NIR (in [Fe\,{\sc ii}], [Si\,{\sc vi}] and H$_{2}$ lines) are properly mapped. 

    \subsubsection{Comparison to the NIFS data}
		\label{sec:comp}

		Two main reasons are given to perform the comparison between the data from SINFONI and NIFS. Firstly, given their equivalent pixel scales, it is worth to highlight the similarity between them after our image processing routine to address the consistency of the deconvolution procedure even after we conclude that the SINFONI data cubes resulted in better spatial resolutions. Secondly, the NIFS observations have nearly twice the exposure time and spectral resolution. From that, fainter emissions and sub-structures in velocity may be detected, complementing our analysis. The data from NIFS presented here is limited to the H$_{2}$ and [Si\,{\sc vi}] emission lines, and were already published by \citet{Thaisa12,Riffel14b,Barbosa14}.
		
		The NIFS data are delimited by the gray square in the RGB compositions of Figs.~\ref{fig:SiVIk} and \ref{fig:h2s} and have the SINFONI RGB mosaic in the background, with similar velocity ranges. The RGB composition was chosen to perform this comparison, instead of intensity maps, because it is able to address differences both in the gas morphology and velocity.

    In the case of the molecular gas, the first noticeable difference is the faint emission spread into the cavity in the NIFS image, which is absent in the SINFONI data. The new structures seem to fill the entire area covered by the [Si\,{\sc vi}] contours and show more filaments possibly connected to the cavity inner wall. We have checked that no fainter emission is significantly found in the rest of the cavity, in the southwest cone. Secondly, the mixing in the RGB colours is naturally more evident with a higher spectral resolution, which is dependent on the line profiles near the spectral interval chosen to produce the images. Lastly, the NIFS image depicts a faint cavity correlated to the [Si\,{\sc vi}] emission, barely seen before. 
		
		For the [Si\,{\sc vi}] emission, it is clear that a better resolution was reached by the SINFONI data cubes. As mentioned before for the molecular emission, fainter emissions start to be visible, as the one in blueshift to the east of the radio emission. The presence of sub-structures in velocity not seen in some of the SINFONI [Si\,{\sc vi}] blobs is also noticeable. 

    The conclusion we can draw from this comparison is that the nuclear main picture remains the same as the one derived from the SINFONI data, with some differences only related to the longer exposure time and the spectral resolution of the NIFS data. This comparison is particularly relevant because some resolved blobs are not seen before our data treatment, and one may ask how real they are. With two distinct instruments and telescopes we were able to show that the deconvolution method, allied to other applied techniques, leads to the same identified structures. 
    
    \section{The nuclear architecture of NGC\,1068}
    \label{sec:arc}

    \subsection{H$_{2}$ vs. the [Si\,{\sc vi}] and [Fe\,{\sc ii}] emissions}
    \label{sec:co}

    There are clear signs of interaction between the [Si\,{\sc vi}] and the molecular gas along its path through the upper molecular wall (Fig.~\ref{fig:h2s}), where two of the [Si\,{\sc vi}] clumps are directly associated with molecular clumps, both in blueshift, and with the highest measured velocities. For instance, this association can be seen between the CLB12 and H$_{2}$B20 blobs (with $v$=-830 km s$^{-1}$~and -586 km s$^{-1}$, shown in Figs.~\ref{fig:Sirbvmaps} and ~\ref{fig:H2rbvmaps}, respectively) and the CLB13 and H$_{2}$B19 blobs (with $v$=-786 km s$^{-1}$~and -470 km s$^{-1}$, respectively). We interpret these blobs as coming from the same fragmented molecular arm, being further accelerated through the small molecular cavity.

    The highly blueshifted H$_{2}$ structures in the northern cone, which are associated with the [Si\,{\sc vi}] emission, define an open angle of $\sim$82\textdegree~and PA$\sim$9\textdegree, equivalent to the [O\,{\sc iii}] cone shown in Fig.~\ref{fig:SiOIII}. This region, delimited by the [O\,{\sc iii}] emission line and now by the molecular gas, could be directly exposed to the radiation of the central source, defining a second ``inner cone'' (perpendicular to the accretion disc), as commented in Sec.~\ref{sec:OIII}. 
				
		\begin{figure*}
    \resizebox{\hsize}{!}{\includegraphics{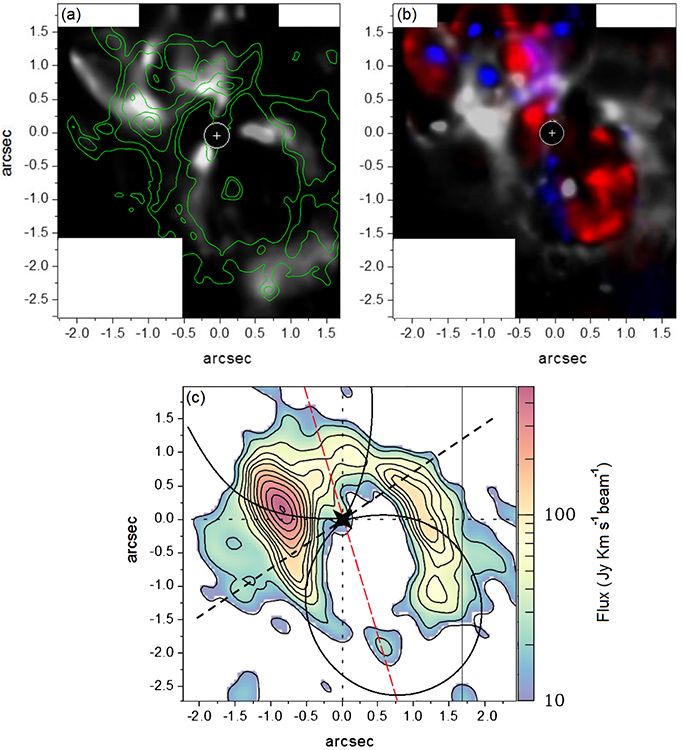}}
    \caption{(a): molecular emission (contours) and the low-velocity walls of the hourglass - phase-1 (white); (b): high-velocity [Fe\,{\sc ii}] emission - phase-2 (same as the right panel of Fig.~\ref{fig:Felh}) and the molecular emission - phase-3 (white); (c): CO(6-5) intensity map of \citet{Burillo14} (with beam size of 0.4 arcsec~$\times$~0.2 arcsec at PA=50\textdegree) and the scheme for the outer walls of the hourglass; the black dashed line is the axis perpendicular to the cones, of 34\textdegree$+$90\textdegree$=$124\textdegree, the red line illustrates the axis of the [O\,{\sc iii}] inner cone and the vertical gray line delimits the area covered by the previous panels. The circle denotes the masked region and the cross indicates the nucleus.}
     \label{fig:3f}
    \end{figure*}

    In Fig.~\ref{fig:3f}, we show the three-phase gas morphology represented by the wall of the hourglass shown by the low-velocity [Fe\,{\sc ii}] emission - phase-1; the high-velocity compact blobs of ionized gas that fill the hourglass volume (here represented only by the [Fe\,{\sc ii}] emission) - phase-2; and the H$_{2}$ distribution defined by the thick and irregular walls of a bubble surrounding a cavity - phase-3. In panel (a) of Fig.~\ref{fig:3f} we compare the phase-1 (represented by Fig.~\ref{fig:Felh} - left panel, in white) with phase-3 (represented by Fig.~\ref{fig:h2s} - left panel, in green contours). Near the southwest apex of the cone there is almost no correlation between both emissions, but it is possible to note that further away the wall of the hourglass partially reach the inner edge of the bubble. The association between the low-velocity [Fe\,{\sc ii}] and H$_{2}$ could represent a photo-dissociation-region (PDR), with the ionized gas exposed to the radiation of the central source and preceding the molecular gas emission. The southernmost wall seems to be more coincident, perhaps because some ``leaking'' of radiation, because we see an extended [Fe\,{\sc ii}] emission as seen in Fig.~\ref{fig:Felh}, right panel. In the northeast cone, both gas phases are nearly coincident for the region occupied by the hourglass, but with two molecular filaments extending outside the cone.  
		
		In panel (b) of Fig.~\ref{fig:3f} we compare the phase-2 with phase-3 gas distributions. These phases are mostly uncorrelated to each other and, moreover, show signs of a complementary morphology. In the southwest cone there is a sharp boundary between the [Fe\,{\sc ii}] and molecular emissions, where the high-velocity [Fe\,{\sc ii}] gas is clearly confined to the molecular cavity. This configuration demands that the dynamics of the low and high-velocity [Fe\,{\sc ii}] gas are strictly distinct.
    In the northeast cone the high-velocity [Fe\,{\sc ii}] gas is mostly not confined inside any molecular structure. In this cone, both emissions tend to indicate a single structure distinguished only by its degree of ionization, with the [Fe\,{\sc ii}] gas mostly distributed in form of blueshifted blobs and redshifted filaments. On the other hand, the H$_{2}$ gas presents fragmented molecular arms closer to the AGN and above them a series of small cavities filled with [Fe\,{\sc ii}] blobs.
    
		Panel (c) of Fig.~\ref{fig:3f} shows the CO(6-5) intensity map taken from \citealt{Burillo14} (with beam size of 0.4 arcsec~$\times$~0.2 arcsec at PA=50\textdegree) as a sub-phase of phase-3, with the contour of the outer limit of the hourglass structure. A higher resolution image is available in \citet{Burillo16}, but our preference was an image with a ring-like morphology as we see for the H$_{2}$ emission. 		
		By comparing to the panels (a) and (b),  we see that almost all CO emission is correlated to the H$_{2}$ gas, but not the contrary. The eastern side of the CO emission is in close agreement with the H$_{2}$ gas, with their emission peak spatially coincident, as well as the southwest molecular wall. However, the fraction of H$_{2}$ not associated with the cold molecular gas is preferably aligned along the hourglass axis and, mainly, along the axis defined by the inner cone of [O\,{\sc iii}] (red line), which is suggestive of temperatures high enough to destroy the CO, but not the H$_{2}$ molecules. An important finding is that most of the CO emission is internal to the area defined by the ``shadow'' of the ionization cones, beyond the outer edge of the hourglass. Furthermore, the CO emission has an asymmetric distribution with respect to the dashed line perpendicular to the hourglass axis, located far from the side of the hourglass wall closest to the inner cone defined by the [O\,{\sc iii}] emission. A similar trend is seen in the HCN\,3-2 and HCO$^{+}$\,3-2 molecular emissions in the work of \citet{Imanishi16}.
		
		The eastern H$_{2}$ filament outside of the northeast cone (as discussed in Sec.~\ref{sec:bubble}) is not seen in the CO map, at least not as bright as the H$_{2}$ emission, indicating a hot filament of molecular gas where the CO molecules were already destroyed through a stronger interaction with the AGN during the ring/bubble expansion. 
			
			The central CO detection will be discussed in light of the work of \citet{Gallimore16} in Sec.~\ref{sec:torus}.

    \subsubsection{The origin of the ionized blobs}
    \label{sec:origin}

    \begin{figure}
    \resizebox{\hsize}{!}{\includegraphics{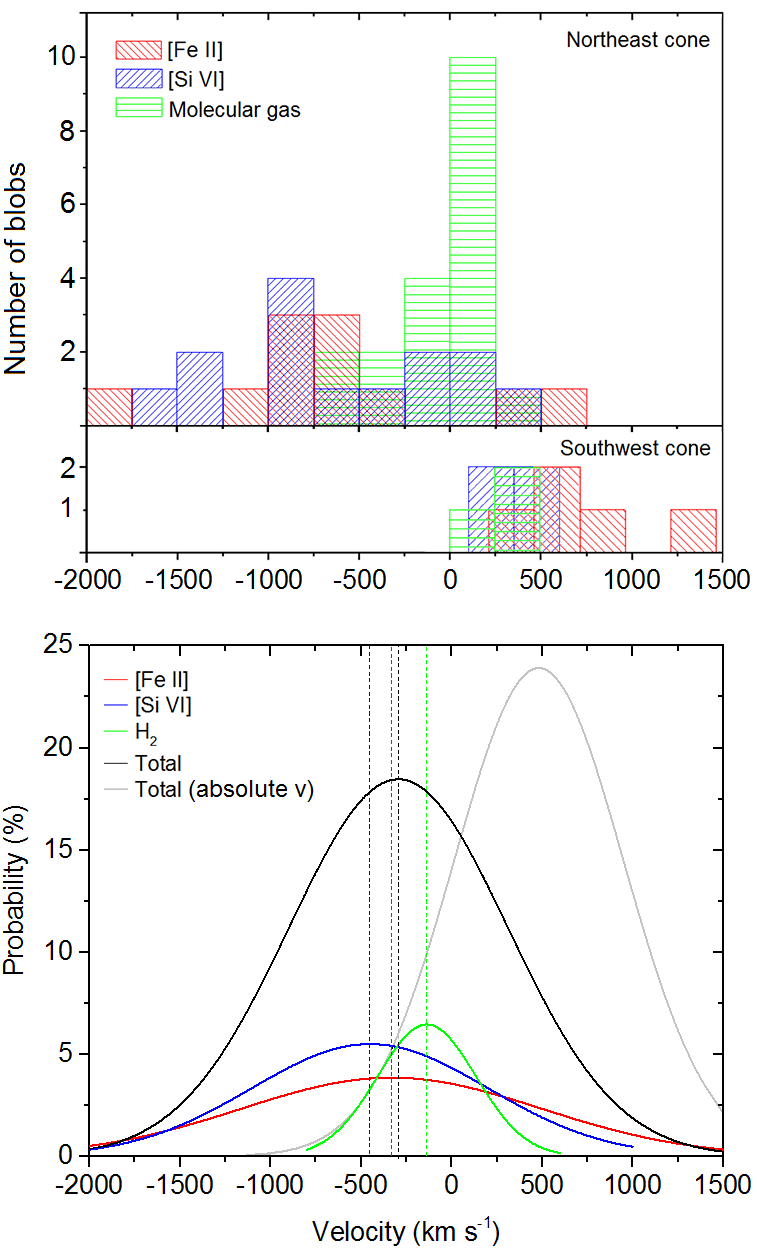}}
    \caption{Top: the frequency of occurrence for the H$_{2}$, [Fe\,{\sc ii}] and [Si\,{\sc vi}] blobs selected in the BRV maps, with their measured properties listed in the respective tables. Bottom: the probability distribution for the detected velocity range, with each set of blobs. The most probable velocity and the standard deviation ($\sigma$) are shown in Table~\ref{table:hist}. The gray curve represents the plot for the absolute velocity for the total number of blobs.}
     \label{fig:histogram}
    \end{figure}

    \begin{table}
    \begin{center}
    \caption[hist]
    {Parameters for each Gaussian function shown in Fig.~\ref{fig:histogram} (bottom panel), based on the velocity distribution of the identified blobs, with $n$ being the respective number of blobs.}
    \begin{tabular}{|c|c|c|c|c|}
    \hline \hline
    $\lambda_{vac}$~(\AA) & Line & $n$ & Peak (km s$^{-1}$) & $\sigma$ (km s$^{-1}$) \\ \hline
    16 440 & [Fe\,{\sc ii}] & 16 & -332 & 829 \\
    19 641 & [Si\,{\sc vi}] & 18 & -452 & 653 \\
    21 218 & H$_{2}$ & 22 & -138 & 272 \\
    \hline
    \multicolumn{3}{c}{Average $v$ of 56} & -294 & 605 \\
    \multicolumn{3}{c}{Average $\mid v\mid$ of 56} & 508 & 437 \\
    \hline
    \end{tabular}
    \label{table:hist}
    \end{center}
    \end{table}

\begin{figure*}
    \resizebox{\hsize}{!}{\includegraphics{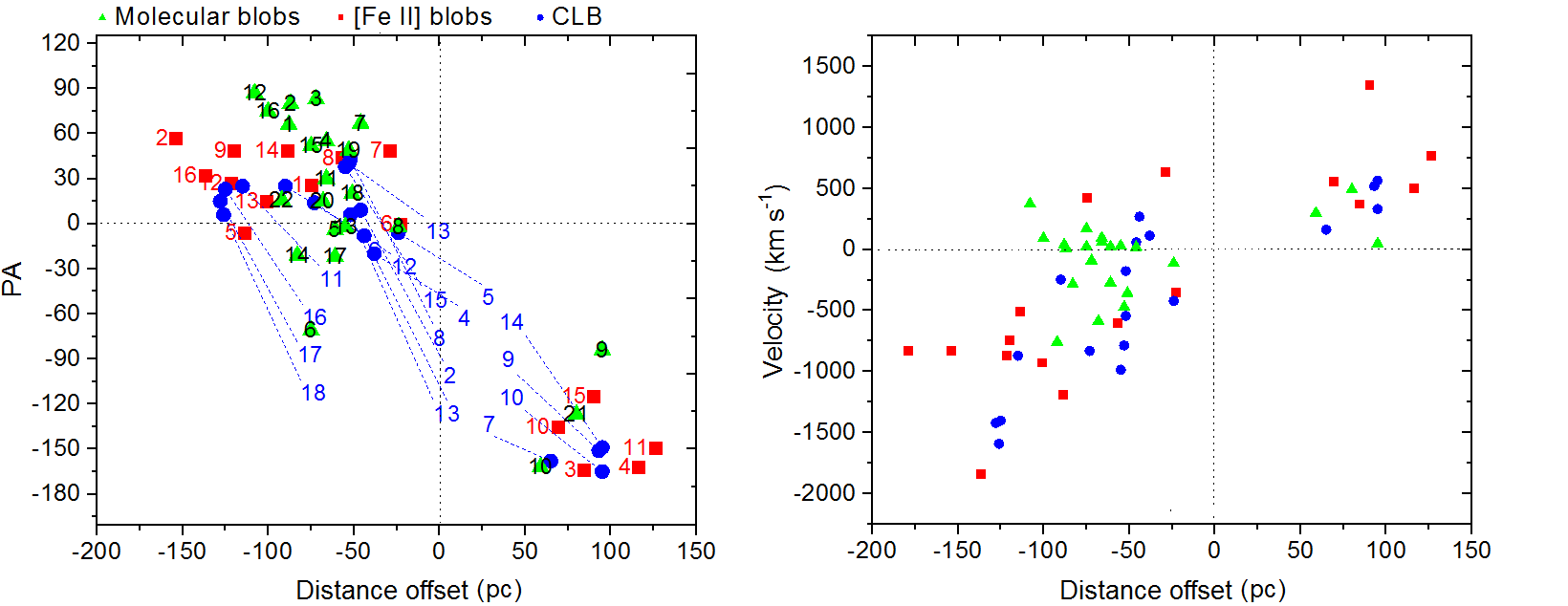}}
    \caption{Left panel: position intersection diagram for the H$_{2}$, [Fe\,{\sc ii}] and [Si\,{\sc vi}] blobs selected in the BRV maps, with their measured properties listed in the respective tables. Right panel: position velocity diagram for the same blobs, showing their radial acceleration.}
     \label{fig:gdpa}
    \end{figure*}

		In this Section we will concentrate on the analysis of possible correlations between the properties of all the identified blobs.
    We have presented the emission line kinematics in the form of BRV maps, as shown in Figs.~\ref{fig:Ferbvmaps1}, ~\ref{fig:Ferbvmaps2}, \ref{fig:Sirbvmaps}, \ref{fig:H2rbvmaps}. The frequency they appear for different velocity ranges, with the probability functions for each set of blobs, are shown in Fig.~\ref{fig:histogram} and the parameters listed in Table~\ref{table:hist}. The most probable velocity, considering all the blobs in blueshift and redshift, is -294 km s$^{-1}$, and if we consider the absolute velocity of the total number of blobs, we arrive at 508 km s$^{-1}$. In Fig.~\ref{fig:gdpa} (left panel), we plot a position intersection diagram of all blobs to highlight those which are spatially correlated, with the identification number listed in Tables~\ref{table:Feb}, ~\ref{table:Sib} and ~\ref{table:H2b}.  

    The overlapping points are co-spatial blobs, and occur mostly for small PAs. We draw attention to the narrow angle covered by the coronal line blobs, extending up to $\sim$130 pc and showing a high degree of collimation, and the two distinct loci, preferably occupied by the H$_{2}$ and [Fe\,{\sc ii}] blobs that are distinguished by the distance they reach (with the [Fe\,{\sc ii}] blobs more extended), and the collimation degree (with the [Fe\,{\sc ii}] blobs more colllimated).

    The most striking feature of the blobs is related to their radial acceleration, which can be seen individually for each emission line in Figs.~\ref{fig:Feblobs}, \ref{fig:Siblobs} and \ref{fig:H2blobs} or, more straightforwardly, all together in Fig.~\ref{fig:gdpa} (right panel). The contribution of the high-velocity gas (as characterized by almost all the compact blobs) in the total flux of each emission line, without considering the masked region, is calculated as 9\%, 2\% and 5\% for the [Fe\,{\sc ii}], H$_{2}$ and [Si\,{\sc vi}] emission lines, respectively.

    Interestingly, the [Fe\,{\sc ii}] blobs present velocities as high as the [Si\,{\sc vi}] ones, even considering that the CL kinematics is clearly more affected by the presence of the jet. We may ask if the jet plays a major role as the radial accelerating mechanism for the highly ionized gas emission, instead of accounting mainly to the lateral expansion of the gas, ionizing elements to higher IPs and sweeping off the material from the jet path, as revealed by the HST long-slit spectroscopy in \citet{Axon98}.

    On the other hand, the occurrence of molecular emission in a limited region above the AGN suggests that the H$_{2}$~molecules survive the central radiation up until a certain distance from the centre, where it stops emitting not because of the lack of an excitation mechanism, but because they are probably destroyed by the continuous exposure to the central radiation, as attested by the presence of ionized blobs. 		

    In Sections~\ref{sec:h2kin} and \ref{sec:clb} we presented the highly collimated CLBs (Fig.~\ref{fig:SiVI}, right panel) and the H$_{2}$B10 and H$_{2}$B20 counterparts as evidence of a certain symmetry between the blown blobs in both cones. In principle, this correlation can be found by looking for the blobs that have PA 180\textdegree~apart and similar distances from the AGN, with nearly the same absolute velocities (assuming only those with blueshift in the northeast cone and with redshift in the southwest cone). As we know that the velocity regime is much broader for the velocities in blueshift, we could relieve the last condition for the velocity (i.e., considering any absolute velocity for the blobs in both sides of the cones). Following this criterion, we found a good candidate for a couple of [Fe\,{\sc ii}] blobs counterparts, namely, the FeB14 and FeB10, and the farthest CLBs 16, 17 and 18 with respect to the CLBs 10, 9 and 14. They differ in velocity by a factor of 2$\times$ for the first two components and 3$\times$ for the CLBs. However, given the intrinsic asymmetry between the cones and, therefore, the blob ejection, we cannot assume they are emerging by a strictly symmetrical process. The high number of [Fe\,{\sc ii}] blobs also prevents us from defining a reliable connection between them. What deserves attention is, in fact, the way that a certain symmetry is preserved considering that, in the northeast cone, the blobs are being fragmented from the molecular arm and, in the southwest cone, they seem to be ``excavated'' and blown away along the inner molecular wall.      

\begin{figure*}
    \resizebox{\hsize}{!}{\includegraphics{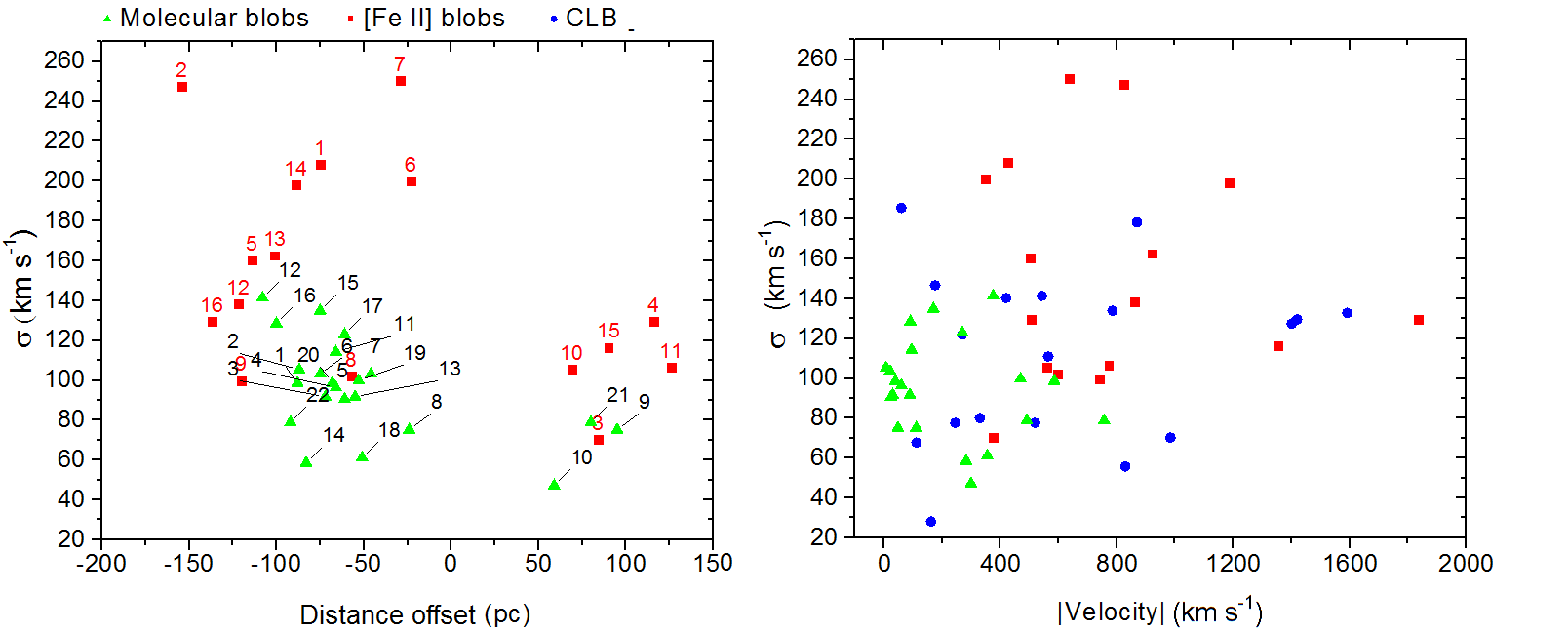}}
    \caption{Left panel: velocity dispersion vs. distance for the H$_{2}$ and [Fe\,{\sc ii}] blobs selected in the BRV maps, with their measured properties listed in the respective tables. Right panel: radial projected velocity vs. velocity dispersion for the same blobs.}
     \label{fig:gds}
    \end{figure*}

    In Fig.~\ref{fig:gds} (left panel) we show the blob's velocity dispersion as a function of the distance from the AGN named according to Tables~\ref{table:Feb} and ~\ref{table:H2b}. To better highlight our point, the [Si\,{\sc vi}] blobs are not shown in this diagram, being spread for all this parameter space. In spite of the FeB8, which lies in one of the H$_{2}$~arms, all the [Fe\,{\sc ii}] blobs either present a much higher velocity dispersion or have similar values for large distances. No molecular blob is found above $\sigma$=141 km s$^{-1}$ or a [Fe\,{\sc ii}] blob with lower velocity dispersion up to a distance of 122 pc. However, in general, there is no correlation between the real vicinity of the blobs and the increase in velocity dispersion. This behaviour strongly suggests that the velocity dispersion could provide a good indicator whether or not the molecule survives and to the later release/ionization of the Fe atoms possible attached to dust grains and/or shielded by the H$_{2}$~molecules. In this sense, the value of $\sigma$=141 km s$^{-1}$ would represent an upper limit to the existence of H$_{2}$ blobs. The smaller velocity dispersion for larger distances seen for some [Fe\,{\sc ii}] blobs could mean or a transition from a molecular blob with smaller velocity dispersion to a [Fe\,{\sc ii}] blob, or a higher susceptibility to the blob ionization.

    The right panel of Fig.~\ref{fig:gds} shows that velocity dispersion is not related to absolute velocity. However once again, the molecular blobs concentrate in a ``[Fe\,{\sc ii}]-free'' region with smaller values for both variables, with the CLBs spread all over the diagram.

\begin{figure*}
    \resizebox{\hsize}{!}{\includegraphics{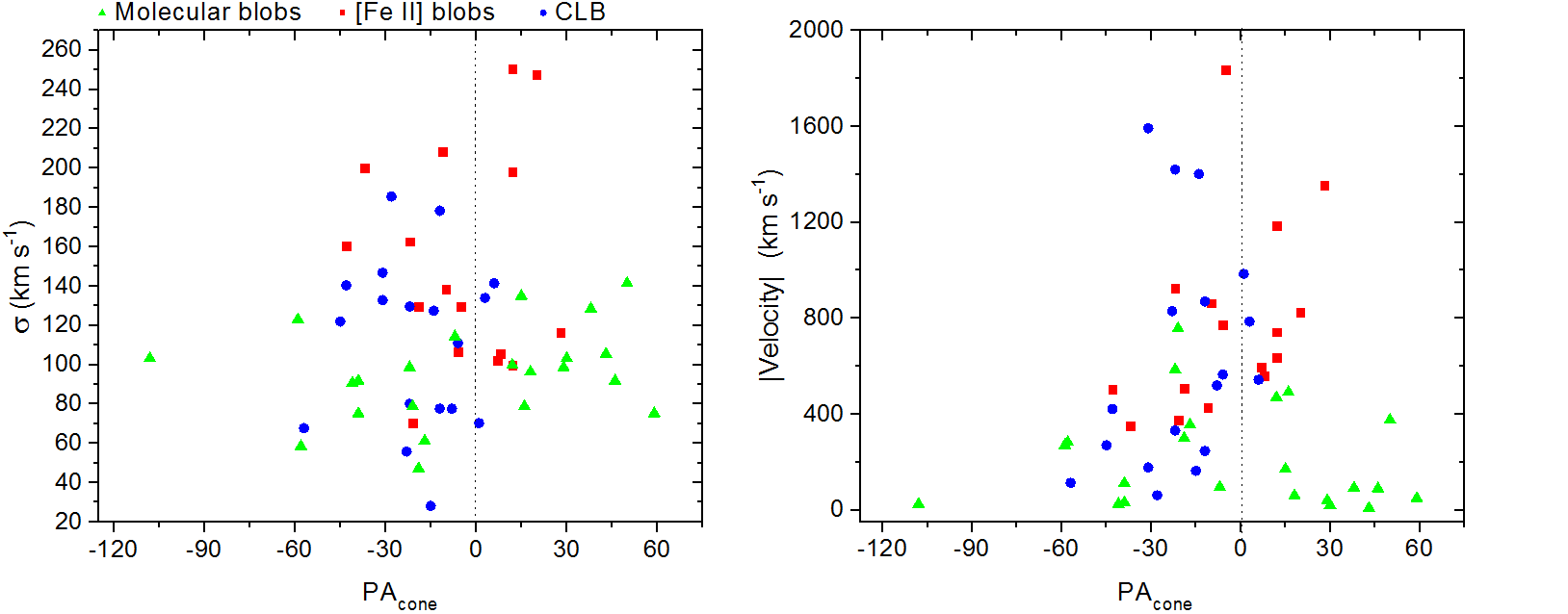}}
    \caption{Left panel: diagram for the position angle with respect to the cone's major axis (34\textdegree) vs. the velocity dispersion for the H$_{2}$, [Fe\,{\sc ii}] and [Si\,{\sc vi}] blobs selected in the BRV maps, with their measured properties listed in the respective tables. Right panel: PA$_{cone}$~vs. the radial projected absolute velocity for the same blobs.}
     \label{fig:gpas}
    \end{figure*}

    We can look closer at the jet's influence, which has a similar PA to that of the cone, by plotting the PA of each blob vs. the absolute velocity and the velocity dispersions in two distinct diagrams, as seen in Fig.~\ref{fig:gpas}. In the left panel there is a trend for the high-velocity dispersion blobs to be closer to the cone's PA (or, equivalent, the jet PA), although in the right diagram this is more evident, with the blobs being located in a narrower area in the vicinity of the cone's major axis as velocities increase. An interesting point seen in the right panel is a trend for an increase in velocity from the western side of the cone to the eastern side.
		
    As reported by optical observations, it is likely that the orientation of the accretion disc, aligned more to the west side of the northeast cone, does not favor an extra acceleration mechanism, such as winds launched from the vicinity of the AGN. Also against this hypothesis is the molecular arm shielding between the AGN and the extended NLR.
    
		Despite the higher velocities close to the jet, its direct influence in the blobs acceleration far from the radio emission is negligible. Furthermore, the blobs are being symmetrically accelerated along the orientation of the northeast cone. In light of these results, there is an indication that the process involved in the jet deflection could be related to a continuous heating spot where strong winds are created and may play a role in the blobs acceleration. This hypothesis will be detailed in Sec.~\ref{sec:discussion}.    

    \subsection{A molecular ring or a bubble?}
    \label{sec:bubble}

    The nuclear molecular gas in NGC\,1068 has been studied with high resolution data and several authors \citep{Galliano02,Muller09,Riffel14b,Thaisa12} argued that it presents a ring morphology, preferably immersed in the same plane of the galactic disc.    

\begin{figure}
    \resizebox{\hsize}{!}{\includegraphics{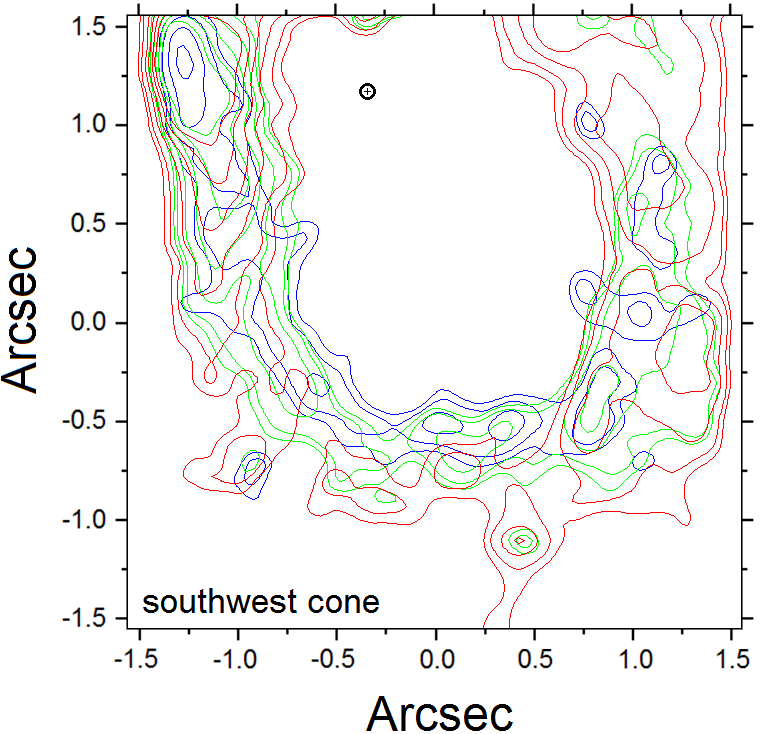}}
    \caption{Three different velocity ranges for the H$_{2}$~southwest wall emission. The blue contours range from -353 km s$^{-1}<v<$-283~km s$^{-1}$~and the red contours from -70 km s$^{-1}$ to the zero velocity. The cross denotes the AGN centre.}
     \label{fig:bubble}
    \end{figure}

    If the molecular gas is, indeed, the source of all the ionized gas emission, then the hypothesis of a ring would hardly be compatible with the well-studied kinematics and geometry of the NLR. Basically, if the ring morphology is correct, we could no longer sustain our scenario for two main reasons. First, the southwest cavity contains the high-velocity emission for almost all the ionized lines, which are accelerating from the spot where the jet hits the inner wall. Thus, the cavity necessarily has to retain the blobs in a way that a ring could not, given the inclination of the NLR with respect the galactic disc. In Fig.~\ref{fig:bubble} we show that the molecular structure is under deceleration, with a higher blueshift associated with the borders of the inner wall (blue contours) and an approach velocity that is decreasing radially up to the outer edge of the structure (red contours). This deceleration would fit the suggested scenario, where the blobs expand faster and are redirected along the inner wall of a bubble. 
		Second, the molecular structure is clearly asymmetric with respect to the AGN, drawing an unusual shape for the eventual formation of a ring in the galactic disc. On the other hand, we have shown that this asymmetry is likely caused by the proximity of the molecular arm in the northeast cone and consequently is submitted to high pressure from the redirected radio plasma, causing its rupture. Such asymmetry is also the cause of the asymmetry seen between the ionization cones, in outflow, demanding the existence of an expanding bubble above the galactic disc. 
		
    In line with this interpretation, both blobs in blueshift and redshift would be seen at the same side and not only close to the jet. The presence of small cavities mainly surrounding the northeast part of the bubble are naturally interpreted as small bubbles, and indicate that a contribution from the AGN radiation escapes among the gaps left by the blobs formation. In this sense, the bubble would have grown more effectively in the southwest cone, where the radiation from the AGN, and possibly the jet counterpart, remains mostly confined.
		
		Interestingly, the H$_{2}$B12 blob (Fig.~\ref{fig:H2rbvmaps}), which is in the ``shadow'' with respect to the ionization cone, may indicate that it was exposed to the AGN radiation in the past. The relatively high-velocity of this blob, in redshift (376 km s$^{-1}$), is still lower than the velocities of the blobs in the southwest cone, or at similar distances, suggesting it reached and expands with a nearly constant radial velocity, which does not occur for the blobs inside the hourglass. Curiously, the H$_{2}$B12 blob has the highest velocity dispersion, 141 km s$^{-1}$, but if we look for those with the highest velocity dispersions, they are all systematically located at the east side of the FoV, meaning that it is not a unique property of this particular blob.

    Our scenario does not rule out that most walls of the bubble could be close to the galactic disc, since it holds a large gas reservoir.
		

		Considering the higher blueshift, -353 km s$^{-1}$, as a constant expanding velocity, the maximum time for the bubble to expand to its current projected distance from the AGN, $\sim$90 pc, can be calculated as 3.6$\times10^{5}yr$.
    The expanding bubble scenario seems to go against the correlation found by \citet{Thaisa12}, between young stars ($\sim$30 Myr) and the molecular emission around its peak. They claimed that, with lower limit to the velocity, the expanding structure would reach its current position much earlier than the age of the young star population, that is, at the time of the starburst the molecular gas ought to be roughly in its current position. These authors interpret the ring kinematics as perturbations caused by supernova explosions in the galactic disc. 
		
		A way to avoid this ``static'' scenario is to consider that the stars, shrouded by the gas reservoir, were already under formation by the time they approached the AGN. This star formation clump would likely be close to the galactic plane and therefore would have a similar velocity range than that found by \citet{Thaisa12}, -80 km s$^{-1}<v<$20~km s$^{-1}$, in the region of the molecular emission peak. In fact, this is very close to the velocity of the clumps we measured for the same region. 

    The hypothesis of a stellar clump rich in molecular gas, which supposedly interacted with the AGN in the past (and thus we would be seeing part of the remnant interaction), requires tracing back the orbit of the clump and recovering a feasible kinematics. Unfortunately, this could not be inferred from our data only, although it is a tempting scenario that fulfills our requirements.
		
		 Finally, the fragmentation of an expanding bubble is consistent both with the molecular gas associated with the outflow close to the nucleus and with the extended emission into the hourglass.    

    \subsection{An overview of the emission lines in DS2}
    \label{sec:ds2}

\begin{figure*}
    \resizebox{\hsize}{!}{\includegraphics{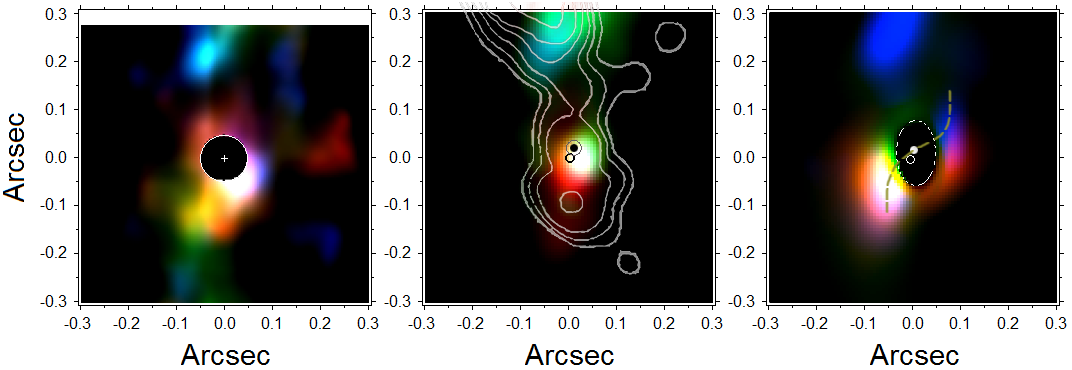}}
    \caption{RGB composition for the [Fe\,{\sc ii}] (left panel), [Si\,{\sc vi}] (middle panel) and H$_{2}$~(right panel) emission lines of DS2. The gray contours represent the MERLIN 5 GHz radio emission, the dashed contour delimits the noisy region and the curved dashed line traces the disc elongation. The RGB overlap results in additional colours. The filled circles in the middle and last panels represent the [Si\,{\sc vi}] kinematic centre. All the images are centred on the dusty emission (denoted by the cross and the open circles) and the north is on top.}
     \label{fig:25mas}
    \end{figure*}

    We will concentrate on a brief analysis of the 25 mas data, highlighting only the points that complement the scenario presented here. Fig.~\ref{fig:25mas} shows the RGB kinematic composition for the three main lines we have detailed so far, namely the [Fe\,{\sc ii}], [Si\,{\sc vi}] and H$_{2}$~lines, for a FoV of $\sim22\times22$\,pc. 

    In the left panel, the [Fe\,{\sc ii}] emission presents the sparsest distribution, without a clear distinction of its kinematics, attributed to the highly perturbed gas motions near the AGN. Among the complex line profiles, it is possible to infer a velocity range of -839 km s$^{-1}<v<$ 565 km s$^{-1}$. Nonetheless, it is possible to distinguish the walls that connect to the hourglass structure seen in DS1, and the cloud where the jet is deflected.

		In the middle panel the RGB composition for the [Si\,{\sc vi}] line is shown, and we see that the first contact of the jet with the ionized and the molecular gas occurs slightly before the \textbf{C} knot, at $\sim$0.15$\pm$0.02 arcsec north of the nucleus, the same location of the jet bending. The PA defined by the central structure and the northern cloud also agrees with the jet PA before bending. Here we see a clear division between the central gas in redshift and in blueshift, in agreement with the 100 mas scale. The velocity ranges between $\pm$916 km s$^{-1}$ and we found that the [Si\,{\sc vi}] emission peak (the black dot above the centre) does not coincide with the reddened continuum centre, or the dust emission (denoted by the black open circle in the centre). The displacement between the line emission peak and the dust emission peak is 0.02 arcsec ($\sim$1.4\,pc), with PA=160\textdegree$\pm$4\textdegree. One should expect that the hot dust emission peak would coincide with the [Si\,{\sc vi}] kinematic centre, since it is natural to place their origins in the inner region of the torus. At this point we assumed that the kinematic centre for the [Si\,{\sc vi}] is the AGN centre and coincides with the \textbf{S1} radio knot. Given the measured spectral range, the hot dust centroid is more susceptible to extinction effects.
		
		Such small displacement is not resolved by the 100\,mas scale images (0.1 arcsec$\sim$7\,pc), so the astrometry between the 100\,mas scale images and the MERLIN 5 GHz radio emission is the same if assuming the AGN centre as the dust continuum peak, as has been done along this work.
				
		Since the molecular structure we detected (Fig.~\ref{fig:25mas}, right panel) does not present enough contrast between the H$_{2}$ and the continuum emission in the nucleus, the ionized and molecular gas distribution and kinematics cannot be properly compared. However, before the analysis of the H$_{2}$ gas, we may compare the kinematic of the central ionized gas with the one found for the CO disc in the work of \citet{Gallimore16}.

		\subsubsection{The central CO and [Si\,{\sc vi}] kinematics}
    \label{sec:sico}
		
		There is a detection of a high-velocity outflow (from $\pm$70 to $\pm$400 km s$^{-1}$) seen in the CO(6-5) emission at distances reaching the borders of the central CO disc \citep{Gallimore16}. These high-velocity components (defined as the centroids of increasing velocity channels) seem to follow roughly the orientation orthogonal to the H$_{2}$O maser disc (PA=33\textdegree), with the blueshifted velocity peak displaced by 1.2\,pc from the redshifted one. These authors interpreted the molecular bipolar outflow in light of the disc-wind scenario \citep{Elitzur06,Elitzur16}, where strong winds are launched in the vicinity of an accretion disc and are responsible to maintain the toroidal structure around AGN.
		
\begin{figure}
    \resizebox{\hsize}{!}{\includegraphics{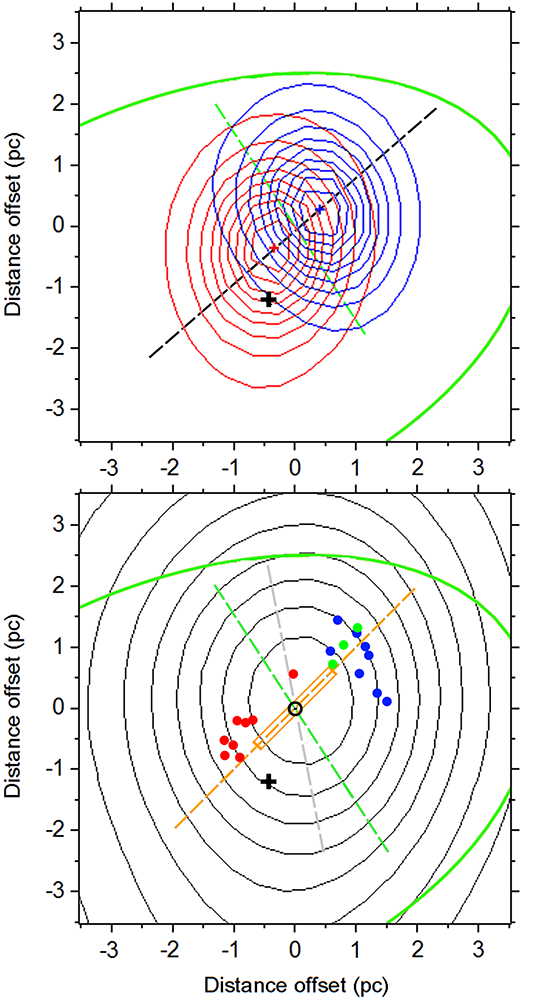}}
    \caption{Upper panel: the contours for the [Si\,{\sc vi}] redshifted and blueshifted emissions within 0.1\,arcsec$^{2}$, from -916 km s$^{-1}<v<-107$~km s$^{-1}$ (blue) and 107 km s$^{-1}<v<916$~km s$^{-1}$ (red); the two centroids present a PA=131\textdegree$\pm$13\textdegree~(black dashed line). The partial green ellipse illustrates the area occupied by the low-velocity CO emission and the green dashed line denotes the PA for the high-velocity CO outflow \citep{Gallimore16}. The cross shows the dust emission peak. Bottom panel: The locations of the [Si\,{\sc vi}] emission peaks marked by coloured circles (according to its kinematics) and the contours for the integrated flux. The open black circle indicates the flux emission peak, the orange dashed line, the orientation of the H$_{2}$O maser, with its size delimited by the rectangle of same colour. The gray line shows the PA of the jet. All the images are centred on the [Si\,{\sc vi}] kinematic centre (which is the same of the line emission peak). The north is on top.}
     \label{fig:sico}
    \end{figure}

		From Fig.~\ref{fig:sico} (upper panel) we represent the [Si\,{\sc vi}] kinematics with an image similar to what was done for the CO high-velocity emission in Fig.1 (right panel) of \citet{Gallimore16}. It is possible to measure that the central [Si\,{\sc vi}] kinematic has a blueshifted velocity peak displaced by 1.8\,pc from the redshifted peak, with PA=131\textdegree$\pm$13\textdegree. Although such centroids may be measured with enough spatial precision, they should be considered with some caution, since the resolution of the 25\,mas scale is 4.3\,pc (corresponding to the difraction limit). Aside that, there is a trend for the velocities in blueshift and redshift being located along a PA almost perpendicular to the orientation of the CO outflow and, therefore, similar to the detected maser disc (Fig.~\ref{fig:25mas}, middle panel). 
		
				To perform a better comparison between the CO outflow found by \citet{Gallimore16} (see their Fig.2, left panel) and the high-velocity [Si\,{\sc vi}], we selected 19 velocity frames, corresponding to the total spectral velocity range we detected, and plotted their centroids together with the contours of the [Si\,{\sc vi}] integrated flux in Fig.~\ref{fig:sico} (bottom panel). The estimated error bars are smaller than 0.05\,pc, even adding random noise typical from the data we have, which are smaller than the coloured filled circles. Our reduced number of components is due to a lower spectral resolution and available velocity channels as compare to those of \citet{Gallimore16}; however a clear spatial distribution for distinct velocity channels is detected. The [Si\,{\sc vi}] components follow a pattern that is distinct from the one found for the CO outflow (with PA given by the green dashed line). The three green components represent the low-velocity, which are located in the same region of the blueshifted ones, a trend seen also in the CO gas. It is worth noticing that the [Si\,{\sc vi}] kinematic could not be mistaken with that one of the maser disc, because the former presents blueshift velocities where the latter shows rotation in redshift.  
				
				The detected [Si\,{\sc vi}] outflow follows an orientation similar to the plane of the maser disc. This result is as much unexpected as the orientation for the detected CO outflow found by \citet{Gallimore16} which is nearly orthogonal to the maser disc. The CO and [Si\,{\sc vi}] emissions peaks are, however, widely distributed along their respective PAs. On the other hand, the orientation of the jet does not coincide neither with the PA measured for the CO or [Si\,{\sc vi}] high-velocity components, being roughly halfway from them. Comparing for the first time both kinematics in this scale we have one more ingredient (the coronal gas) to find a feasible explanation for the very central part of NGC\,1068. In Sec.~\ref{sec:2phase}, after analyzing the H$_{2}$ emission for DS2, we present our interpretation in light of these new findings.
				     
    \subsubsection{The H$_{2}$ disc-like structure}
    \label{sec:torus}

    The image of the molecular gas kinematics is shown in the right panel of Fig.~\ref{fig:25mas}. Again, the circles denote the same centre shown in the [Si\,{\sc vi}] image. The white contours encircle the region where the continuum subtraction resulted in negative flux, given the very noisy spectra associated with the nuclear brightness.
    The velocity range for the H$_{2}$ is between -494 km s$^{-1}<v<$ 184 km s$^{-1}$, with the northern cloud presenting $v\leq$70 km s$^{-1}$.  
		
		We see a disc-like structure with PA for the major axis of 132\textdegree$\pm$10\textdegree, $r=8\pm$2\,pc, in close agreement with the detected CO disc in \citet{Gallimore16} and \citet{Burillo16} ($r\sim$5\,pc and PA=112\textdegree), but presents no sign of rotation. The peak velocity distribution for the gas in the disc ranges from $\pm$50 km s$^{-1}$. \citet{Burillo16} also detected a dusty emission with PA=142\textdegree$\pm$23\textdegree~and $r$=7\,pc, and called their detections of ``torus''. On the other hand, \citet{Gallimore16} claim that only the high-velocity CO is part of the torus, which is slightly more compact than the CO disc. The VLBA 5\,GHz central radio continuum image, with radius of $\sim$0.8\,pc, has a PA$\sim$110\textdegree~\citep{Gallimore04}, similar to the ones measured for the maser emission (135\textdegree~\citealt{Greenhill96}) and for the dusty torus found by \citet{Gratadour15} (118\textdegree$\pm$5\textdegree). The H$_{2}$O maser, for instance, is detected up to a radius of 1\,pc, which is beyond the internal radius for the torus in NGC\,1068 ($\sim$0.5\,pc for L$_{Bol}\sim10^{45}$ erg s$^{-1}$). 
		
		In the present work, the existence of a torus was indicated by two findings: the point-like source of hot dust continuum, successfully used as PSF; and the well-defined ionization cones seen by the low-velocity [Fe\,{\sc ii}] emission, which requires a central collimating structure. The estimated size for the torus, according to \citet{Raban09}, is 1.35\,pc, with PA=138\textdegree$\pm2$\textdegree. Since this PA and the cone's major axis, of 34\textdegree$\pm4$\textdegree, agrees within the errors with the H$_{2}$ major axis (42\textdegree$\pm$10\textdegree) one can reasonably assume that we are seeing a H$_{2}$ disc-like structure, 12$\times$ bigger than the dusty torus, with similar orientation.
		
		The mean PA for all the central presented structures, including our measurement for the H$_{2}$ gas, is $\sim$123\textdegree, clearly not perpendicular to the jet position angle before bending, of $\sim$11\textdegree. On the contrary, 123\textdegree-90\textdegree=33\textdegree~is in close agreement with the PA of the hourglass structure seen in [Fe\,{\sc ii}] and with the jet after bending (34\textdegree). As already noted and shown by \citet{Gallimore04} (their Fig.\,8), the accretion disc, whose orientation is inferred by the jet PA, is misaligned by $\sim$22\textdegree~with respect to the resolved central structures.   

    The spatially resolved kinematics of the 14\,pc disc found by \citet{Burillo16} with ALMA observations displays the same preference for redshift and blueshift velocities at the southwest and northeast sides, respectively, of the molecular structure. Here, the velocity dispersion is $\sim$85 km s$^{-1}$, certainly higher than any weak evidence of rotation. 

	  According to the elongations seen in the H$_{2}$ disc-like structure (Fig.~\ref{fig:25mas}, right panel) and the subtle displacement between the apexes of each cone (Fig.~\ref{fig:FeII}), it seems that the central radiation is being collimated with a slight asymmetry. The physical reason for that may be the interaction between the jet/outflow and the internal walls of the H$_{2}$ disc-like structure, ``excavating'' the northwest part of the northern wall and the southeast part of the southern inner wall. 		

    In NGC\,6951, \citet{Dmay16} found an edge-on molecular disc with diameter of 47 pc, which differs from the jet PA by only 32\textdegree. Such misalignment is the cause of a high-velocity dispersion measured at the edge of the disc, showing elongations similar to the one we found here. However, here the H$_{2}$~disc is smaller and, if there is some misalignment and an interaction between the outflow and the inner portion of the disc, this is not well resolved by our data.

    \section{Discussion}
    \label{sec:discussion}

    Relying on the spatial distribution of the different gas components and their correlations, described mainly by the coronal line of [Si\,{\sc vi}], the partially ionized regions of [Fe\,{\sc ii}] and the molecular gas, through the H$_{2}$~lines, we propose a self-consistent scenario where distinct degrees of gas ionization originate primarily from the interaction between the AGN radiation/jet and the circum-nuclear molecular gas.
    Our analysis can be summarized in an attempt to answer the following questions: Why do we see an asymmetry in the molecular cavity around the AGN? How and where did the high ionization/coronal lines originate? What is the role of the jet in powering the outflow and how does the jet-galaxy disc interaction takes place and explains the extended velocity components? Such questions may be answered on the basis of a close comparison between the derived properties of each emission line and the discrimination of several blobs as byproducts from different stages of the jet-wind/galaxy disc interaction. 
		
		\subsection{\textbf{DS1}: the answer is blowing in the wind}

    The relation between the jet and the NLR in NGC\,1068 has long being clarified by the concomitant analysis of radio and optical observations \citep{Cecil90,Gallimore96,Capetti99,Das06} and, more recently, with IFUs in the NIR by \citep{Mazzalay13a,Riffel14b,Barbosa14}. In general, there is an agreement in the literature with respect to the role of the jet in the kinematics of the CLs in the NLR. Correlations between the ionized gas and the radio emission are seen from the knot \textbf{S1} (the nucleus) to point C (where the jet is deflected \citealt{Gallimore96}), and far from this knot with the lateral expansion of the gas, as described in \citet{Axon98} and confirmed here.
		
		The blobs acceleration not related to the jet is commonly attributed to a wind, launched by the central source and collimated by the thick torus geometry. This wind will flow freely through the filaments and the clouds, and will be continuously redirected by means of the pressure gradient along a denser material (such as the galactic disc). Simulations of wind-cloud interaction, near the AGN, show that, if such wind encounters a denser material, sparse filaments will form and the gas clouds will be fragmented \citep{Gallimore96}, in disagreement with the number of blobs found here. Furthermore, the driving mechanism for the high-velocity and distant clouds is still more uncertain.

    The FoV covered by our data is not sufficient to confirm the clouds' deceleration at a distance greater than $\sim$150 pc, reported by \citet{Das06}. Apparently, we have mapped the blobs until the acceleration turnover is reached. However, we must draw attention to the velocity ranges we found, significantly larger than the ones found in previous studies (columns 6 and 8 of Table~\ref{table:vel}), reaching values up to twice as large as the ones shown in \citet{Barbosa14} channel maps. This additional result challenges even more the efficiency of an underlying mechanism capable of accelerating the compact blobs.

    \subsubsection{The two-stage outflow}

    Briefly speaking, in the astrophysical context, ``wind'' means gas flowing due to an underlying acceleration mechanism. The driving force, for instance, may be momentum transfer by photons (radiation-driven wind) or electrons (magnetically-driven). The challenge has been to effectively couple this energy transfer to the ambient gas, and a proper understanding of such interaction is still missing. Basically, the wind can be energy- or momentum-conserving, depending on the cooling time of the shocked wind. In fact, the momentum cannot be radiated away and is always conserved, unlike thermal energy. The case of NGC\,1068 could be associated to both scenarios, with the expanding bubble having its global energy nearly conserved and the blobs accelerating only by momentum transfer, as argued by the model of \citet{Zubovas14}.   
     
		Based on better data quality, it is better shown that the hourglass structure (phase-1), depicted by the low-velocity [Fe\,{\sc ii}] emission and excited by X-rays photons from the central source (creating partially ionized regions), has its resulting geometry defined by the PA of the torus axis.
    The orientation and the southwest walls of the bubble defined by the molecular cavity (Figs.~\ref{fig:h2} and \ref{fig:3f}) are nearly coincident with those of the low-velocity glowing wall of the hourglass. If this bubble is being expanded by the wind coming from the accretion disc, it would have preferably the direction of the jet before bending, which is perpendicular to the accretion disc. Therefore, based on this geometrical argument, and given the physical association between the molecular and the [Fe\,{\sc ii}] gas, the central wind cannot be the main driving mechanism for the acceleration seen in the wide-open angle of the cones.

	  If a central wind was the only driving mechanism, it would be hard to explain the blobs' acceleration at longer distances outside the northeast part of the bubble, which is partially shielded by a molecular arm; the larger velocities up to $\sim$150 pc from the nucleus and the trend for the [Fe\,{\sc ii}] blobs being accelerated more from the west to the east in the NLR (see the discussion of Fig.~\ref{fig:gpas}, right panel, in Sec.~\ref{sec:origin}). These evidences may point out to the fact that a secondary acceleration mechanism is taking place in the same direction of the bent jet.
		
		At this point, we draw attention to the orientation of the jet after bending, which is very similar to that of the hourglass structure. There is no physical reason for this coincidence. The change in the jet PA is attributed only to the physics of the jet and of the impacted molecular cloud, and the physics of the ionized cones is associated with the central source and its collimation by the torus.
		
		In essence, the X-ray photons, collimated by the torus, penetrate the molecular barrier and excite the [Fe\,{\sc ii}] lines. The plane of the accretion disc, misaligned with respect to the torus, concentrates the central radiation more to the northwest part of the cone defined by the hourglass (see Fig.~\ref{fig:h2s}), where the [O\,{\sc iii}] emitting cone is seen.

    Simulations performed by \citet{Elvis10} demonstrate that a compact cloud near the AGN, and exposed to the central wind, has its cross-section increased by instabilities caused by perpendicular shocks, enhancing the feedback from the same radiated energy by a factor of $\sim$10. This could be the case in the initial stage of the blob ejection. However, the blobs remain with a compact morphology, indicating that this process would not persist over larger distances. In turn, the existence of distant and compact blobs is an indicator of a hot surrounding gas present inside the cones.      
  
    \begin{figure}
    \resizebox{\hsize}{!}{\includegraphics{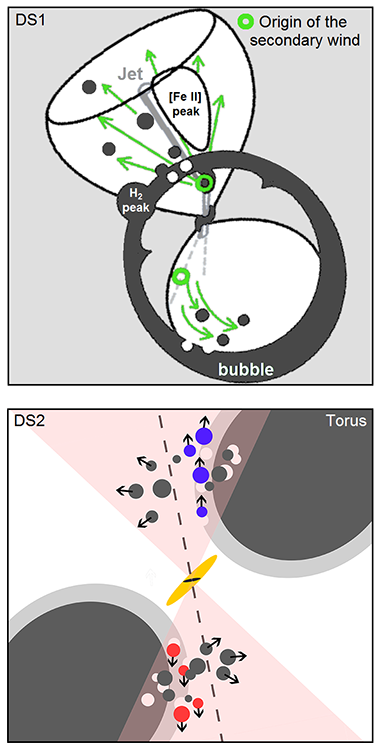}}
    \caption{Upper panel: a sketch of the interaction between the jet and the molecular bubble for DS1. The production of the secondary wind is marked by two green circles and the arrows illustrate its direction. Bottom panel: sketch for DS2, showing the interaction between the central radiation with a two-phase gas density torus. The coloured and gray clouds denote the [Si\,{\sc vi}] and the molecular outflows, respectively. In turn, arrows indicate the direction of the outflow. The maser and the accretion disc are shown in yellow and black, respectively. The jet orientation is traced by the dashed line and the light red area shows the ionization cones.}
     \label{fig:model}
    \end{figure}

    \citet{Elvis10} named this effect ``secondary outflow'', driven by the same primary wind. What we propose is that a ``secondary wind'' is excited by the collision of the jet with a molecular cloud, creating a second stage for the outflow. The difference is that in their model there is no jet, and here it is the cause of the secondary wind.
		
		The region where the jet hits the molecular gas and changes its direction, a well-known scenario described by \citet{Gallimore96,Gallimore04}, leads to a continuous heating spot from where the secondary wind is launched. In Fig.~\ref{fig:model} there is a sketch of the geometrical model we propose. 
		The two-stage outflow is uniquely related to the high-velocity gas (phase-2). The widely distributed [Fe\,{\sc ii}] blobs in the northeast cone seem to be drilling holes in the molecular shielding and agree with the orientation of the proposed secondary wind. Although we do not see the jet interacting with the southwest cone, the [Fe\,{\sc ii}] and [Si\,{\sc vi}] emissions are good indicators of its presence, ejecting and accelerating the blobs from the spot where the jet hits the molecular inner wall.		
		
		The secondary wind is responsible for redirecting and spreading the collimated energetics of the jet, and this process could take place always when a dense molecular cloud is impacted by a jet without being completely wiped out. Even if the jet is not deflected, the ejected material may follow a different orientation from that of the jet, which may be the case in NGC\,6951 \citep{Dmay16}. This combination of effects likely accounts for the high velocities reached by the blobs into the cone for distances up to $\sim$150\,pc from the nucleus.
		The combination of the AGN feedback in the form of a jet and a central wind, with the secondary wind generated in the spots where the jet hits and is deflected by the molecular gas, is what we call a two-stage outflow. This could account for the most evident feedback effect seen for this galaxy.
				    
    It is worth noting that the word ``feedback'' here is subtle and has quite a different meaning from the context in which it is generally evoked. Normally, it reffers to the quasar and radio modes feedback, with dramatic effects for the host galaxy and its environment. There is, indeed, a ``minor-feedback'', but not in the same sense as it occurred in the past during galaxy formation. The role of this small version of feedback is still uncertain and surely does not affect the galaxy evolution beyond the nuclear region. 
		
		The main difference between our results and those in the literature is that the outflow in NGC\,1068 is described by the interaction between the AGN and the galactic disc in the form of a bow-shock. What we propose is that all the ionized lines originating from the elements previously contained in the initial molecular gas supply, which dissociated by means of the primary interaction, suffered around the AGN.
This initial molecular clump may configure a unique case at the centre of this galaxy. Its existence is directly related to the recent starburst found by \citet{Thaisa12} and is greatly favored by our interpretation, namely, that the remaining configuration after its interaction with the AGN feedback led to the current scenario of an expanding bubble. 
This hypothesis is a direct consequence of a nuclear starburst interacting with the galactic nucleus. 
      
    \subsubsection{Blowing the bubble}

    The hypothesis of an expanding bubble, at this point, is greatly favored by our results, but estimating the energy involved in this process requires extra assumptions. The starting point is the evaluation of the molecular mass contained in the emitting southwest wall of the bubble, which is not disrupted. Assuming thermal equilibrium during the expansion, according to the calculations performed in \citet{Scoville82,Riffel08,Dmay16}:

    \begin{equation}
    \begin{aligned}
    M_{H_{2}}&=\frac{2m_{p}F_{H_{2}\lambda 2.1218}4\pi D^{2}}{f_{\nu =1,J=3}A_{S(1)}h\nu}\nonumber \\
    &=5.0776\times 10^{13}~\left(\frac{F_{H_{2}\lambda 2.1218}}{erg s^{-1}cm^{-2}}\right)\left(\frac{D}{Mpc}\right)^{2}
    \label{h2mass}
    \end{aligned}
    \end{equation}

    where $m_{p}$ is the proton mass, $F_{H_{2}\lambda 2.1218}$ is the line flux, $D$ is the adopted galaxy distance, of 14.4\,Mpc, and $f_{\nu =1,J=3}$ is the fraction of hot H$_{2}$ in the level $\nu=1$ and $J=3$ , with $M_{H_{2}}$ given in solar masses. The linear dependence of the H$_{2}$ emissivity on density assumes T=2000\,K and $n_{H_{2}}>10^{4.5}$ cm$^{-3}$. This implies a population fraction of $1.22\times10^{-2}$ with transition probability $A_{S(1)}=3.47\times10^{-7}$ s$^{-1}$. We delimited the southwest wall by measuring the H$_{2}$ flux below a line crossing the AGN centre at the same orientation of the H$_{2}$ disc-like structure (132\textdegree), which gives $F_{H_{2}\lambda 2.1218}=2.8 \times 10^{-14} erg~s^{-1} cm^{-2}$, obtaining $M_{H_{2}}=294 \Msun$. It is worth emphasizing that this is an inferior limit for the amount of gas, which accounts only for the projected bubble as viewed by the hot H$_{2}$ emitting ring.

    The average projected velocity in blueshift, taken from the velocity range measured for the bubble, is -142 km\,s$^{-1}$, which provides a kinetic energy $E_{k}=6 \times 10^{49}$ erg. Using the same velocity, the time scale to reach the distance of 90\,pc is $6.2 \times 10^{5}$ yr, leading to a kinetic luminosity of $\sim 10^{44}$ erg s$^{-1}$. The estimated jet outflow power, based on the northeast radio emission in 1.4\,GHz, implies $P_{jet}=1.8\times 10^{43}$ erg s$^{-1}$ \citep{Birzan08}, below the necessary energy to inflate the molecular bubble. This estimation requires that some energy comes from the radiative wind of the central source. Despite the quadratic dependence on the velocity, the main uncertainty is likely to come from the estimate of the mass, since we are not accounting for the cold molecular and atomic gas. Taking into account any of these assumptions would increase the necessary power to blow out the bubble.    

    \subsubsection{Cold gas flowing onto the torus: feeding the monster in the shadow}
		\label{sec:shadow}

		If the molecular bubble expanded to its current shape (or still expands), this scenario implies that in the initial stage of the expansion a huge amount of molecular material was closer to the AGN and, in principle, was a time of more favorable conditions to trigger the gas accretion into the SMBH. Thus, a natural question arises: Is there still some of this material inflowing to the centre?   

    Looking at the DS1 (100\,mas scale), there are two prominent filaments of molecular emission coupled to the central masked region and to the internal wall of the bubble (Fig.~\ref{fig:h2}), where highly ionized emission is also seen. It has been suggested by \citet{Muller09}, based on the 25 mas scale observations, that the H$_{2}$~``tongues'' would be inflowing gas to the AGN, but this hypothesis was later challenged by \citet{Thaisa12}, who argue that this feature is seen in outflow, given the extended NLR dynamics. We agree that both molecular filaments are being pushed away by the jet and, furthermore, that we are seeing the apex of the ionization cones in these structures.

		In accord to Sec.~\ref{sec:torus} and to the DS2 (25\,mas scale), there is no evidence of gas inflowing through the innermost detected H$_{2}$ structure, which is an extension of the maser disc. A natural question that may arise at this point is: How is the cold gas related to the H$_{2}$~emission? As a good tracer of cold gas ($\sim$100\,K), we reproduced the intensity map of the CO(6-5) 432\,$\mu$m emission, shown in \citet{Burillo14}, in the panel (c) of Fig.~\ref{fig:3f}. In Sec.~\ref{sec:co} we compared the CO to the H$_{2}$~emissions and concluded that most of the cold gas is correlated to the H$_{2}$ emission and is located outside the outer limits of the hourglass walls, in the ``shadow'' of the ionization cones.	Although DS1 does not detect any sign of H$_{2}$ emission in this region, in Fig.~\ref{fig:h2s} (right panel) it is possible to see a weak emission associated with this region in the NIFS MDC (and in the channel maps of \citealt{Barbosa14}), given a higher exposure time. Such region, in the plane of the torus, is likely heated by the kinetic energy deposited and confined inside the cavity. But once the molecules dissociate, the gas could remain in its atomic form (with a small fraction of hot H$_{2}$ gas) and flow to the centre in the shadow of the cones, feeding the gas reservoir in the central structure. 
			
			Looking at Fig.~\ref{fig:Felh} (left panel) one may see that the redshifted portion of the hourglass wall to the west of the nucleus could represent a spatially constant emission, but with a moving gas overpassing through the partially ionized [Fe\,{\sc ii}] wall into the shadow of the cone, pushed by the energy injected into the cavity. A map of H\,{\sc i} emission with similar resolution could clarify this hypothesis. 		
				
				\subsection{\textbf{DS2}: the torus sailing into the wind}
				\label{sec:2phase}
				
		In Sec.~\ref{sec:sico} we presented the comparison between the central CO \citep{Gallimore16} and [Si\,{\sc vi}] emissions, with both lines showing unexpected results with regard to their outflow's orientations: the high-velocity components of cold molecular gas nearly orthogonal to the maser disc and of the coronal gas closer to the maser disc plane (Fig.~\ref{fig:sico}, bottom panel).
		
		In an attempt to explain these unusual results together, in Fig.~\ref{fig:model} (bottom panel) we propose a scenario, although somewhat speculative, which seems to be consistent with both outflows. We represent the high-velocity [Si\,{\sc vi}] components by filled circles coloured according to its kinematics and the CO components by gray circles, with the arrows pointing to the direction of each outflow. The ionized gas is accelerated radially by the radiation pressure from the central source and the cold gas is blown away from the region heated by the same radiation. Both outflows arise from distinct responses of a two-phase gas density torus exposed to the ionization cones. These phases, for the sake of simplicity, are represented by the light gray (lower density) and dark gray (higher density) colours. 
		
		Based on the misalignment between the jet and the maser disc (which has a PA similar to the torus), we can distinguish two outflows: the first would be characterized by the lower density gas and, therefore, a higher ionization parameter, with the [Si\,{\sc vi}] components ionized and blown away by the central radiation pressure; and the second is represented by the interaction between the radiation and the higher density gas (with lower ionization parameter), which warms it up and releases molecular clouds accelerated by internal thermal pressure, ejecting them in the direction normal to the surface of the torus. The nature of both ionized and molecular outflows, therefore, would be consequences of two distinct processes triggered by the radiation from the central source, delimited by the cone with major axis orthogonal to the accretion disc. One straightforward prediction of this hypothesis is that the CO components should be found with some asymmetry with respect to the cone axis, since they are coming from different sides of the torus.

    \section{Conclusions}
    \label{sec:conclusions}

    We have presented a new scenario for the active galactic nucleus of NGC\,1068, supported by high-quality data in the NIR. Some findings are in agreement with previous works and are simply shown with better resolution here, such as the hourglass structure in the [Fe\,{\sc ii}] emission, seen by \citet{Riffel14b} and \citet{Barbosa14}; the [O\,{\sc iii}] and jet association to the CLs seen by \citep{Mazzalay13a}; and the NLR accelerating gas seen by \citep{Cecil90} and \citet{Das06}. However, for the first time, the hourglass asymmetry is presented and characterized both in morphology and in velocity. We also mapped the NLR blobs from emission lines with three distinct ionization degrees ([Fe\,{\sc ii}], [Si\,{\sc vi}] and H$_{2}$ emissions) in light of the proposed molecular bubble. Therefore, together with the new findings, we distinguish three phases for the gas morphology: (1) the low-velocity [Fe\,{\sc ii}] emission representing the glowing wall of an hourglass structure, (2) the high-velocity compact blobs of low and high ionization emission filling the hourglass volume, and (3) a distribution of H$_{2}$ molecular gas defines the thick and irregular walls of a bubble surrounding a cavity. These phases are a consequence of the interaction between the jet and the molecular gas. We also propose an accelerating mechanism for the widely distributed blobs. 
		
		With regard to DS2, the misalignment between the accretion disc and the torus leads to a strong interaction between the central radiation field and the molecular gas. This process ejects high-velocity [Si\,{\sc vi}] and CO components, which follow distinct kinematics, according to the proposed two-phase gas density torus.
		
    In light of these new scenarios, the main conclusions are the following:

    \begin{enumerate}
    \item The low-velocity [Fe\,{\sc ii}] emission shows the glowing wall of an hourglass structure (phase-1). The hourglass defines an axis with a position angle of PA=34\textdegree$\pm4$\textdegree~(Fig.~\ref{fig:Felh}, left panel).	
    \\
		\item The [Fe\,{\sc ii}] emission with high velocities (-1951 km s$^{-1}<v<-401$~km s$^{-1}$~and 328 km s$^{-1}<v<1514$~km s$^{-1}$) is mainly characterized by compact blobs (part of phase-2). In the southwest cone, the emission is confined within the low-velocity hourglass, which is mostly within the molecular walls, showing a surprisingly sharp limit to the confined blobs (Fig.~\ref{fig:3f}).  
		\\
\item The H$_{2}$~emission (phase-3) is quite asymmetric: while the northeast part, closer to the nucleus, shows signs of a disrupted molecular wall, the southwest part seems to have allowed the formation of a bubble of ionized and possibly neutral atomic gas. The PA of this bubble, 30\textdegree$\pm$2\textdegree, is in agreement with that of the hourglass structure. With respect the high-velocity emission, two H$_{2}$ ``bullets'' were identified symmetrically located with respect to the central source, with velocities of -586$\pm$12 km s$^{-1}$~and 300$\pm$10 km s$^{-1}$.
    \\
		\item The H$_{2}$ emitting ring around the hot bubble is probably a photo-dissociation region (PDR), suggesting an association between the PDR and the partial ionization zone, responsible for the low ionization emission.
		\\
		\item The [Si\,{\sc vi}] blobs (part of phase-2) display the most compact morphologies among all blobs. Their alignment in both cones suggests that the UV ionization beams are highly collimated and shifted with respect to the hourglass, which also has a wider opening angle, defined by the torus. This difference is explained as a non-alignment between the torus and the inner accretion disc.
		Furthermore, the [Si\,{\sc vi}] emitting blobs seem to be blowing holes in the northern side of the H$_{2}$ bubble.
		\\
    \item We identified a total of 56 compact blobs of [Fe\,{\sc ii}], [Si\,{\sc vi}] and H$_{2}$ emission lines. These blobs have velocities that are correlated to the distance to the central source.
    \\
    \item Most of the Br$\gamma$~emission can be associated with the [Si\,{\sc vi}] emission blobs (Fig.~\ref{fig:hbr}), but it also presents a more diffuse emission near the blobs. The Pa$\alpha$ emission is also mostly related to the [Si\,{\sc vi}] line, but with a weak emission associated with the low-velocity walls of the [Fe\,{\sc ii}] hourglass. In this sense, the Pa$\alpha$ emission was the only one that displayed both the structures related to the [Si\,{\sc vi}] and [Fe\,{\sc ii}] lines.
    \\ 
		\item We established a common origin for all the ionized blobs as coming from the interaction between the jet/AGN wind and the molecular gas. This interaction occurs at the northeast molecular arm and in the inner walls of the bubble in the southwest cone. In this sense, the H$_{2}$ morphology also describes the outflow for this galaxy together with the ionized gas, instead of being a fraction of gas unaffected by the outflow in the galactic disc.  
		\\
		\item The PA of the jet after bending is very similar to the PA of the torus' axis, which defines the hourglass geometry. Their orientations, however, have no physical relation: the torus' axis accounts for the overall shape of the hourglass structure and the jet PA, for the wide coverage of the acceleration seen in the blobs.  
		\\
    \item We propose that a strong wind is formed in the northeast cone where the jet hits one of the H$_{2}$ molecular clouds. This secondary wind probably changes the direction and accelerates the [Si\,{\sc vi}] emitting blobs, and also changes the direction and enhances the radio emitting jet. Furthermore, it is probably the origin of the north-south asymmetry in the ionized gas emission, seeing that the same process likely occurs in the southwest cone, but with the exact location of the secondary wind production less clear. In both places this wind would be responsible for accelerating the NLR blobs and inflating the bubble. 
    \\
    \item The lower limit estimated for the kinetic luminosity of the southwest part of the molecular bubble is $\sim 10^{44}$ erg s$^{-1}$, which is above the jet power of $1.8\times 10^{43}$ erg s$^{-1}$. We concluded that a fraction of the energy radiated from the central wind is required to inflate the bubble.
    \\
		\item We compared the cold CO(6-5) emission \citep{Burillo14} to that of the hot H$_{2}$ gas (Fig.~\ref{fig:3f}, panel c), and found that most of the CO is associated with the H$_{2}$ distribution, but not all the H$_{2}$ emission is associated with the cold gas. This lack of association is mainly seen inside the hourglass structure, where the CO molecules are exposed and destroyed by the radiation from the central source or due to excessive heating. On the other hand, most of the CO is located outside the outer edger of the hourglass, at the ``shadow'' of the ionization cones. It also has an asymmetry vis-\`a-vis the axis perpendicular to the cones, with the emitting regions closer to the walls that are farther from the inner cone defined by the [O\,{\sc iii}] emission (Fig.~\ref{fig:SiOIII}).     
		\\
		\item Contrary to what is expected, the central [Si\,{\sc vi}] kinematics is oriented with a PA similar to that of the maser disc (or to the H$_{2}$ disc-like structure/dusty torus), and orthogonal to the axis of the high-velocity CO emission found by \citet{Gallimore16}. The PA of the jet, in turn, is located roughly halfway from both outflows. We interpret these unexpected observations in light of the interaction between the central radiation with a two-phase gas density torus (Fig.~\ref{fig:model}, bottom panel). In the first phase the lower density gas (higher ionization parameter) is blown in form of [Si\,{\sc vi}] clouds in the direction of the radiation pressure from the central source. In the second phase the energy which penetrates through the higher density gas (lower ionization parameter) leads to the occurrence of a thermal molecular wind which ejects material in the direction normal to the surface of the torus (ending up nearly perpendicular to the maser disc).
		\\
		\item We detected a central H$_{2}$ disc-like structure, with PA=132\textdegree$\pm$10\textdegree~and $r$=8$\pm$2\,pc, without signs of rotation. Its orientation agrees with the one found for the H$_{2}$O maser emission, with PA=135\textdegree~and others PAs associated with the torus (Sec.~\ref{sec:torus}). 
		Given the jet PA before bending (11\textdegree), which is supposedly perpendicular to the accretion disc, the planes occupied by the H$_{2}$ emission and the accretion disc are misaligned by 31\textdegree. 
		\\
		\item Differently from the molecular gas flow feeding the nucleus proposed by \citet{Muller09}, we suggest that there is an additional mechanism provided mostly by the flow of neutral atomic gas in the shadow of the molecular/dusty torus, that transfers gas from the hot bubble to the torus. 
		\end{enumerate}

    \section*{Acknowledgments}

    A general acknowledgment is made of all institutions and the maintainers of the massive public data archive currently available to the scientific community. Based on observations collected at the European Organisation for Astronomical Research in the Southern Hemisphere under ESO programme 076.B-0098(A). Additional data were also obtained from the Gemini Observatory, which is operated by the Association of Universities for Research in Astronomy, Inc., under a cooperation agreement with the NSF on behalf of the Gemini partnership. We also made use of NASA's Astrophysics Data System Bibliographic Services and the NASA/IPAC Extragalactic Database (NED), which is operated by the Jet Propulsion Laboratory, California Institute of Technology, under contract with the National Aeronautics and Space Administration. This work is also based on observations made with the NASA/ESA \textsc{Hubble Space Telescope} obtained from the Space Telescope Institute, which is operated by the Association of Universities for Research in Astronomy, Inc., under NASA contract NAS5-26555. Finally, we would like to thank Menezes R.B. for the carefully reading and insightful suggestions, the referee, who led us to a deeper and more consistent discussion and FAPESP (Funda\'c\~ao de Amparo \`a Pesquisa do Estado de S\~ao Paulo) for support under grants 2011/19824-8 (DMN) and 2011/51680-6 (JES).

    \bibliographystyle{mnras}
    \bibliography{biblio}

\begin{thebibliography}{}
\makeatletter
\relax
\def\mn@urlcharsother{\let\do\@makeother \do\$\do\&\do\#\do\^\do\_\do\%\do\~}
\def\mn@doi{\begingroup\mn@urlcharsother \@ifnextchar [ {\mn@doi@}
  {\mn@doi@[]}}
\def\mn@doi@[#1]#2{\def\@tempa{#1}\ifx\@tempa\@empty \href
  {http://dx.doi.org/#2} {doi:#2}\else \href {http://dx.doi.org/#2} {#1}\fi
  \endgroup}
\def\mn@eprint#1#2{\mn@eprint@#1:#2::\@nil}
\def\mn@eprint@arXiv#1{\href {http://arxiv.org/abs/#1} {{\tt arXiv:#1}}}
\def\mn@eprint@dblp#1{\href {http://dblp.uni-trier.de/rec/bibtex/#1.xml}
  {dblp:#1}}
\def\mn@eprint@#1:#2:#3:#4\@nil{\def\@tempa {#1}\def\@tempb {#2}\def\@tempc
  {#3}\ifx \@tempc \@empty \let \@tempc \@tempb \let \@tempb \@tempa \fi \ifx
  \@tempb \@empty \def\@tempb {arXiv}\fi \@ifundefined
  {mn@eprint@\@tempb}{\@tempb:\@tempc}{\expandafter \expandafter \csname
  mn@eprint@\@tempb\endcsname \expandafter{\@tempc}}}

\bibitem[\protect\citeauthoryear{{Antonucci}}{{Antonucci}}{1993}]{Antonucci93}
{Antonucci} R.,  1993, \mn@doi [\araa] {10.1146/annurev.aa.31.090193.002353},
  \href {http://adsabs.harvard.edu/abs/1993ARA%26A..31..473A} {31, 473}

\bibitem[\protect\citeauthoryear{{Antonucci} \& {Miller}}{{Antonucci} \&
  {Miller}}{1985}]{Antonucci85}
{Antonucci} R.~R.~J.,  {Miller} J.~S.,  1985, \mn@doi [\apj] {10.1086/163559},
  \href {http://adsabs.harvard.edu/abs/1985ApJ...297..621A} {297, 621}

\bibitem[\protect\citeauthoryear{{Axon}, {Marconi}, {Capetti}, {Macchetto},
  {Schreier}  \& {Robinson}}{{Axon} et~al.}{1998}]{Axon98}
{Axon} D.~J.,  {Marconi} A.,  {Capetti} A.,  {Macchetto} F.~D.,  {Schreier} E.,
    {Robinson} A.,  1998, \mn@doi [\apjl] {10.1086/311249}, \href
  {http://adsabs.harvard.edu/abs/1998ApJ...496L..75A} {496, L75}

\bibitem[\protect\citeauthoryear{{Barbosa}, {Storchi-Bergmann}, {McGregor},
  {Vale}  \& {Rogemar Riffel}}{{Barbosa} et~al.}{2014}]{Barbosa14}
{Barbosa} F.~K.~B.,  {Storchi-Bergmann} T.,  {McGregor} P.,  {Vale} T.~B.,
  {Rogemar Riffel} A.,  2014, \mn@doi [\mnras] {10.1093/mnras/stu1637}, \href
  {http://adsabs.harvard.edu/abs/2014MNRAS.445.2353B} {445, 2353}

\bibitem[\protect\citeauthoryear{{Bicknell}, {Dopita}, {Tsvetanov}  \&
  {Sutherland}}{{Bicknell} et~al.}{1998}]{Bicknell98}
{Bicknell} G.~V.,  {Dopita} M.~A.,  {Tsvetanov} Z.~I.,   {Sutherland} R.~S.,
  1998, \mn@doi [\apj] {10.1086/305336}, \href
  {http://adsabs.harvard.edu/abs/1998ApJ...495..680B} {495, 680}

\bibitem[\protect\citeauthoryear{{B{\^i}rzan}, {McNamara}, {Nulsen}, {Carilli}
  \& {Wise}}{{B{\^i}rzan} et~al.}{2008}]{Birzan08}
{B{\^i}rzan} L.,  {McNamara} B.~R.,  {Nulsen} P.~E.~J.,  {Carilli} C.~L.,
  {Wise} M.~W.,  2008, \mn@doi [\apj] {10.1086/591416}, \href
  {http://adsabs.harvard.edu/abs/2008ApJ...686..859B} {686, 859}

\bibitem[\protect\citeauthoryear{{Bock} et~al.,}{{Bock} et~al.}{2000}]{Bock00}
{Bock} J.~J.,  et~al., 2000, \mn@doi [\aj] {10.1086/316871}, \href
  {http://adsabs.harvard.edu/abs/2000AJ....120.2904B} {120, 2904}

\bibitem[\protect\citeauthoryear{{Bonnet} et~al.,}{{Bonnet}
  et~al.}{2004}]{Bonnet04}
{Bonnet} H.,  et~al., 2004, The Messenger, \href
  {http://adsabs.harvard.edu/abs/2004Msngr.117...17B} {117, 17}

\bibitem[\protect\citeauthoryear{{Capetti}, {Axon}, {Macchetto}, {Sparks}  \&
  {Boksenberg}}{{Capetti} et~al.}{1996}]{Capetti96a}
{Capetti} A.,  {Axon} D.~J.,  {Macchetto} F.,  {Sparks} W.~B.,   {Boksenberg}
  A.,  1996, \mn@doi [\apj] {10.1086/177804}, \href
  {http://adsabs.harvard.edu/abs/1996ApJ...469..554C} {469, 554}

\bibitem[\protect\citeauthoryear{{Capetti}, {Axon}, {Macchetto}, {Marconi}  \&
  {Winge}}{{Capetti} et~al.}{1999}]{Capetti99}
{Capetti} A.,  {Axon} D.~J.,  {Macchetto} F.~D.,  {Marconi} A.,   {Winge} C.,
  1999, \mn@doi [\apj] {10.1086/307099}, \href
  {http://adsabs.harvard.edu/abs/1999ApJ...516..187C} {516, 187}

\bibitem[\protect\citeauthoryear{{Cecil}, {Bland}  \& {Tully}}{{Cecil}
  et~al.}{1990}]{Cecil90}
{Cecil} G.,  {Bland} J.,   {Tully} R.~B.,  1990, \mn@doi [\apj]
  {10.1086/168742}, \href {http://adsabs.harvard.edu/abs/1990ApJ...355...70C}
  {355, 70}

\bibitem[\protect\citeauthoryear{{Crenshaw} \& {Kraemer}}{{Crenshaw} \&
  {Kraemer}}{2000}]{Crenshaw00}
{Crenshaw} D.~M.,  {Kraemer} S.~B.,  2000, \mn@doi [\apjl] {10.1086/312581},
  \href {http://adsabs.harvard.edu/abs/2000ApJ...532L.101C} {532, L101}

\bibitem[\protect\citeauthoryear{{Das}, {Crenshaw}, {Kraemer}  \& {Deo}}{{Das}
  et~al.}{2006}]{Das06}
{Das} V.,  {Crenshaw} D.~M.,  {Kraemer} S.~B.,   {Deo} R.~P.,  2006, \mn@doi
  [\aj] {10.1086/504899}, \href
  {http://adsabs.harvard.edu/abs/2006AJ....132..620D} {132, 620}

\bibitem[\protect\citeauthoryear{{Davies} et~al.,}{{Davies}
  et~al.}{2014}]{Davies14792}
{Davies} R.~I.,  et~al., 2014, \mn@doi [\apj] {10.1088/0004-637X/792/2/101},
  \href {http://adsabs.harvard.edu/abs/2014ApJ...792..101D} {792, 101}

\bibitem[\protect\citeauthoryear{{Dopita} \& {Sutherland}}{{Dopita} \&
  {Sutherland}}{1996}]{Dopita96}
{Dopita} M.~A.,  {Sutherland} R.~S.,  1996, \mn@doi [\apjs] {10.1086/192255},
  \href {http://adsabs.harvard.edu/abs/1996ApJS..102..161D} {102, 161}

\bibitem[\protect\citeauthoryear{{Eisenhauer} et~al.,}{{Eisenhauer}
  et~al.}{2003}]{Eisenhauer03}
{Eisenhauer} F.,  et~al., 2003, in {Iye} M.,  {Moorwood} A.~F.~M.,  eds,
  \procspie Vol. 4841, Instrument Design and Performance for Optical/Infrared
  Ground-based Telescopes. pp 1548--1561 (\mn@eprint {} {astro-ph/0306191}),
  \mn@doi{10.1117/12.459468}

\bibitem[\protect\citeauthoryear{{Elitzur} \& {Netzer}}{{Elitzur} \&
  {Netzer}}{2016}]{Elitzur16}
{Elitzur} M.,  {Netzer} H.,  2016, \mn@doi [\mnras] {10.1093/mnras/stw657},
  \href {http://adsabs.harvard.edu/abs/2016MNRAS.459..585E} {459, 585}

\bibitem[\protect\citeauthoryear{{Elitzur} \& {Shlosman}}{{Elitzur} \&
  {Shlosman}}{2006}]{Elitzur06}
{Elitzur} M.,  {Shlosman} I.,  2006, \mn@doi [\apjl] {10.1086/508158}, \href
  {http://adsabs.harvard.edu/abs/2006ApJ...648L.101E} {648, L101}

\bibitem[\protect\citeauthoryear{{Evans}, {Ford}, {Kinney}, {Antonucci},
  {Armus}  \& {Caganoff}}{{Evans} et~al.}{1991}]{Evans91}
{Evans} I.~N.,  {Ford} H.~C.,  {Kinney} A.~L.,  {Antonucci} R.~R.~J.,  {Armus}
  L.,   {Caganoff} S.,  1991, \mn@doi [\apjl] {10.1086/185951}, \href
  {http://adsabs.harvard.edu/abs/1991ApJ...369L..27E} {369, L27}

\bibitem[\protect\citeauthoryear{{Filippenko}}{{Filippenko}}{1982}]{Filippenko82}
{Filippenko} A.~V.,  1982, \mn@doi [\pasp] {10.1086/131052}, \href
  {http://adsabs.harvard.edu/abs/1982PASP...94..715F} {94, 715}

\bibitem[\protect\citeauthoryear{{Fraquelli}, {Storchi-Bergmann}  \&
  {Levenson}}{{Fraquelli} et~al.}{2003}]{Fraquelli03}
{Fraquelli} H.~A.,  {Storchi-Bergmann} T.,   {Levenson} N.~A.,  2003, \mn@doi
  [\mnras] {10.1046/j.1365-8711.2003.06397.x}, \href
  {http://adsabs.harvard.edu/abs/2003MNRAS.341..449F} {341, 449}

\bibitem[\protect\citeauthoryear{{Galliano} \& {Alloin}}{{Galliano} \&
  {Alloin}}{2002}]{Galliano02}
{Galliano} E.,  {Alloin} D.,  2002, in {Bergeron} J.,  {Monnet} G.,  eds,
  Scientific Drivers for ESO Future VLT/VLTI Instrumentation. p.~175,
  \mn@doi{10.1007/10857019_26}

\bibitem[\protect\citeauthoryear{{Gallimore}, {Baum}, {O'Dea}  \&
  {Pedlar}}{{Gallimore} et~al.}{1996a}]{Gallimore96c}
{Gallimore} J.~F.,  {Baum} S.~A.,  {O'Dea} C.~P.,   {Pedlar} A.,  1996a,
  \mn@doi [\apj] {10.1086/176798}, \href
  {http://adsabs.harvard.edu/abs/1996ApJ...458..136G} {458, 136}

\bibitem[\protect\citeauthoryear{{Gallimore}, {Baum}, {O'Dea}, {Brinks}  \&
  {Pedlar}}{{Gallimore} et~al.}{1996b}]{Gallimore96b}
{Gallimore} J.~F.,  {Baum} S.~A.,  {O'Dea} C.~P.,  {Brinks} E.,   {Pedlar} A.,
  1996b, \mn@doi [\apj] {10.1086/177187}, \href
  {http://adsabs.harvard.edu/abs/1996ApJ...462..740G} {462, 740}

\bibitem[\protect\citeauthoryear{{Gallimore}, {Baum}  \& {O'Dea}}{{Gallimore}
  et~al.}{1996c}]{Gallimore96}
{Gallimore} J.~F.,  {Baum} S.~A.,   {O'Dea} C.~P.,  1996c, \mn@doi [\apj]
  {10.1086/177311}, \href {http://adsabs.harvard.edu/abs/1996ApJ...464..198G}
  {464, 198}

\bibitem[\protect\citeauthoryear{{Gallimore}, {Henkel}, {Baum}, {Glass},
  {Claussen}, {Prieto}  \& {Von Kap-herr}}{{Gallimore}
  et~al.}{2001}]{Gallimore01}
{Gallimore} J.~F.,  {Henkel} C.,  {Baum} S.~A.,  {Glass} I.~S.,  {Claussen}
  M.~J.,  {Prieto} M.~A.,   {Von Kap-herr} A.,  2001, \mn@doi [\apj]
  {10.1086/321616}, \href {http://adsabs.harvard.edu/abs/2001ApJ...556..694G}
  {556, 694}

\bibitem[\protect\citeauthoryear{{Gallimore}, {Baum}  \& {O'Dea}}{{Gallimore}
  et~al.}{2004}]{Gallimore04}
{Gallimore} J.~F.,  {Baum} S.~A.,   {O'Dea} C.~P.,  2004, \mn@doi [\apj]
  {10.1086/423167}, \href {http://adsabs.harvard.edu/abs/2004ApJ...613..794G}
  {613, 794}

\bibitem[\protect\citeauthoryear{{Gallimore} et~al.,}{{Gallimore}
  et~al.}{2016}]{Gallimore16}
{Gallimore} J.~F.,  et~al., 2016, \mn@doi [\apjl] {10.3847/2041-8205/829/1/L7},
  \href {http://adsabs.harvard.edu/abs/2016ApJ...829L...7G} {829, L7}

\bibitem[\protect\citeauthoryear{{Garc{\'{\i}}a-Burillo}
  et~al.,}{{Garc{\'{\i}}a-Burillo} et~al.}{2014}]{Burillo14}
{Garc{\'{\i}}a-Burillo} S.,  et~al., 2014, \mn@doi [\aap]
  {10.1051/0004-6361/201423843}, \href
  {http://adsabs.harvard.edu/abs/2014A%26A...567A.125G} {567, A125}

\bibitem[\protect\citeauthoryear{{Garcia-Burillo} et~al.,}{{Garcia-Burillo}
  et~al.}{2016}]{Burillo16}
{Garcia-Burillo} S.,  et~al., 2016, preprint, \href
  {http://adsabs.harvard.edu/abs/2016arXiv160400205G} {} (\mn@eprint {arXiv}
  {1604.00205})

\bibitem[\protect\citeauthoryear{{Genzel} et~al.,}{{Genzel}
  et~al.}{2014}]{Genzel14796}
{Genzel} R.,  et~al., 2014, \mn@doi [\apj] {10.1088/0004-637X/796/1/7}, \href
  {http://adsabs.harvard.edu/abs/2014ApJ...796....7G} {796, 7}

\bibitem[\protect\citeauthoryear{{Gratadour}, {Rouan}, {Grosset}, {Boccaletti}
  \& {Cl{\'e}net}}{{Gratadour} et~al.}{2015}]{Gratadour15}
{Gratadour} D.,  {Rouan} D.,  {Grosset} L.,  {Boccaletti} A.,   {Cl{\'e}net}
  Y.,  2015, \mn@doi [\aap] {10.1051/0004-6361/201526554}, \href
  {http://adsabs.harvard.edu/abs/2015A%26A...581L...8G} {581, L8}

\bibitem[\protect\citeauthoryear{{Greenhill}, {Gwinn}, {Antonucci}  \&
  {Barvainis}}{{Greenhill} et~al.}{1996}]{Greenhill96}
{Greenhill} L.~J.,  {Gwinn} C.~R.,  {Antonucci} R.,   {Barvainis} R.,  1996,
  \mn@doi [\apjl] {10.1086/310346}, \href
  {http://adsabs.harvard.edu/abs/1996ApJ...472L..21G} {472, L21}

\bibitem[\protect\citeauthoryear{{Hopkins} \& {Elvis}}{{Hopkins} \&
  {Elvis}}{2010}]{Elvis10}
{Hopkins} P.~F.,  {Elvis} M.,  2010, \mn@doi [\mnras]
  {10.1111/j.1365-2966.2009.15643.x}, \href
  {http://adsabs.harvard.edu/abs/2010MNRAS.401....7H} {401, 7}

\bibitem[\protect\citeauthoryear{{Imanishi}, {Nakanishi}  \&
  {Izumi}}{{Imanishi} et~al.}{2016}]{Imanishi16}
{Imanishi} M.,  {Nakanishi} K.,   {Izumi} T.,  2016, \mn@doi [\apjl]
  {10.3847/2041-8205/822/1/L10}, \href
  {http://adsabs.harvard.edu/abs/2016ApJ...822L..10I} {822, L10}

\bibitem[\protect\citeauthoryear{{Izumi}, {Nakanishi}, {Imanishi}  \&
  {Kohno}}{{Izumi} et~al.}{2016}]{Izumi16}
{Izumi} T.,  {Nakanishi} K.,  {Imanishi} M.,   {Kohno} K.,  2016, \mn@doi
  [\mnras] {10.1093/mnras/stw324}, \href
  {http://adsabs.harvard.edu/abs/2016MNRAS.459.3629I} {459, 3629}

\bibitem[\protect\citeauthoryear{{Lepp} \& {McCray}}{{Lepp} \&
  {McCray}}{1983}]{Lepp83}
{Lepp} S.,  {McCray} R.,  1983, \mn@doi [\apj] {10.1086/161062}, \href
  {http://adsabs.harvard.edu/abs/1983ApJ...269..560L} {269, 560}

\bibitem[\protect\citeauthoryear{{L{\'o}pez-Gonzaga}, {Jaffe}, {Burtscher},
  {Tristram}  \& {Meisenheimer}}{{L{\'o}pez-Gonzaga} et~al.}{2014}]{Gonzaga14}
{L{\'o}pez-Gonzaga} N.,  {Jaffe} W.,  {Burtscher} L.,  {Tristram} K.~R.~W.,
  {Meisenheimer} K.,  2014, \mn@doi [\aap] {10.1051/0004-6361/201323002}, \href
  {http://adsabs.harvard.edu/abs/2014A%26A...565A..71L} {565, A71}

\bibitem[\protect\citeauthoryear{{May}, {Steiner}, {Ricci}, {Menezes}  \&
  {Andrade}}{{May} et~al.}{2016}]{Dmay16}
{May} D.,  {Steiner} J.~E.,  {Ricci} T.~V.,  {Menezes} R.~B.,   {Andrade}
  I.~S.,  2016, \mn@doi [\mnras] {10.1093/mnras/stv2929}, \href
  {http://adsabs.harvard.edu/abs/2016MNRAS.457..949M} {457, 949}

\bibitem[\protect\citeauthoryear{{Mazzalay}, {Rodr{\'{\i}}guez-Ardila},
  {Komossa}  \& {McGregor}}{{Mazzalay} et~al.}{2013}]{Mazzalay13a}
{Mazzalay} X.,  {Rodr{\'{\i}}guez-Ardila} A.,  {Komossa} S.,   {McGregor}
  P.~J.,  2013, \mn@doi [\mnras] {10.1093/mnras/stt064}, \href
  {http://adsabs.harvard.edu/abs/2013MNRAS.430.2411M} {430, 2411}

\bibitem[\protect\citeauthoryear{{McGregor} et~al.,}{{McGregor}
  et~al.}{2003}]{McGregor03}
{McGregor} P.~J.,  et~al., 2003, in {Iye} M.,  {Moorwood} A.~F.~M.,  eds,
  Society of Photo-Optical Instrumentation Engineers (SPIE) Conference Series
  Vol. 4841, Instrument Design and Performance for Optical/Infrared
  Ground-based Telescopes. pp 1581--1591, \mn@doi{10.1117/12.459448}

\bibitem[\protect\citeauthoryear{{Menezes}, {Steiner}  \& {Ricci}}{{Menezes}
  et~al.}{2014}]{Menezes14}
{Menezes} R.~B.,  {Steiner} J.~E.,   {Ricci} T.~V.,  2014, \mn@doi [\mnras]
  {10.1093/mnras/stt2381}, \href
  {http://adsabs.harvard.edu/abs/2014MNRAS.438.2597M} {438, 2597}

\bibitem[\protect\citeauthoryear{{Menezes}, {da Silva}, {Ricci}, {Steiner},
  {May}  \& {Borges}}{{Menezes} et~al.}{2015}]{Menezes15}
{Menezes} R.~B.,  {da Silva} P.,  {Ricci} T.~V.,  {Steiner} J.~E.,  {May} D.,
  {Borges} B.~W.,  2015, \mn@doi [\mnras] {10.1093/mnras/stv629}, \href
  {http://adsabs.harvard.edu/abs/2015MNRAS.450..369M} {450, 369}

\bibitem[\protect\citeauthoryear{{Moore} \& {Cohen}}{{Moore} \&
  {Cohen}}{1996}]{Moore96}
{Moore} D.,  {Cohen} R.~D.,  1996, \mn@doi [\apj] {10.1086/177868}, \href
  {http://adsabs.harvard.edu/abs/1996ApJ...470..301M} {470, 301}

\bibitem[\protect\citeauthoryear{{M{\"u}ller S{\'a}nchez}, {Davies}, {Genzel},
  {Tacconi}, {Eisenhauer}, {Hicks}, {Friedrich}  \& {Sternberg}}{{M{\"u}ller
  S{\'a}nchez} et~al.}{2009}]{Muller09}
{M{\"u}ller S{\'a}nchez} F.,  {Davies} R.~I.,  {Genzel} R.,  {Tacconi} L.~J.,
  {Eisenhauer} F.,  {Hicks} E.~K.~S.,  {Friedrich} S.,   {Sternberg} A.,  2009,
  \mn@doi [\apj] {10.1088/0004-637X/691/1/749}, \href
  {http://adsabs.harvard.edu/abs/2009ApJ...691..749M} {691, 749}

\bibitem[\protect\citeauthoryear{{M{\"u}ller-S{\'a}nchez}, {Prieto}, {Hicks},
  {Vives-Arias}, {Davies}, {Malkan}, {Tacconi}  \&
  {Genzel}}{{M{\"u}ller-S{\'a}nchez} et~al.}{2011}]{Muller11}
{M{\"u}ller-S{\'a}nchez} F.,  {Prieto} M.~A.,  {Hicks} E.~K.~S.,  {Vives-Arias}
  H.,  {Davies} R.~I.,  {Malkan} M.,  {Tacconi} L.~J.,   {Genzel} R.,  2011,
  \mn@doi [\apj] {10.1088/0004-637X/739/2/69}, \href
  {http://adsabs.harvard.edu/abs/2011ApJ...739...69M} {739, 69}

\bibitem[\protect\citeauthoryear{{Raban}, {Jaffe}, {R{\"o}ttgering},
  {Meisenheimer}  \& {Tristram}}{{Raban} et~al.}{2009}]{Raban09}
{Raban} D.,  {Jaffe} W.,  {R{\"o}ttgering} H.,  {Meisenheimer} K.,   {Tristram}
  K.~R.~W.,  2009, \mn@doi [\mnras] {10.1111/j.1365-2966.2009.14439.x}, \href
  {http://adsabs.harvard.edu/abs/2009MNRAS.394.1325R} {394, 1325}

\bibitem[\protect\citeauthoryear{{Riffel}, {Storchi-Bergmann}, {Winge},
  {McGregor}, {Beck}  \& {Schmitt}}{{Riffel} et~al.}{2008}]{Riffel08}
{Riffel} R.~A.,  {Storchi-Bergmann} T.,  {Winge} C.,  {McGregor} P.~J.,  {Beck}
  T.,   {Schmitt} H.,  2008, \mn@doi [\mnras]
  {10.1111/j.1365-2966.2008.12936.x}, \href
  {http://adsabs.harvard.edu/abs/2008MNRAS.385.1129R} {385, 1129}

\bibitem[\protect\citeauthoryear{{Riffel}, {Vale}, {Storchi-Bergmann}  \&
  {McGregor}}{{Riffel} et~al.}{2014}]{Riffel14b}
{Riffel} R.~A.,  {Vale} T.~B.,  {Storchi-Bergmann} T.,   {McGregor} P.~J.,
  2014, \mn@doi [\mnras] {10.1093/mnras/stu843}, \href
  {http://adsabs.harvard.edu/abs/2014MNRAS.442..656R} {442, 656}

\bibitem[\protect\citeauthoryear{{Rodr{\'{\i}}guez-Ardila}, {Pastoriza},
  {Viegas}, {Sigut}  \& {Pradhan}}{{Rodr{\'{\i}}guez-Ardila}
  et~al.}{2004}]{Ardila04}
{Rodr{\'{\i}}guez-Ardila} A.,  {Pastoriza} M.~G.,  {Viegas} S.,  {Sigut}
  T.~A.~A.,   {Pradhan} A.~K.,  2004, \mn@doi [\aap]
  {10.1051/0004-6361:20034285}, \href
  {http://adsabs.harvard.edu/abs/2004A%26A...425..457R} {425, 457}

\bibitem[\protect\citeauthoryear{{Rodr{\'{\i}}guez-Ardila}, {Riffel}  \&
  {Pastoriza}}{{Rodr{\'{\i}}guez-Ardila} et~al.}{2005}]{Ardila05}
{Rodr{\'{\i}}guez-Ardila} A.,  {Riffel} R.,   {Pastoriza} M.~G.,  2005, \mn@doi
  [\mnras] {10.1111/j.1365-2966.2005.09638.x}, \href
  {http://adsabs.harvard.edu/abs/2005MNRAS.364.1041R} {364, 1041}

\bibitem[\protect\citeauthoryear{{Rodr{\'{\i}}guez-Ardila}, {Prieto},
  {Portilla}  \& {Tejeiro}}{{Rodr{\'{\i}}guez-Ardila} et~al.}{2011}]{Ardila11}
{Rodr{\'{\i}}guez-Ardila} A.,  {Prieto} M.~A.,  {Portilla} J.~G.,   {Tejeiro}
  J.~M.,  2011, \mn@doi [\apj] {10.1088/0004-637X/743/2/100}, \href
  {http://adsabs.harvard.edu/abs/2011ApJ...743..100R} {743, 100}

\bibitem[\protect\citeauthoryear{{Scoville}, {Hall}, {Ridgway}  \&
  {Kleinmann}}{{Scoville} et~al.}{1982}]{Scoville82}
{Scoville} N.~Z.,  {Hall} D.~N.~B.,  {Ridgway} S.~T.,   {Kleinmann} S.~G.,
  1982, \mn@doi [\apj] {10.1086/159618}, \href
  {http://adsabs.harvard.edu/abs/1982ApJ...253..136S} {253, 136}

\bibitem[\protect\citeauthoryear{{Steiner}, {Menezes}, {Ricci}  \&
  {Oliveira}}{{Steiner} et~al.}{2009}]{Steiner09}
{Steiner} J.~E.,  {Menezes} R.~B.,  {Ricci} T.~V.,   {Oliveira} A.~S.,  2009,
  \mn@doi [\mnras] {10.1111/j.1365-2966.2009.14530.x}, \href
  {http://adsabs.harvard.edu/abs/2009MNRAS.395...64S} {395, 64}

\bibitem[\protect\citeauthoryear{{Storchi-Bergmann}, {Riffel}, {Riffel},
  {Diniz}, {Borges Vale}  \& {McGregor}}{{Storchi-Bergmann}
  et~al.}{2012}]{Thaisa12}
{Storchi-Bergmann} T.,  {Riffel} R.~A.,  {Riffel} R.,  {Diniz} M.~R.,  {Borges
  Vale} T.,   {McGregor} P.~J.,  2012, \mn@doi [\apj]
  {10.1088/0004-637X/755/2/87}, \href
  {http://adsabs.harvard.edu/abs/2012ApJ...755...87S} {755, 87}

\bibitem[\protect\citeauthoryear{{Tully} \& {Fisher}}{{Tully} \&
  {Fisher}}{1988}]{Tully88}
{Tully} R.~B.,  {Fisher} J.~R.,  1988, {Catalog of Nearby Galaxies}

\bibitem[\protect\citeauthoryear{{Urry} \& {Padovani}}{{Urry} \&
  {Padovani}}{1995}]{Urry95}
{Urry} C.~M.,  {Padovani} P.,  1995, \mn@doi [\pasp] {10.1086/133630}, \href
  {http://adsabs.harvard.edu/abs/1995PASP..107..803U} {107, 803}

\bibitem[\protect\citeauthoryear{{Veilleux}, {Tully}  \&
  {Bland-Hawthorn}}{{Veilleux} et~al.}{1991}]{Veilleux91}
{Veilleux} S.,  {Tully} R.~B.,   {Bland-Hawthorn} J.,  1991, in Bulletin of the
  American Astronomical Society. pp 1428--+

\bibitem[\protect\citeauthoryear{{Wang}, {Fabbiano}, {Karovska}, {Elvis}  \&
  {Risaliti}}{{Wang} et~al.}{2012}]{Wang12}
{Wang} J.,  {Fabbiano} G.,  {Karovska} M.,  {Elvis} M.,   {Risaliti} G.,  2012,
  \mn@doi [\apj] {10.1088/0004-637X/756/2/180}, \href
  {http://adsabs.harvard.edu/abs/2012ApJ...756..180W} {756, 180}

\bibitem[\protect\citeauthoryear{{Wilson} \& {Ulvestad}}{{Wilson} \&
  {Ulvestad}}{1983}]{Wilson83}
{Wilson} A.~S.,  {Ulvestad} J.~S.,  1983, \mn@doi [\apj] {10.1086/161507},
  \href {http://adsabs.harvard.edu/abs/1983ApJ...275....8W} {275, 8}

\bibitem[\protect\citeauthoryear{{Zubovas} \& {Nayakshin}}{{Zubovas} \&
  {Nayakshin}}{2014}]{Zubovas14}
{Zubovas} K.,  {Nayakshin} S.,  2014, \mn@doi [\mnras] {10.1093/mnras/stu431},
  \href {http://adsabs.harvard.edu/abs/2014MNRAS.440.2625Z} {440, 2625}

\makeatother
\end{thebibliography}



    \label{lastpage}

    \end{document}